\documentclass[11pt]{article}
\usepackage{amssymb,amsmath}
\usepackage{bm} 
\usepackage{booktabs} 
\usepackage{array}
\usepackage{latexsym}
\usepackage{graphicx}
\usepackage{color}
\usepackage{datetime}
\usepackage[nosort]{cite}
\usepackage{verbatim}
\usepackage{enumerate}
\usepackage{enumitem}
\usepackage{chngpage} 
\usepackage{mathrsfs}
\usepackage{euscript}
\usepackage{psfrag}
\usepackage{bbold}

\usepackage{mciteplus}

\usepackage[colorlinks=true,      linkcolor=blue,      urlcolor=blue,
filecolor=blue,      citecolor=blue,       pdfstartview=FitH,
pdfpagemode=UseNone,      bookmarksopen=true]{hyperref}  
\usepackage[all]{hypcap}     

\usepackage{tensor}
\usepackage{physics}
\usepackage{tikz}
\usepackage{mathtools}
\usepackage[normalem]{ulem}

\definecolor{myPurple}{rgb}{0.6,0.0,0.6}

\renewcommand{\comm}[1]{} 

\DeclareMathOperator*{\arctanh}{arctanh}

\def\({\left(}
\def\){\right)}
\def\[{\left[}
\def\]{\right]}

\def\coeff#1#2{{\textstyle \frac{#1}{#2}}}

\def\One{{\hbox{ 1\kern-.8mm l}}}

\def\barray{\begin{array}}
	\def\earray{\end{array}}
\def\be{\begin{equation}}
	\def\ee{\end{equation}}
\def\bea{\begin{eqnarray}}
	\def\eea{\end{eqnarray}}
\def\bal{\begin{align}}
	\def\eal{\end{align}}

\def\nBPS#1{$\frac{1}{#1}$-BPS}

\numberwithin{equation}{section} 

\definecolor{cardinal}{rgb}{0.6,0,0}
\definecolor{darkgreen}{rgb}{0,0.4,0}
\definecolor{golden}{rgb}{0.92, 0.7, 0}
\definecolor{midnight}{rgb}{0, 0, 0.5}
\definecolor{darkblue}{rgb}{0, 0, 0.7}
\definecolor{darkred}{rgb}{0.6, 0, 0}
\definecolor{purple}{rgb}{0.5, 0, 0.5}

\def\oneone{\rlap 1\mkern4mu{\rm l}}

\def\Neql#1{{\cal N}\!=\!{#1}}

\def\IR{\mathbb{R}}

\def\IT{\mathbb{T}}

\def\ZZ{\mathbb{Z}}

\def\cD{{\cal D}}

\def\cL{{\cal L}}

\def\cX{{\cal X}}

\def\scrC{{ \mathscr{C}}}

\def\nBPS#1{$\frac{1}{#1}$-BPS}

\def\RAdS{{R_{AdS}}}
\def\RAdSsq{{R^2_{AdS}}}

\def\contparama{{\zeta_1}}
\def\contparamb{{\zeta_2}}
\def\contparam#1{{\zeta_{#1}}}

\def\muscal{{\rho}}
\def\param1{{\sigma}} 

\allowdisplaybreaks
\begin{document}
	\font\cmss=cmss10 \font\cmsss=cmss10 at 7pt
	
	\hfill
	\vspace{18pt}
	\begin{center}
		{\Large 
			\textbf{  Microstrata }}
		
	\end{center}

	\vspace{8pt}
	\begin{center}
		{\textsl{Bogdan Ganchev$^{\,a}$, Stefano Giusto$^{\,b, c}$, Anthony Houppe$^{\,a}$, \\
				Rodolfo Russo$^{\,d}$ and   Nicholas P. Warner$^{\,a\,, e\,, f}$}}
		
		\vspace{1cm}
		
		\textit{\small ${}^a$Université Paris-Saclay, CNRS, CEA, Institut de physique théorique,\\
			91191, Gif-sur-Yvette, France.} \\  \vspace{6pt}

		\textit{\small ${}^b$Dipartimento di Fisica,  Universit\`a di Genova, Via Dodecaneso 33, 16146, Genoa, Italy.} \\  \vspace{6pt}
		
		\textit{\small ${}^c$I.N.F.N. Sezione di Genova,
			Via Dodecaneso 33, 16146, Genoa, Italy.}\\
		\vspace{6pt}
		
		\textit{\small ${}^d$Centre for Theoretical Physics, Department of Physics and Astronomy,\\
			Queen Mary University of London,
			Mile End Road, London, E1 4NS,
			United Kingdom.}\\
		\vspace{6pt}
		
		\textit{\small {${}^e$Department of Physics and Astronomy}
			and ${}^f$Department of Mathematics,\\ 
			University of Southern California,  
			Los Angeles, CA 90089, USA}

	\end{center}

	\vspace{12pt}
	
	\begin{center}
		\textbf{Abstract}
	\end{center}
	
	\vspace{4pt} {\small
		\noindent 
		Microstrata are the non-extremal analogues of superstrata: they are smooth, non-extremal (non-BPS) solitonic solutions to IIB supergravity whose deep-throat limits approximate black holes. Using perturbation theory and numerical methods, we construct families of  solutions  using a consistent truncation to three-dimensional supergravity. The most general families presented here  involve two continuous parameters, or amplitudes, and four quantized parameters that set the angular momenta and energy levels.  Our solutions are asymptotic to the vacuum of the D1-D5 system: AdS$_3$ $\times S^3 \times \IT^4$. Using holography, we show that the they are dual to multi-particle states in the D1-D5 CFT involving a large number of mutually non-BPS supergravitons and we determine the anomalous dimensions of these states from the binding energies in supergravity.  These binding energies are uniformly negative and depend non-linearly on the amplitudes of the states. In one family of solutions, smoothness restricts some of the fields to lie on a special locus of the parameter space. Using precision holography we show that this special locus can be identified with the multi-particle states constructed via the standard OPE of the single-particle constituents. Our numerical analysis shows that microstrata are robust at large amplitudes and the solutions can be obtained to very high precision.}

		\vspace{1cm}

		\thispagestyle{empty}

		\vfill
		\vskip 5.mm
		\hrule width 5.cm
		\vskip 2.mm
		{
			\noindent {\scriptsize e-mails: bogdan.ganchev@ipht.fr, stefano.giusto@ge.infn.it, anthony.houppe@ipht.fr,   r.russo@qmul.ac.uk, warner@usc.edu }
		}

		\setcounter{footnote}{0}
		\setcounter{page}{0}

		
		\baselineskip=17pt
		\parskip=5pt
		
		\newpage
		
		\setcounter{tocdepth}{2}
		\tableofcontents
		
		\baselineskip=15pt
		\parskip=3pt
		
		
		\section{Introduction}
		\label{sec:Intro}
		
		While the original motivation for the microstate geometry program came from the black-hole information problem, the study of microstate geometries has led to one of the most remarkable and well-mapped-out holographic dualities, so much so, that this has become a major enterprise in its own right.  This involves the D1-D5 CFT (or the S-dual F1-NS5 CFT) with momentum excitations, whose dual geometries are known as superstrata and microstrata.   Over the last 15 years,  progress has been driven  by extensive interplay between both sides of the holographic dictionary, see \cite{Bena:2022ldq,Bena:2022rna} and references therein.  Matching of VEVs and correlators in the CFT against the fluxes and two-point functions of the geometries has been achieved to a startling level of detail.  As one might expect, much of this has been achieved for supersymmetric geometries and CFT excitations.  
		
		From  the perspective of the black-hole information problem and for holography, it is very important to move beyond supersymmetry and construct non-extremal microstate geometries and map out the CFT states that they capture.  This is a much more difficult enterprise even when focusing on states that can be described within supergravity. At the technical level, one has to solve the full non-linear equations of the supergravity and, at the conceptual level, these states will no longer be ``protected'' by supersymmetry, and will be inherently more complicated and potentially unstable.  Indeed, one ultimately expects to see them decay into some analogue of Hawking radiation.
		
		Despite these challenges, there have been two important breakthroughs in the construction of non-supersymmetric microstate geometries.  First, there are non-BPS bubbled geometries  \cite{Heidmann:2021cms,Bah:2022yji,Bah:2022wot,Bah:2023ows}, which are generically far from BPS, and even Schwarzschild-like. These non-BPS solutions are somewhat similar in spirit to the original bubbled supersymmetric microstate geometries of \cite{Bena:2005va,Berglund:2005vb,Bena:2007kg} in that they are supported by multiple topological bubbles.  While extremely interesting, the holographic dictionary for multi-bubbled backgrounds remains unclear.  This paper will focus on, and develop, the second breakthrough: the construction of ``microstrata'' \cite{Ganchev:2021pgs, Ganchev:2021ewa}.  The beauty of this approach is that it builds on the  success of superstrata  and the precision holography that maps out the CFT states of the D1-D5 system to which they are dual.
		
		Since superstrata and microstrata are dual to coherent string excitations of the D1-D5 system, they are fundamentally solutions of IIB supergravity.  The D5 is then wrapped on a $\IT^4$ on which there are no excitations.  This means they can be reduced to six-dimensional supergravity, and this is where most of the work has been done \cite{Bena:2015bea,Bena:2016agb,Bena:2016ypk,Bena:2017xbt,
			Bakhshaei:2018vux, Ceplak:2018pws, Heidmann:2019zws,Heidmann:2019xrd,Ceplak:2022wri,Ceplak:2022pep}.   The conformally invariant vacuum of the D1-D5 background is  dual to the global AdS$_3$ $\times S^3$ geometry, and is mapped under spectral flow to the maximally spinning supertube.  The supergravity excitations that underlie superstrata and microstrata may therefore be considered as large collective excitations of the six-dimensional {\it supergraviton gas} \cite{deBoer:1998kjm}. Their CFT duals are  multi-trace operators whose elementary single-trace constituents are supergravity operators.  
			
			Within the supergraviton gas, superstrata are the microstate geometries preserving four supercharges -- that will also be referred to as \nBPS{8} -- that are generated by very specific left-moving momentum excitations. The right-moving sector remains in the ground state,  preserving four right-moving supersymmetries. Microstrata are the non-supersymmetric geometries dual to states that, in the Ramond sector, carry both left-moving superstrata excitations and right-moving “anti-superstrata” excitations. From the CFT perspective, the elementary constituents of the multi-particle operator dual to the geometry are mutually supersymmetric in superstrata, but for microstrata they are not.   However, at least at leading order in perturbation theory, we should be able to interpolate the holographic dictionary of microstrata from our detailed knowledge of supertubes and superstrata. 
		
		Since we now have a system with coherent left-moving and right-moving open-string excitations, their collisions should result in copious quantities of closed string modes, and, in supergravity, this means radiation into the gravity multiplet.  Indeed, this should be how such microstates decay into some analogue of Hawking radiation. However, such decay means the solution is time dependent, and thus very challenging to construct. To simplify the problem, we seek time-independent solutions with AdS$_3$ boundary conditions: this effectively puts the system in a box, and time-independence means that the open string collisions will be in equilibrium with the radiation they produce.  (One can then be more adventurous and couple the microstratum back to flat space and watch it decay, but this is for the future.)  Even then, the most general microstratum will have fields and a metric that depend on all five spatial variables. As such, it would seem hopelessly out of reach.  However, there are two huge simplifications that can be made in such a manner that one still obtains microstrata with extremely interesting physical properties.
		
		If one restricts ones attention to  particular families of left-moving and right-moving excitations, then it turns out that they can be compactified on the $S^3$ to yield a {\it consistent truncation} as a particular $SO(4)$ gauged supergravity in three dimensions \cite{Mayerson:2020tcl, Houppe:2020oqp}.  The crucial aspect of the consistent truncation is that if one solves the three-dimensional theory, it produces {\it an exact} solution of the six-dimensional theory, and hence of IIB supergravity. All the complexities of the dependence on the $S^3$ coordinates are handled by the machinery of the consistent truncation.  The problem is then reduced to the three-dimensional excitations of the AdS$_3$.  Indeed, since we seek time-independent solutions, the problem now only depends on the two spatial coordinates of the AdS$_3$.  However, one can now make a further  simplification using the ``Q-ball'' trick \cite{Coleman:1985ki}.
		
		In three dimensions, all the complicated supergravity fields and interactions reduce to a metric, some scalars and some vector fields.  In essence, the ``Q-ball'' trick is to allow some of the scalars to depend on time, $\tau$, through a single phase and a frequency parameter, $\omega$, but do so in such a manner that the phases all cancel in the currents and in the energy-momentum tensor.  To further simplify this we will use the ``coiffuring trick''   \cite{Bena:2013ora,Bena:2014rea,Bena:2015bea,Bena:2017xbt} (a spatial analogue of the ``Q-ball'' trick) to allow the same scalars to have only certain specified phase-dependences on the angular coordinate, $\psi$, of the AdS$_3$, so that these phases also cancel in both the currents and the energy-momentum tensor.  As a result of this, the equations of motion can be reduced to a coupled non-linear system of functions that depend solely upon the radial coordinate, $\xi$.  The only residue of the time and angular dependence come through the frequencies, $\omega$, and the angular quantum numbers, $n$.  While there are quite a number of functions in this system of equations, it was solved in \cite{Ganchev:2021pgs,Ganchev:2021ewa} both numerically and to very high orders in perturbation theory.
		
		There are two technical points that are worth highlighting here.  First, because the three-dimensional supergravity is gauged and descends from string theory, the global symmetries are gauged.  This means that any single phase dependence can be ``gauged away.''  However, this involves a constant shift in the electric and magnetic potentials.  The physics lies in the gauge invariant combinations of frequency and electrostatic potentials and $\psi$-mode numbers and magnetostatic potentials.  In particular, the supersymmetry breaking is naively induced by the time dependence of the scalars, but is really the result of making the gauge-invariant ``$\omega + \Phi$'' terms limit to  values at infinity that are inconsistent with supersymmetry.  In this paper, unlike the earlier ones  \cite{Ganchev:2021pgs,Ganchev:2021ewa}, we will choose a gauge that removes all space and time dependence but introduces the corresponding constant parameters in the electromagnetic potentials.  While extremely convenient computationally, this choice can conceal some of the important physics, as we will discuss in Section \ref{sec:Holography}.
		
		Returning to the gauge with explicit time dependences, one finds that to implement the Q-ball trick in the supergravity potential requires some scalar fields, encoded in a matrix $m_{AB}$,  to have exactly twice the frequency of other scalar fields, encoded in a vector $\chi_A$.  This ``phase-locking'' is not a problem, {\it a priori}, but it appears that it may be the source of singularities in the fully back-reacted system.  As has been discussed in  \cite{Ganchev:2021pgs,Ganchev:2021ewa}, the non-linear interactions and supersymmetry breaking of the microstrata result in fields developing anomalous dimensions, which, in supergravity, means smoothness of the solution requires  frequency shifts away from the normal modes of the supersymmetric system. These frequency shifts do not have to respect the 2:1 phase-locking required by the Q-ball Ansatz. Thus there are two options: The Q-ball Ansatz results in singular solutions (plagued by log-terms at infinity), or the smooth solution requires one to abandon the 2:1 phase-locking.  The latter will then violate the Q-ball Ansatz as there will be new, non-trivial time-dependences.  Simply put: The Q-ball solution can be singular once interactions begin to enter into the higher orders in perturbation theory, and any possible smooth solutions will be necessarily time-dependent. 
		
We find two ways to avoid this problem, leading to two families of microstrata, which we call the  $\alpha$-class and  $\beta$-class.   These names come from the amplitude parameters,  $\alpha$ and $\beta$, of the scalars in  $\chi_A$, and $m_{AB}$ respectively.  The $\beta$-class is the simplest because the  scalars, $\chi_A$, are all set to zero.   From the  perspective of six, or ten, dimensions, these scalars describe metric shape modes on the $S^3$ and, in the $S$-dual NS5-F1 frame,  the solution lives entirely in the NS sector of IIB supergravity.  The resulting microstrata are smooth  and the only restriction on the parameter space is a ``CTC bound'' on the $\beta$-parameters. 

The $\alpha$-class solutions start with  $\chi_A$ excitations. As noted above, they are phase-locked to the scalars in  $m_{AB}$ and have twice the mode-numbers of the scalars in  $\chi_A$, but with arbitrary amplitudes, $\beta$. It was found in \cite{Ganchev:2021pgs,Ganchev:2021ewa} that singular  log terms started to appear at third or fourth order in perturbation theory.  However, it  was shown in  \cite{Ganchev:2021ewa} that the log divergences would disappear completely under one of two circumstances: a) if the solution were supersymmetric,  or b) if the supersymmetry was broken but the amplitudes of the scalars in $m_{AB}$ are quadratically locked to the amplitudes of the scalars in $\chi_A$.  This constraint on the amplitudes  defines what we refer to as the  ``{\it special locus''}.  It is only on this locus that  we obtain a family of microstrata with non-trivial left-moving and right-moving modes for $\chi_A$ and for which  the Q-ball Ansatz (and its 2:1 phase-locking)  results in a completely smooth solution.  As with the $\beta$-class, there is a ``CTC bound'' on the parameters.   

One of the purposes of this paper is to examine this special locus much more closely, classify some of its properties in supergravity, and give some insight into what it means, in terms of gases of supergraviton states, from the perspective of the dual CFT.  We will also generalize and extend the special locus to other sectors of the theory.

		The first papers on microstrata \cite{Ganchev:2021pgs,Ganchev:2021ewa} were focused on ``proof of concept:'' finding simplest  example of a microstratum, exploring its phase space and pinning down the holography as far as possible.  This paper has multiple purposes.  First, it is to exhibit and explore a larger multi-parameter space of smooth microstrata that still fall into the Q-ball Ansatz.  
		We will  map out the holographic dictionary on this larger space of microstrata and exhibit another part of the special locus. We will also describe several new classes of microstrata and compute the corresponding frequency shifts of their normal modes.  As was seen in  \cite{Ganchev:2021pgs,Ganchev:2021ewa}, these frequency shifts depend non-linearly on the amplitudes of the excitations, and this is expected to lead to the chaotic spectrum in more typical non-BPS microstate geometries. 
		
		In the language of superstrata, the microstrata of \cite{Ganchev:2021pgs,Ganchev:2021ewa} lived in the left-moving and right-moving $(1,0,n)$ sector\footnote{These quantum numbers label modes on AdS$_3$ and $S^3$ that are being excited, and encode precisely which CFT operators are involved in creating the CFT state.}.   The larger  families of  microstrata that we explore here involve both the $(1,0,n)$ sector and the $(1,1,n)$ sectors.  This doubles the number of  parameters available, and ``doubles'' the special locus.  Even more significantly, the $(1,0,n)$ and $(1,1,n)$ sectors are ``weakly coupled'' to each other, interacting only through the gravitational and electromagnetic sectors.    This  opens the way to new explorations of microstrata.   Indeed, the really interesting physics of superstrata emerges from ``deep geometries'' in which the redshifts in the AdS$_2$ throat can be made arbitrarily large.  Such superstrata involve ``large'' values of the parameters and are hard to access perturbatively.  Fortunately, such superstratum solutions can be constructed exactly and analytically.  Microstrata are necessarily constructed perturbatively or numerically, and so microstrata with very high red-shifts would be hard to access using the single sector employed in \cite{Ganchev:2021pgs,Ganchev:2021ewa}.  However, by doubling the sectors and exploiting their indirect coupling to one another, one has a new route to microstrata at high red-shift:  One can construct an exactly-known analytic superstratum in the $(1,1,n)$ sector and then add supersymmetry-breaking right-moving microstratum excitations through the $(1,0,n)$ sector.  Indeed, there is a two parameter family of $(1,1,n)$  superstrata on which one can superpose another one-parameter family of smooth (special locus) microstrata in the $(1,0,n)$ sector.

		Most of the families of microstrata that we  construct in this paper can be interpreted through the framework of ``almost-BPS''  solutions \cite{Goldstein:2008fq,Bena:2009ev,Bena:2009en,Bobev:2009kn,DallAgata:2010srl,Vasilakis:2011ki}.  As with a  superstratum, the majority of the  microstrata  constructed here only involve  left-moving momentum waves.   The supersymmetry is broken {\it ab initio} by spinning up the underlying supertube in a manner incompatible with the supersymmetry of the momentum wave.  Thus, as in the almost-BPS story, these solutions are constructed out of competing BPS elements that possess mismatched supersymmetries. For some of these solutions we find that these almost-BPS microstrata have some special properties and some qualitative features in common with the older almost-BPS solutions. 
		
		We begin in Section \ref{sec:3D-Sugr} by reviewing the three-dimensional supergravity that results from the consistent truncation procedure outlined above.   In  Section \ref{sub:qball}, we implement a version of the Q-ball/coiffuring Ansatz in the combined $(1,0,n) + (1,1,n)$ sectors of the theory and arrive at a reduced system that is determined by 19 functions of the radial variable.  Since there is a residual gauge invariance in our reduced system, these 19 functions can be parametrized in different ways and we describe the two  gauge choices that will be used in this paper.   In Section \ref{sub:BPSsystem} we review and catalog some of the supersymmetric solutions, superstrata, that lie within our reduced system, including a new family that comes from two independent metric and flux deformations of the standard $(1,0,n)$ superstratum.  In Section 	\ref{ss:speciallocus} we define the ``special locus'' from the supergravity perspective.

		In Section \ref{sec:nonBPS} we construct a number of different perturbative microstrata.  We start by describing the boundary conditions in detail and how we select such that break supersymmetry.  We then explain our parametrization and construction of the perturbative solutions, and especially the $\alpha$- and  $\beta$-class solutions. The former are more complicated and require the special locus. We give the $\alpha$- and  $\beta$-class solutions to third order in perturbation theory and examine the accuracy and convergence by comparing one such solution against high precision numerics.  We conclude the section by doing WKB analysis of non-BPS perturbations about an arbitrarily deep superstratum solution. 
		
		In Section \ref{sec:CFT} we give an overview of the CFT states that are dual to the supergravity geometries that we will construct. We set out the essential leading-order duality between the CFT states and the supergravity excitations and outline how one extracts the holographic CFT data from the asymptotic behavior of the supergravity solution. In Section \ref{ssech:preshol} we take a quick dive into precision holography so as to obtain one of the central results of this paper:  the relationship between the ``special locus'' of supergravity and a gas of multi-particle states in the CFT.     Sections  \ref{sec:3D-Sugr} and \ref{sec:CFT} are intended to orient the reader from the supergravity and CFT perspectives and can be read in either order.
		
Section \ref{sec:Holography} contains a deeper exploration of the  duality between the CFT and the supergravity.  We rotate the supergravity solution to the canonical gauges required by holography and read off the asymptotic charges.   We then interpret these in terms of the quantum numbers of the CFT and obtain the relationships between the multiplicities, or the number of copies, of the CFT states and the supergravity amplitudes.  From this we determine how the energies of the microstratum states vary with these multiplicities.  In CFT these yield the anomalous dimensions of the states, while in supergravity they represent binding energies.
		
 In Section \ref{sec:Nums} we obtain a very high precision numerical solution to one of our $\beta$-class microstrata.  This not only provides an excellent check on our perturbative results but also allows us to explore microstrata at large parameter values with deep throats and large red-shifts.  We determine the anomalous dimensions, or binding energies, and examine their limiting behavior as the geometries approach the maximal depth. Our numerical analysis also shows   that   solutions with deep throats and large red-shifts do indeed exist and are readily accessible to numerical exploration.   
		
Section \ref{sec:Conclusions} contains our final remarks and a discussion of many future directions in which one can explore microstrata.   There are also three appendices.  In Appendix  \ref{app:spLocusEqs} we catalog some of the geometric properties of the special locus.  In Appendix \ref{app:uplift} we construct some of the details of the uplift of our microstrata to six dimensions and in Appendix \ref{app:betaNumericsEqs} we give a list of the equations of motion for the $\ZZ_2$-invariant $\beta$-class solutions that are used in some of our most accurate and extensive numerical analyses.

		\section{The three-dimensional supergravity}
		\label{sec:3D-Sugr}

		Here we summarize the essential features of the underlying three-dimensional supergravity.  More details may be found in  \cite{Houppe:2020oqp,Ganchev:2021pgs}. Our presentation here will be closely parallel to that of \cite{Ganchev:2022exf}, except we will consider a more general truncation. 
		
		\subsection{The field theory content}
		\label{sec:fields}
		
		We will be working with an extended $\Neql{4}$ gauged supergravity theory in three dimensions.   This theory has eight supersymmetries and the spectrum consists of a  graviton, four gravitini, $\psi_\mu^A$, two sets of six vector fields, $A_\mu^{IJ}= -A_\mu^{JI}$ and $B_\mu^{IJ}= -B_\mu^{JI}$, $20$ fermions, $\chi^{\dot A r}$, and  $20$ scalars parametrized by the coset: 
		\begin{equation}
			\frac{G}{H} ~\equiv~ \frac{SO(4,5)}{SO(4) \times SO(5)} \,.
			\label{coset1}
		\end{equation}
		The gauge group is an $SO(4) \ltimes \IT^6$ subgroup of $G$ and the $\IT^6$ gauge invariance  is typically fixed by setting six of the scalars to zero.   The  associated gauge fields, $B_\mu^{IJ}$, can then be integrated out.  The result is an action for the graviton, the gravitini, the fermions, the  $SO(4)$ gauge fields, $A_\mu^{IJ}$, and $14$ scalars. 
		
		Capital Latin indices, $I,\,J$, denote the vector representation of $SO(4)$ and we use small Greek letters, $\mu,\,\nu$, for spacetime indices. The scalars can be described by the non-compact generators of a  $GL(4,\IR)$ matrix, ${P_I}^J$, and an $SO(4)$ vector,  $\chi_I$.  The theory then has a remaining gauge symmetry $SO(4) \subset G$ that acts on the $\chi_I$ and on the left of  ${P_I}^J$, along with a composite, local symmetry $SO(4) \subset H$ that acts on the fermions and on the right of ${P_I}^J$. The composite local symmetry is typically fixed by eliminating the compact generators of  $P$, thereby rendering it symmetric.  Indeed, once all this gauge fixing is done, the scalars are usually parametrized by a manifestly symmetric matrix, $m_{IJ}$, and its inverse, $m^{IJ}$: 
		\begin{equation}
			m_{IJ}   ~\equiv~   \big(P  \, P^T\big)_{IJ} \,, \qquad m^{IJ}   ~=~  \big ( (P^{-1})^T\,P^{-1}  \big)^{IJ}  \,.
			\label{Mdefn}
		\end{equation}

		The gauge covariant derivatives are defined in terms of the $SO(4)$-dual vectors:
		\begin{equation}
			{\widetilde A_\mu}{}^{IJ}  ~\equiv~ \coeff{1}{2} \,\epsilon_{IJKL}\,{A_\mu}^{KL} \,,  
			\label{dualGFs}
		\end{equation}
		with minimal couplings 
		\begin{equation}
			\cD_\mu \, \cX_{I}  ~=~ \partial_\mu \, \cX_{I}   ~-~  2\, g_0\,\widetilde A_\mu{}^{IJ} \, \cX_{J} \,. 
			\label{covderiv}
		\end{equation}
		In particular, one has:
		\begin{equation}
			\cD_\mu m_{IJ}   ~=~   \partial_\mu m_{IJ}  ~-~   2\, g_0\,\widetilde A_\mu{}^{IK} m_{KJ }~-~   2\, g_0\,\widetilde A_\mu{}^{JK} m_{IK }   \,   \,.
			\label{Dmform}
		\end{equation}

		The field strengths are thus
		\begin{equation}
			F_{\mu \nu}{}^{IJ}  ~=~ \coeff{1}{2}\, \epsilon_{IJKL} \,  \widetilde F_{\mu \nu}{}^{KL}    ~=~  \partial_{\mu}   A_{\nu}{}^{IJ}   ~-~  \partial_{\nu}   A_{\mu}{}^{IJ}   ~-~ 2 \,  g_0 \, \big(   A_{\mu} {}^{IL} \,\widetilde  A_{\nu} {}^{LJ}  ~-~ A_{\mu} {}^{JL} \,\widetilde  A_{\nu} {}^{LI}\big)  \,.
			\label{fieldstrength}
		\end{equation}
		The gauge coupling has dimensions of inverse length, and is related to the charges of the D1-D5 system via: 
		\begin{equation}
			g_0 ~\equiv~ (Q_1 Q_5)^{-\frac{1}{4}} \,.
			\label{g0reln}
		\end{equation}

		We  also define the following combinations of fields:
		\begin{equation}
			\begin{aligned}
				Y_{\mu \, IJ}  ~\equiv~&  \chi_J \,\cD_\mu \chi_I ~-~  \chi_I \,\cD_\mu\chi_J \,, \qquad  \qquad   C_{\mu}^{IJ}  ~\equiv~  g_{\mu \rho} \,  \varepsilon^{\rho \sigma \nu} \, m_{IK} m_{JL}\,F_{\sigma \nu}^{KL}  \,,   \\
				\scrC_{\mu}^{IJ}   ~\equiv~ & P^{-1}{}_{I}{}^K \,  P^{-1}{}_{J}{}^L \, C_\mu^{KL}~=~  g_{\mu \rho}  \, \varepsilon^{\rho \sigma \nu} \,    P{}_{I}{}^K \,  P{}_{J}{}^L \,F_{\sigma \nu}^{KL}\,.
			\end{aligned}
			\label{cs_definition}
		\end{equation}
		Our  conventions for $\varepsilon$ can be found in Appendix A of \cite{Ganchev:2021iwy}.

		\subsection{The bosonic action}
		\label{sec:3Daction}

		Following \cite{Houppe:2020oqp},  we will use a metric signature\footnote{Section 2.7 of \cite{Houppe:2020oqp} discusses how to convert to the mostly positive signature.} of $(+- -) $. The action may now be written as  \cite{Mayerson:2020tcl,Houppe:2020oqp}\footnote{Note that the first reference uses different metric conventions.}: 
		\begin{equation}
			\begin{aligned}
				\cL ~=~ & -\coeff{1}{4} \,e\,R    ~+~ \coeff{1}{8}\,e \, g^{\mu \nu} \, m^{IJ}   \, (\cD_\mu\, \chi_{I})  \, (\cD_\nu\, \chi_{J})  ~+~  \coeff{1}{16}\,e \, g^{\mu \nu} \,  \big( m^{IK} \, \cD_\mu\, m_{KJ}  \big)   \big( m^{JL} \, \cD_\nu\, m_{LI}  \big)  \\
				&-~   \coeff{1}{8}\, e \, g^{\mu \rho}  \, g^{\nu \sigma} \, m_{IK} \,m_{JL}\,  F_{\mu \nu }^{IJ}  \, F_{\rho \sigma }^{KL}  ~-~e\, V   \\
				&+~  \coeff{1}{2}\,e  \, \varepsilon^{\mu \nu \rho} \, \Big[  g_0 \,\big(A_\mu{}^{IJ}\, \partial_\nu  \widetilde A_\rho{}^{IJ}  ~+~\coeff{4}{3}\,  g_0 \, A_\mu{}^{IJ} \,  A_\nu{}^{JK}\, A_\rho{}^{KI} \,\big) ~+~  \coeff{1}{8}\,  {Y_\mu}{}^{IJ}  \, F_{\nu \rho}^{IJ} \Big] \,,
			\end{aligned}
			\label{eq:3Daction}
		\end{equation}

		The scalar potential is given by:
		\begin{equation}
			V ~=~  \coeff{1}{4}\, g_0^2   \,  \det\big(m^{IJ}\big) \, \Big [\, 2 \,\big(1- \coeff{1}{4} \,  (\chi_I \chi_I)\big)^2    ~+~ m_{IJ} m_{IJ}  ~+~\coeff{1}{2} \,  m_{IJ} \chi_I \chi_J  ~-~\coeff{1}{2} \,  m_{II}  \,  m_{JJ}\, \Big] \,.
			\label{potential1}
		\end{equation}

		It is frequently convenient to  fix the local $SO(4)$ gauge symmetry by diagonalizing $P$ in terms of four scalar fields, $\muscal_i$:
		\begin{equation}
			P  ~=~  {\rm diag} \big(\,  e^{\muscal_1} \,, \,  e^{\muscal_2} \,, \,  e^{\muscal_3} \,, \,  e^{\muscal_4} \, \big) \,.
			\label{Pdiag}
		\end{equation}
		The potential then reduces to:
		\begin{equation}
			\begin{aligned}
				V~=~ & \coeff{1}{4}\, g_0^2   \,e^{-2\, (\muscal_1 +\muscal_2+\muscal_3+\muscal_4)} \, \Big [\, 2 \,\big(1- \coeff{1}{4} \,  (\chi_I \chi_I)\big)^2    ~+~ \big( e^{4\, \muscal_1}+e^{4\, \muscal_2}+e^{4\, \muscal_3}+e^{4\, \muscal_4}  \big)  \\
				& \qquad\qquad\qquad\qquad \qquad\qquad~+~\coeff{1}{2}\, \big( e^{2\, \muscal_1}\, \chi_1^2 +e^{2\, \muscal_2}\, \chi_2^2 + e^{2\, \muscal_3}\, \chi_3^2+e^{2\, \muscal_4}\, \chi_4^2  \big) \\
				& \qquad\qquad\qquad\qquad \qquad\qquad~-~\coeff{1}{2}\, \big( e^{2\, \muscal_1} +e^{2\, \muscal_2} + e^{2\, \muscal_3} + e^{2\, \muscal_4}  \big)^2  \, \Big] 
				\,.
			\end{aligned}
			\label{potential2}
		\end{equation}

		The supersymmetry leads to a superpotential:
		\begin{equation}
			\begin{aligned}
				W  ~\equiv~ & \coeff{1}{4} \, g_0 \,  (\det(P))^{-1}  \,   \Big [\, 2 \,\Big(1- \coeff{1}{4} \,  (\chi_A \chi_A)\Big) ~-~ {\rm Tr}\big(P\, P^T\big)  \, \Big] \\
				~=~ & \coeff{1}{4} \, g_0 \,e^{-\muscal_1 -\muscal_2-\muscal_3-\muscal_4} \, \Big [\, 2 \,\Big(1- \coeff{1}{4} \,  (\chi_A \chi_A)\Big) ~-~ \Big( e^{2\, \muscal_1}+e^{2\, \muscal_2}+e^{2\, \muscal_3}+e^{2\, \muscal_4}  \Big) \, \Big]   \,,
			\end{aligned}
			\label{superpot}
		\end{equation}
		and  the potential may be written as
		\begin{equation}
			V~=~ \delta^{ij} \frac{\partial W}{\partial \muscal_i}  \frac{\partial W}{\partial \muscal_j}  ~+~ 2\, m^{IJ} \, \frac{\partial W}{\partial \chi_I} \frac{\partial W}{\partial \chi_J}   ~-~2\, W^2  \,.
			\label{potential3}
		\end{equation}

		The potential has a supersymmetric critical point\footnote{There are also non-supersymmetric flat directions extending from this supersymmetric critical point.} for  $\muscal_j = \chi_I =0$,  where $V$ takes the value
		\begin{equation}
			V_0 ~=~    -  \coeff{1}{2}\, g_0^2   \,.
			\label{susypt}
		\end{equation}
		Setting all the other fields to zero, the Einstein equations give: 
		\begin{equation}
			R_{\mu \nu} ~=~ - 4 \, V_0 \, g_{\mu \nu} ~=~    2 \, g_0^2 \, g_{\mu \nu}  \,.
			\label{susyvac}
		\end{equation}
		and the supersymmetric vacuum\footnote{One should note that because we are using a metric signature $(+ - - )$ the cosmological constant of AdS is positive, contrary to the more standard and rational choice of signature.} is an AdS$_3$ of radius, $g_0^{-1}$.  We therefore define
		\begin{equation}
			\RAdS ~=~ \frac{1}{g_0} \,,
			\label{RAdSscale}
		\end{equation}
		and we will henceforth use this to set the overall scale of the metric.

		\subsection{The metric}
		\label{sec:metric}
		
		The coordinate conventions are inherited from the study of superstrata in six dimensions. In particular, $(u,v)$ 
		are double null coordinates that are related to the time and circle coordinates, $(t,y)$, via:
		\begin{equation}
			u ~\equiv~\frac{1}{\sqrt{2}} \, \big( t ~-~y  \big) \,, \qquad  v ~\equiv~\frac{1}{\sqrt{2}} \, \big(t ~+~y )\,, 
			\label{uvtyreln}
		\end{equation}
		where $y$ is periodically identified as 
		\begin{equation}
			y ~\equiv~ y ~+~ 2 \pi \,R_y\,.
			\label{yperiod}
		\end{equation}
		It is  convenient to compactify the radial coordinate, $\rho$, of AdS$_3$ and use  the scale-free coordinates:
		\begin{equation}
			\xi ~=~\frac{\rho}{\sqrt{\rho^2+1}} \,,  \qquad  \tau~=~ \frac{t}{R_y}\,  \,,  \qquad  \psi~=~ \frac{\sqrt{2}\, v }{R_y}\equiv\tau+\sigma\,, 
			\label{xidef}
		\end{equation}
		where $ 0 \le \xi < 1$, $\psi$ inherits the periodicity $\psi \equiv \psi + 2 \pi$ from (\ref{yperiod}), and we have introduced the coordinate $ \sigma~\equiv~ \sigma + 2\,\pi$, for later convenience.

		As noted in \cite{Houppe:2020oqp,Ganchev:2021pgs}, the most general three-dimensional metric can then be recast in the form:
		\begin{equation}
			ds_{3}^{2}  ~=~  \RAdSsq \, \bigg[ \,\Omega_1^{2} \, \bigg(d \tau +   \frac{k}{(1- \xi^{2})} \, d\psi \bigg)^2~-~\,\frac{\Omega_0^{2}}{(1-\xi^{2} )^{2}} \, \big( d \xi^2 ~+~ \xi^2 \, d \psi^2 \big) \, \bigg] \,,
			\label{genmet1}
		\end{equation}
		for three arbitrary functions $\Omega_0$,  $\Omega_1$ and $k$ of the three coordinates, $(\tau ,\xi,\psi)$.   
		
		Part of metric regularity is to require that there are no closed time-like curves (CTC's).  In particular, this means that the $\psi$-circle must always be space-like, and hence 
		\begin{equation}
			\xi^2\, \Omega_0^{2}  ~-~  \Omega_1^{2}  \,k^2 ~\ge~  0 \,.
			\label{CTC-bound}
		\end{equation}

		If one returns to the coordinates $(t,\rho,v)$, one obtains the more canonical superstratum metric:
		\begin{equation}
			ds_{3}^{2}  ~=~   \RAdSsq \, \bigg[ \, \frac{\Omega_1^{2}}{R_y^2} \, \Big(dt  + \sqrt{2} \, (\rho^2  + 1 ) \, k  \, dv  \Big)^2~-~  \Omega_0^{2}\,\bigg(\frac{d\rho^2}{\rho^2 + 1} ~+~\frac{2}{R_y^2} \,\rho^2\,(\rho^2 + 1) \, dv^2 \bigg)  \, \bigg]\,.
			\label{genmet2}
		\end{equation}
		If one further sets:
		\begin{equation}
			\Omega_0 ~=~ \Omega_1 ~=~ 1\,, \qquad  k ~=~ \xi^2    ~=~  \frac{  \rho^2 }{ (\rho^2+ 1) }  \,,
			\label{AdSvals}
		\end{equation}
		then (\ref{genmet2}) becomes the metric of global AdS$_3$:
		\begin{equation}
			ds_{3}^{2}  ~=~ \RAdSsq \, \bigg[ \,   \big(\rho^2 +1\big)\,  d\tau^2~-~  \frac{d\rho^2}{\rho^2 + 1}  ~-~ \rho^2\,  d\sigma^2 \, \bigg] \,.
			\label{AdSmet}
		\end{equation}
		%

		\subsection{The generalized microstratum truncation}
		\label{sub:qball}
		
		In order to arrive at solutions that only depend on the radial coordinate, we are going to impose invariance under $\tau$-translations and $\psi$-translations.  To simplify the Ansatz even further, we will also impose a $\ZZ_2$ symmetry.   Specifically, we impose the following  symmetries on the fields:
		\begin{itemize}
			\item[(i)] Translation  invariance under:  $\tau\to \tau -\alpha $ and   $\psi \to \psi - \beta$
			\item[(ii)] Reflection invariance under $\psi \to - \psi $, $\tau \to - \tau $, accompanied by a discrete internal $SO(4)$ rotation,  $2 \to -2$, $4 \to -4$.
		\end{itemize}
		One should  note that since we are reducing the fields to the singlet sector of a symmetry action, the resulting truncation is  necessarily consistent with the equations of motion.   Our Ansatz is more general than that used in \cite{Ganchev:2022exf}, and, as we will see,  it contains single modes of the ``double superstratum'' that will be introduced shortly. 
		
		First, the invariance under $\tau$- and $\psi$-translations  means that $\Omega_0$,  $\Omega_1$  and  $k$, can only depend on $\xi$. 
		
		For the fields, $\chi_I$, these symmetries imply
		\begin{equation}
			\chi_1 ~=~    \chi_1 (\xi)  \,, \quad \chi_3 ~=~    \chi_3 (\xi) \,, \qquad  \chi_2 ~=~   \chi_4 ~=~ 0 \,.   \label{trunc1}
		\end{equation}
		Following \cite{Ganchev:2021pgs,Ganchev:2022exf,Ganchev:2021ewa}, we parametrize these scalars as:
		\begin{equation}
			\chi_1   ~=~ \hat\nu_1(\xi)  ~=~   \sqrt{1 - \xi^2} \,   \nu_1(\xi)  \,, \qquad  \chi_3 ~=~ \hat\nu_2(\xi)   ~=~  \sqrt{1 - \xi^2} \,   \nu_2(\xi)  \,.
			\label{chiansatz}
		\end{equation}
		We have introduced  $\nu_1(\xi), \nu_2(\xi)$ so as to handle  the branch cuts in these fields at infinity ($\xi =1$).
		
		The scalar matrix, $m$, must take the form:
		\begin{equation}
			m ~=~\begin{pmatrix}
				m_1 &0  & m_5 & 0\\
				0 & m_2 & 0 & m_6\\
				m_5 & 0 & m_3 & 0\\
				0 & m_6 & 0 & m_4
			\end{pmatrix} \,,
			\label{mmatrix}
		\end{equation}
		and the $m_j$ can only be  functions of $\xi$.
		
		The gauge fields, $\tilde A^{IJ}$, are classified as to whether they are even or odd under $2 \to -2$, $4 \to -4$:  if they are odd, they can only have components along $d\tau$ or $d\psi$, and if they are even then they can only have $d\xi$ components.  We therefore have:
		\begin{equation}
			\begin{aligned}
				\tilde A^{12} ~=~& \frac{1}{g_0} \,\big[\,  \Phi_1(\xi)  \, d\tau ~+~  \Psi_1(\xi)  \, d\psi \, \big]\,, \qquad  \tilde A^{34} ~=~ \frac{1}{g_0} \,\big[\,\Phi_2(\xi)  \, d\tau ~+~  \Psi_2(\xi)  \, d\psi    \, \big] \,,\\
				\tilde A^{23} ~=~& \frac{1}{g_0} \,\big[\,  \Phi_3(\xi)  \, d\tau ~+~  \Psi_3(\xi)  \, d\psi \, \big]\,, \qquad  \tilde A^{14} ~=~ \frac{1}{g_0} \,\big[\,\Phi_4(\xi)  \, d\tau ~+~  \Psi_4(\xi)  \, d\psi    \, \big] \,, \\
				\tilde A^{13} ~=~& \frac{1}{g_0}  \, \Psi_5(\xi)  \, d\xi \,, \qquad \qquad\qquad\qquad\quad \tilde A^{24}  ~=~ \frac{1}{g_0}  \, \Psi_6(\xi)  \, d\xi   \,.
			\end{aligned}
			\label{gauge_ansatz}
		\end{equation}
		Note that we have  introduced explicit factors of $g_0^{-1}$ so as to cancel the $g_0$'s in the minimal coupling and thus render the fields and interactions scale independent. 
		
		There are residual gauge invariances that allow one to make $\xi$-dependent $U(1)$ rotations in both the $(1,3)$ and $(2,4)$ directions. There are two natural ways to fix the gauge:
		
		\medskip
		\leftline{\underline{Diagonal gauge}} 
		Here one fixes the gauge by setting $m_5 = m_6 =0$ in (\ref{mmatrix}).  We will find it useful to parametrize the matrix, $P$, in this gauge according to 
		\begin{equation}
			P  ~=~  {\rm diag} \big(\,  e^{\frac{1}{2}(\mu_1 + \lambda_1)} \,, \,  e^{\frac{1}{2}(\mu_1 - \lambda_1)} \,, \,  e^{\frac{1}{2}(\mu_2 + \lambda_2)}   \,, \, e^{\frac{1}{2}(\mu_2 - \lambda_2)} \big)  \quad {\rm with } \quad m    ~=~   P  \, P^T \,,
			\label{Pdiag-Dgauge}
		\end{equation}
		as in (\ref{Mdefn}).  The gauge with diagonal $P$ is more convenient for the analysis of the supersymmetry. In this gauge, the Ansatz  involves  the following nineteen arbitrary functions of  one variable, $\xi$:
		\begin{equation}
			{\cal F} ~\equiv~ \big\{\, \nu_1  \ \,, \nu_2  \,,\  \mu_1   \,, \  \mu_2  \,, \ \lambda_1   \,, \ \lambda_2 \,,  \  \Phi_1\,,  \dots, \Phi_4 \,, \ \Psi_1\,, \dots\,,   \Psi_6   \,, \  \Omega_0  \,, \  \Omega_1    \,, \ k \, \big\} \,.
			\label{functionlist1}
		\end{equation}
		%

		\medskip
		\leftline{\underline{Axial gauge}} 
		Here one fixes the residual gauge invariance by setting $\Psi_5 = \Psi_6 =0$ in (\ref{gauge_ansatz}).  We will find it useful to parametrize the matrix, $m$, in this gauge according to 
		\begin{equation}
			m ~=~\begin{pmatrix}
				e^{\mu_1 + \lambda_1}  &0  & m_5 & 0\\
				0 & e^{\mu_1 - \lambda_1} & 0 & m_6\\
				m_5 & 0 & e^{\mu_2 + \lambda_2} & 0\\
				0 & m_6 & 0 & e^{\mu_2 - \lambda_2} 
			\end{pmatrix} \,,
			\label{mmatrix-axgauge}
		\end{equation}
		Note that if $m_5 = m_6 =0$, then this parametrization matches (\ref{Pdiag-Dgauge}).
		As we will see, this gauge is more convenient for the perturbative solutions to the equations of motion. The Ansatz now  involves  the following nineteen arbitrary functions of  one variable, $\xi$:
		\begin{equation}
			{\cal F} ~\equiv~ \big\{\, \nu_1  \ \,, \nu_2  \,,\  \mu_1   \,, \  \mu_2  \,, \ \lambda_1   \,, \ \lambda_2  \,, \ m_ 5  \,, \ m_6 \,,  \  \Phi_1\,,  \dots, \Phi_4 \,, \ \Psi_1\,, \dots\,,   \Psi_4   \,, \  \Omega_0  \,, \  \Omega_1    \,, \ k \, \big\} \,.
			\label{functionlist2}
		\end{equation}

		Our goal in this paper is to construct some of the non-BPS solutions to the equations of motion of the action (\ref{eq:3Daction}) using the nineteen functions of (\ref{functionlist1})  or (\ref{functionlist2}).  Before doing this, we first review some of the BPS equations.

		\subsection{ A set of BPS  equations and BPS solutions}
		\label{sub:BPSsystem}

		The supersymmetry transformations of the Lagrangian (\ref{eq:3Daction}) were described in detail in \cite{Houppe:2020oqp} and we will use that as a basis of our discussion here. The resulting BPS equations for specific truncations were analyzed in  \cite{Houppe:2020oqp,Ganchev:2021iwy,Ganchev:2022exf}.  Our goal here is to extend this BPS analysis to a subsystem of the truncation defined in the previous section.  This will make a valuable reference point for the non-BPS solutions we construct later in this paper.

		The BPS analysis starts by deciding which of the supersymmetries are to be preserved, and a standard way to do this is to use projectors. Indeed, following   \cite{Houppe:2020oqp}  we will require that any residual supersymmetries obey\footnote{The $\gamma^a$ are space-time gamma-matrices, while the $\Gamma^{I}$ are internal $SO(4)$ gamma matrices. The details may be found in \cite{Houppe:2020oqp}. }:
		\begin{equation}
			\Big(\oneone ~-~ \gamma^{12}\, \Gamma^{12}  \Big)\, \epsilon ~=~  0 \,, \qquad      \Big(\oneone ~+~ \gamma^{12}\, \Gamma^{34} \Big)\, \epsilon ~=~  0 \,,
			\label{basic-proj}
		\end{equation}
		which implies:
		\begin{equation}
			\Big(\oneone ~-~ \Gamma^{1234}  \Big)\, \epsilon ~=~  0 \,.
			\label{int-proj}
		\end{equation}
		Indeed, any two of the  projectors from (\ref{basic-proj}) and (\ref{int-proj})  imply the third.
		
		We also need to allow phase dependences in the supersymmetries:
		\begin{equation}
			\qquad \dd \epsilon   ~=~  \Gamma^{12}\,(\eta_\tau \dd \tau  +  \eta_\psi \dd \psi )\, \epsilon   \,,
			\label{phases}
		\end{equation}
		for some constants $\eta_\tau$ and $\eta_\psi$.  Note that this phase dependence is consistent with the discrete symmetries of Section \ref{sub:qball}.
		
		It is important to note that the projectors (\ref{basic-proj}) and phase dependences {\it are not invariant} under internal $SO(4)$ rotations, and, in particular, not invariant under the $1 \leftrightarrow 3$ and $2 \leftrightarrow 4$ rotations that remain in our truncation after the symmetries of Section \ref{sub:qball} have been imposed.  Thus the projectors and phase dependences that we are imposing are not the most general possibilities allowed by our truncation, and so we will only derive a subsystem of all the BPS solutions.  This subsystem includes the ``double superstrata'' and some generalizations that will be important to our analysis later.  A complete BPS analysis of our truncation requires one to allow  $\xi$-dependent,  $1 \leftrightarrow 3$ and $2 \leftrightarrow 4$ gauge rotations of   (\ref{basic-proj}).  Such an analysis is technically complicated and takes us far from the non-BPS focus of this paper.  We therefore leave it to future work.
		
		\subsubsection{Preserving and breaking supersymmetry through boundary conditions}
		\label{ss:boundaryconds}

		The BPS equations are a combination of algebraic constraints and first order differential equations.  Moreover, at the boundaries ($\xi =0,1$) these differential equations can reduce to further algebraic constraints involving the boundary conditions.  Since we are dealing with a gauged supergravity, the fields and the supersymmetries are minimally coupled, and so such algebraic constraints will involve the gauge potentials.  For example, it was noted in \cite{Ganchev:2021iwy}  that the gauge invariant combinations appearing in the covariant derivatives of $\epsilon$ at the boundaries are:
		\begin{equation}
			\eta_\tau  ~+~ \Phi_2(1)  ~-~ \Phi_1(1) \qand \eta_\psi ~+~  \Psi_2(0)~-~ \Psi_1(0) \,.
		\end{equation}
Smoothness combined with the vanishing of the gravitino variation  imposes the constraints:
		\begin{equation}
			\eta_\tau ~+~ \Phi_2(1)  ~-~ \Phi_1(1)  ~=~  - \Omega_1  \quad \qand \quad \eta_\psi ~+~  \Psi_2(0)~-~ \Psi_1(0) ~=~  \frac{1}{2}   \,,
			\label{etaconstr}
		\end{equation}
	where, as we will discuss, the BPS equations require $\Omega_1$ to be a constant.	
		We are going to impose specific mode numbers for fields by fixing  $\Psi_j(0)$, (or  $\Psi_j(1)$ since these modes are integer quantized) and thus  (\ref{etaconstr}) will generically determine $\eta_\psi$.  
		
		Of more relevance here are the gaugino variations, which  relate covariant derivatives of scalar fields to variations of the superpotential.    Since we are working in a gauge in which all the fields are independent of $\tau$, the $\tau$-derivatives only involve the electrostatic potentials $\Phi_j$.   At infinity, ($\xi =1$), and for the charged gaugini, the vanishing of the  variations  typically imposes  relationships   between the values of the electric potentials, $\Phi_j(1)$, and the asymptotic values of the superpotential terms.   If there are several charged gaugini, with different minimal couplings, one will obtain independent constraints and,  together with   (\ref{etaconstr}), they restrict, or perhaps  determine, the $\Phi_j (1)$.   
		
		Thus requiring supersymmetry imposes boundary conditions on the $\Phi_j (1)$, and, in Section \ref{sec:nonBPS}, we will break supersymmetry by violating these conditions.

		\subsubsection{The BPS equations}
		\label{ss:bps_eqs}
		
		Because the supersymmetry transformations involve the scalar matrix, $P$, which is the ``square-root'' of $m$, it is simpler to use the diagonal gauge, and work with (\ref{Pdiag-Dgauge}).  
		
		The BPS equations lead to a set of algebraic constraints and first-order differential equations.  The simplest of the differential equations is $\Omega_1' =0$, and so $\Omega_1$ is constant. This reflects the freedom to rescale the time coordinate, $\tau$, and in some of the earlier papers on three-dimensional superstrata and microstrata \cite{Ganchev:2021iwy},  $\Omega_1$ was simply set to $1$.  However, as was pointed out in \cite{Ganchev:2021ewa}, this is not the best choice for holography, and so we will keep $\Omega_1$ as an arbitrary constant for now.
		The algebraic BPS equations then become:
		\begin{equation}
			\begin{aligned}
				\eta_\tau ~=~&  0\,, \qquad  \Phi_1   ~=~  \frac{\Omega_1}{2} \, e^{-\mu_2}   \,, \qquad \Phi_2  ~=~ - \frac{\Omega_1}{2} \, e^{-\mu_1}   \,, \qquad \Phi_3 ~=~ \Phi_4 ~=~0  \,, \\
				\Psi_5   ~=~ &-  \frac{\sinh (\lambda_2) \, \Psi_3+\sinh (\lambda_1) \, \Psi_4 }{\xi \sinh (\lambda_1-\lambda_2)} \,, \qquad \Psi_6 ~=~\frac{\sinh (\lambda_1) \, \Psi_3+\sinh (\lambda_2) \, \Psi_4 }{\xi \sinh (\lambda_1-\lambda_2)} 
			\end{aligned}
			\label{algBPS}
		\end{equation}
		To write the first order equations, it is convenient to define:
		\begin{equation}
			H_0   ~\equiv~  \frac{\xi^2 \Omega_0^2}{(1- \xi^2)^2 \Omega_1^2}\,, \,\, H ~\equiv~ H_0 \, \Omega_1^2 \, e^{-(\mu_1+\mu_2)} = - 4 \,H_0 \,\Phi_1 \Phi_2 \,, \,\,  F_j   ~\equiv~ \Psi_j - \frac{k}{1- \xi^2}\,  \Phi_j \,, \,\, j =1,\dots,4 \,. 
			\label{FHdefns}
		\end{equation}
		Then one has the scalar equations: 
		\begin{equation}
			\begin{aligned}
				& \xi \, \partial_\xi \lambda_1  ~-~  4 \, \sinh(\lambda_1) \, F_1 ~=~ 0 \,, \qquad \xi \, \partial_\xi \lambda_2  ~+~  4 \, \sinh(\lambda_2) \, F_2 ~=~ 0 \,, \\
				&  \xi \, \partial_\xi \hat\nu_1  ~-~  2 \, e^{\lambda_1} F_1 \,\hat\nu_1  ~+~ \frac{2\,  e^{-\lambda_2} \, \sinh (\lambda_1) \, }{ \sinh (\lambda_1-\lambda_2)} \, (e^{\lambda_1}   F_3 + e^{\lambda_2}  F_4 )\,\hat\nu_2~=~ 0 \,,  \\
				&  \xi \, \partial_\xi\hat\nu_2  ~+~  2 \, e^{\lambda_2} F_2 \hat\nu_2  ~-~ \frac{2\,  e^{-\lambda_1} \, \sinh (\lambda_2) \, }{ \sinh (\lambda_1-\lambda_2)} \, (e^{\lambda_1}   F_3 + e^{\lambda_2}  F_4 )\,\hat\nu_1~=~ 0 \,;
			\end{aligned}
			\label{BPS1}
		\end{equation}
		the electromagnetic equations: 
		\begin{equation}
			\begin{aligned}
				& \xi \, \partial_\xi F_1  ~+~ 2\,(F_3 \Psi_5 - F_4 \Psi_6)  ~+~ H \,\cosh (\lambda_2)   ~=~ 0 \,,  \\
				&  \xi \, \partial_\xi F_2  ~+~ 2\,(F_4 \Psi_5 - F_3 \Psi_6)  ~-~ H \,\cosh (\lambda_1)   ~=~ 0  \,, \\
				&\xi \, \partial_\xi F_3  ~-~ 2\,(F_1 \Psi_5 - F_2 \Psi_6)   ~=~ 0  \,,  \\
				&\xi \, \partial_\xi F_4  ~-~ 2\,(F_2 \Psi_5 - F_1 \Psi_6)   ~=~ 0  \,,  
			\end{aligned}
			\label{BPS2}
		\end{equation}
		and metric equations: 
		\begin{equation}
			\begin{aligned}
				& \xi \, \partial_\xi  \Omega_1   ~=~ 0 \,, \\
				& \xi \, \partial_\xi \bigg( \frac{k}{1-\xi^2} \bigg) ~-~ \frac{2}{ \Omega_1} \, \Big( e^{\mu_1} \cosh (\lambda_1) +e^{\mu_2} \cosh (\lambda_2) - \big(1- \coeff{1}{4}(\hat\nu_1^2+ \hat\nu_2^2) \big) \Big) \, H   ~=~ 0 \,,  \\
				& \xi \, \partial_\xi \log(H) ~-~ 4\, \Big(\eta_\psi -   \cosh (\lambda_1) \, F_1  +   \cosh (\lambda_2)\, F_2   \Big)   ~=~ 0 \,.
			\end{aligned}
			\label{BPS3}
		\end{equation}
		The phase parameter, $\eta_\psi$, only appears in the last equation and is determined by the boundary conditions for  $F_1$, $F_2$  and $H$.
		
		There is also an elementary first integral of these equations 
		\begin{equation}
			H \, \sinh (\lambda_1) \, \sinh (\lambda_2)~=~  c_0 \,  \xi^{4 \eta_{\psi}}     \,, 
			\label{integral1}
		\end{equation}
		for some constant, $c_0$.
		
		We are going the require that the scalar fields vanish at infinity so that  the solution limits to the supersymmetric AdS vacuum.  This means that supersymmetry imposes the constraints:
		\begin{equation}
			\eta_\tau ~=~ 0 \,, \qquad \Phi_1(1)  ~=~  \coeff{1}{2} \,\Omega_1 \,, \qquad \Phi_2(1)  ~=~ - \coeff{1}{2} \,\Omega_1 \,.
			\label{infbcs}
		\end{equation}
		%
		
		\subsubsection{The double superstratum}
		\label{sec:doublestratum}

		The double superstratum was obtained in \cite{Heidmann:2019xrd} and recast in three-dimensional supergravity in \cite{Mayerson:2020tcl}.  The fundamental scalars  that create this solution involve setting 
		\begin{equation}
			\chi_1 + i \chi_2 ~=~   \sqrt{1 - \xi^2} \,  G_0 (\zeta)  \,,  \qquad    \chi_3 -  i \chi_4 ~=~   \sqrt{1 - \xi^2} \,  G_1 (\zeta)    \,, 
			\label{chi_dblsuperstrata}
		\end{equation}
		where $\zeta \equiv \xi e^{i \psi}$ and $G_0$ and $G_1$ are any holomorphic functions.  The remaining fields are then determined in terms of the $G_j$. 
		
		To fit within the Ansatz of section \ref{sub:qball}, one must first restrict to ``single modes,'' by taking $ G_0 (\zeta) = \alpha_1 \zeta^{n_1}$   and $ G_1 (\zeta) = \alpha_1 \zeta^{n_2}$, for some positive integers $n_1,n_2$ and some arbitrary constants $\alpha_1$ and $\alpha_2$.  One must them make a gauge transformation so as to remove the $\psi$-dependence.  This introduces  $n_1$ and $n_2$ as constants in $\Psi_1$ and $\Psi_2$. One therefore starts from (\ref{chiansatz}) with:
		\begin{equation}
			\nu_1 ~=~  \alpha_1\, \xi^{n_1}  \,,  \qquad     \nu_2 ~=~  \alpha_2\, \xi^{n_2}    \,, \qquad  \Psi_1(0)  ~=~  \coeff{1}{2} \, {n_1}   \,, \qquad  \Psi_2(0)  ~=~  -\coeff{1}{2} \, {n_2}  \,.
			\label{nu_dblss0}
		\end{equation}
		For convenience, we will also use a gauge transformation so as to make $\alpha_1$ and $\alpha_2$ real.
		As in \cite{Houppe:2020oqp}, it is easiest to verify the supersymmetries with the projectors    (\ref{basic-proj}) if one makes a further $SO(4)$ rotation in the $(1,3)$ plane so as to obtain: 
		\begin{equation}
			\nu_1 ~=~ \sqrt{ \alpha_1^2\, \xi^{2\,n_1} ~+~ \alpha_2^2\, \xi^{2\,n_2}}  \,,  \qquad     \nu_2 ~=~  0   \,.
			\label{nu_dblss1}
		\end{equation}

		This solution has  $\lambda_1 =  \lambda_2  =  0$, and so BPS equations  degenerate, however the BPS solution is given by:
		\begin{equation}
			\begin{aligned}
				\eta_\tau ~=~&  0\,, \quad \eta_\psi ~=~  \coeff{1}{2} \, (1+n_1+n_2)\,,  \quad \lambda_1  ~=~ 0  \,, \quad \lambda_2  ~=~ 0  \,, 
				\\
				\mu_1   ~=~ & \log \big(1 ~-~   \coeff{1}{4} \,(1- \xi^2)\,\nu_1^2\big)  \,, \quad \mu_2  ~=~ 0     \,, \quad \nu_1 ~=~ \sqrt{ \alpha_1^2\, \xi^{2\,n_1} ~+~ \alpha_2^2\, \xi^{2\,n_2}}  \,,  \quad     \nu_2 ~=~  0 \,,
				\\
				\quad \Phi_1   ~=~ & \coeff{\Omega_1}{2} \, e^{-\mu_2}  ~=~  \coeff{\Omega_1}{2}  \,, \quad \Phi_2  ~=~ - \coeff{\Omega_1}{2} \, e^{-\mu_1}   \,, \quad \Phi_3 ~=~ \Phi_4 ~=~0  \,,
				\\
				\Psi_1    ~=~& \coeff{1}{2}\, n_1 -  \coeff{1}{2}\,(n_1-n_2)\,  \frac{\alpha_2^2 \, \xi^{2\,n_2} }{\nu_1^2}  \,, \quad \Psi_2  = - \coeff{1}{2}\, n_2 -  \coeff{1}{2}\,(n_1-n_2)\,  \frac{\alpha_2^2 \, \xi^{2\,n_2} }{\nu_1^2} + \frac{\xi^2}{2\,(1- \xi^2)} \, \big(1 - e^{-\mu_1}  \big) \,,
				\\
				\Psi_3   ~=~ &\Psi_4 ~= - \xi \,\Psi_5 ~=~  \xi \,\Psi_6  ~=~ \coeff{1}{2}\, (n_1-n_2)\,   \frac{\alpha_1 \alpha_2 \,\xi^{n_1+n_2} }{\nu_1^2}   \,,
				\\
				\Omega_0 ~=~& \sqrt{1 - \frac{1}{4} (1-\xi^2) \nu_1^2} \,, \quad k ~=~ \frac{\xi^2}{\Omega_1} \,,
			\end{aligned}
			\label{algBPS1}
		\end{equation}
		where we have assumed for simplicity $n_1\leq n_2$. The standard single superstratum is obtained by setting $\alpha_2=0$ (or $\alpha_1 =0$ and relabelling).
		
		We note that the CTC bound (\ref{CTC-bound}) is satisfied if   
		\begin{equation}
			\alpha_1^2 ~+~ \alpha_2^2 ~\le~  4 \,.
			\label{CTC1}
		\end{equation}
		Indeed, we also note that if the bound is strictly satisfied then the radius of $\psi$-circle diverges as $(1-\xi)^{-1} \sim \rho^2$, and the metric is asymptotic to AdS$_3$.  If (\ref{CTC1}) is an equality, the radius of the $\psi$-circle limits to $+ \sqrt{n} \, R_y$, and the metric is asymptotic to AdS$_2$ $\times S^1$.

		An important aspect of the double superstratum  is that a non-trivial value of $\eta_\psi$ is essential to its smoothness.  Setting $\eta_\psi = 0$ gives $n_2 =-(n_1+1)$, which  produces a smooth solution if and only if $\alpha_1 =0$ or $\alpha_2=0$.
		
		\subsubsection{Abelian sphere deformations}
		\label{ss:Deformations}
		
		The BPS equations can also be used to develop a new set of superstratum solutions based on a two-parameter family of deformations of the $S^3$ that is used to obtain the three-dimensional supergravity from six dimensions.  These $S^3$ deformations are parametrized by the scalar matrix, $m_{AB}$, and the family is ``Abelian'' because the only gauge fields we allow are $\Phi_j$ and $\Psi_j$, $j=1,2$.   We therefore set  $\Psi_j \equiv 0$, $j=3,4,5,6$. 
		
		The algebraic BPS equations become:
		\begin{equation}
			\begin{aligned}
				\eta_\tau ~=~&  0\,, \qquad \Phi_1   ~=~  \frac{\Omega_1}{2} \, e^{-\mu_2}   \,,  \qquad \Phi_2  ~=~ - \frac{\Omega_1}{2} \, e^{-\mu_1}   \,, \\
				\Phi_3 ~=~&  \Phi_4 ~=~0  \,, \qquad  \Psi_3   ~=~\Psi_4   ~=~ \Psi_5   ~=~ \Psi_6 ~=~0\,.
			\end{aligned}
			\label{simpBPS1}
		\end{equation}
		The remaining first order equations are :
		\begin{equation}
			\begin{aligned}
				& \xi \, \partial_\xi \lambda_1  ~-~  4 \, \sinh(\lambda_1) \, F_1 ~=~ 0 \,, \qquad \xi \, \partial_\xi \lambda_2  ~+~  4 \, \sinh(\lambda_2) \, F_2 ~=~ 0 \,,\qquad  \xi \, \partial_\xi \nu_1  ~-~  2 \, e^{\lambda_1} F_1 \,\nu_1  ~=~ 0  \,,\\
				&\xi \, \partial_\xi \nu_2  ~+~  2 \, e^{\lambda_2} F_2 \,\nu_2  ~=~ 0  \,, \qquad  \xi \, \partial_\xi F_1  ~+~ H \,\cosh (\lambda_2)   ~=~ 0 \,,  \qquad    \xi \, \partial_\xi F_2  ~-~ H \,\cosh (\lambda_1)   ~=~ 0  \,, \\
				&  \xi \, \partial_\xi \bigg( \frac{k}{(1-\xi^2)} \bigg) ~-~ 2\, \Big( e^{\mu_1} \cosh (\lambda_1) +e^{\mu_2} \cosh (\lambda_2) - \big(1- \coeff{1}{4}(\nu_1^2+ \nu_2^2) \big) \Big) \, H   ~=~ 0 \,.
			\end{aligned}
			\label{simpBPS2}
		\end{equation}
		where $H$ is determined via the integral of the motion (\ref{integral1}).  The functions $\mu_1$ and $\mu_2$ are not fixed by the BPS equations, but are determined by their  equations of motion.
		
		This simplified system of BPS equations has two more integrals of the motion:
		\begin{equation}
			\nu_1^2 ~=~  c_1 \,   (e^{2 \lambda_1} ~-~1)   \,, \qquad   \nu_2^2 ~=~  c_2 \,   (e^{2 \lambda_2} ~-~1)  \,, 
			\label{integral2}
		\end{equation}
		for some constants, $c_1$ and $c_2$.  In this sense these BPS solutions are  driven by the evolution of the $\lambda_j$.
		
		To determine the evolution of the $\lambda_j$, note that:
		\begin{equation}
			F_1 ~=~ \coeff{1}{4}\,  \xi \, \partial_\xi \log \big(\tanh(\coeff{1}{2}\,\lambda_1 )\big)    \,,  \qquad F_2 ~=~ - \coeff{1}{4}\,  \xi \, \partial_\xi \log \big(\tanh(\coeff{1}{2}\,\lambda_2 )\big)    \,.
			\label{Fexpr1}
		\end{equation}
		The first five BPS equations in (\ref{simpBPS2}) thus reduce to (\ref{integral2}) and two coupled, second order equations in $\lambda_1$ and $\lambda_2$. 
		
		It turns out that one can actually decouple the system. Define:
		\begin{equation}
			\sigma_\pm  ~\equiv~  \log \big(\tanh(\coeff{1}{2}\,\lambda_1 )\big)  ~\pm~  \log \big(\tanh(\coeff{1}{2}\,\lambda_2 )\big)  \,, \qquad  U ~\equiv~  - 4\, c_0\, \xi^{4 \eta_{\psi}} \big(   \cosh \sigma_+ ~+~ \cosh  \sigma_- \big)     \,
			\label{sigdefn}
		\end{equation}
		The the equations for $\lambda_j$ reduce to the decoupled system:
		\begin{equation}
			\xi \, \partial_\xi \big( \xi \, \partial_\xi  \sigma_\pm \big)  ~=~ - \frac{\partial  U} {\partial  \sigma_\pm} ~=~    4\, c_0\, \xi^{4 \eta_{\psi}}   \sinh \sigma_\pm  \,.
			\label{sigeqn}
		\end{equation}
		If $\eta_{\psi} =0$ then these equations can be integrated trivially, but   smoothness at $\xi=0$ generically requires $\eta_{\psi} \ne 0$, and these equations are not straightforward to integrate.
		
		To obtain the supersymmetric solutions,  one first solves  (\ref{sigeqn}), and determines the $\lambda_j$ and  $\nu_j$  from (\ref{sigdefn}) and (\ref{integral2}).  One then determines the $F_j$ from (\ref{Fexpr1}). The $\mu_j$ have to be determined from the equations of motion, and this determines the $\Phi_j$ from (\ref{simpBPS1}).  One can then disentangle the  $\Psi_j$ and $\Omega_0$ from the $F_j$ and $H$.  One then obtains $k$ from (\ref{simpBPS2}).  This process is rather similar to the one in \cite{Ganchev:2022exf}.
		
		One should note that the solutions described here are not deformations of {\it double} superstrata.  This is because generic double superstrata necessarily have non-trivial gauge fields $\Phi_j, \Psi_j$ for  $j=3,4$ (in both diagonal and axial gauge) but here we have set them to zero.  On the other hand, the {\it single superstratum},  with $\nu_2 \equiv 0$ (or $\nu_1 \equiv 0$) has trivial gauge fields  $\Phi_j$,   $j=3,4$ and $\Psi_j$ ,  $j=3,4,5,6$ and so falls within the Ansatz here.  Thus, setting $c_2 =0$ (or  $c_1 =0$) will lead to a two parameter deformation (described by the $\lambda_j$) of the single superstratum.  This is a new class of superstrata that extend the solutions of \cite{Ganchev:2021iwy,Ganchev:2022exf} by including an extra deformation of the $S^3$.
		
		There are also interesting families of ``$\beta$-class'' BPS solutions with $\nu_1 \equiv \nu_2 \equiv 0$ obtained by setting $c_1 =c_2 =0$.  These involve only the scalars of $m_{AB}$  and correspond to S-duals of ``pure NS sector'' solutions of IIB supergravity, generalizing the family of such solutions obtained in \cite{Ganchev:2022exf}.  We will review ``pure NS'' solution of  \cite{Ganchev:2021iwy,Ganchev:2022exf}  in Section \ref{ss:pureNS2} and make use of this family of BPS solutions later in this paper.
		
		\subsubsection{Some observations about  BPS  series solutions}
		
		It is instructive to examine the series solutions  of the equations (\ref{sigeqn}) as it further reveals the crucial role of $\eta_\psi$ in smoothness of double superstrata. 
		
		Suppose that as $\xi \to 0$ one has:
		\begin{equation}
			\lambda_j ~\sim~ \beta_j \, \xi^{2\,n_j}    \,.
			\label{lamasymp1}
		\end{equation}
		The metric function, $\Omega_0$, and the scalars, $\mu_j$  must be finite as $\xi \to 0$, and so it follows from (\ref{FHdefns}) and (\ref{integral1}) that one must take:
		\begin{equation}
			\eta_\psi  ~=~ \coeff{1}{4}( n_1 + n_2 +2)   \,.
			\label{etapsi1}
		\end{equation}
		Assuming that $n_1 > n_2 >0$ one then finds series solutions for $\sigma_\pm$ that take the following form: 
		\begin{equation}
			\begin{aligned}
				\sigma_+  ~=~& \log\big(\coeff{1}{4}\beta_1\beta_2 \big)  ~+~ (n_1 + n_2) \log \xi ~+~ \xi^{2}(\gamma_0 + \gamma_1 \xi^2 ~+~ \dots )\,, \\
				\sigma_-  ~=~& \log\big(\beta_1\beta_2^{-1} \big)  ~+~ (n_1 - n_2) \log \xi ~+~ \xi^{2(n_2+1)}(\tilde \gamma_0 +\tilde \gamma_1 \xi^2 ~+~ \dots )\,.
			\end{aligned}
			\label{sigasymp1}
		\end{equation}
		(If $n_2 > n_1 >0$, the second series starts at $\xi^{2(n_1+1)}$.) These solutions then translate into regular power series for $\lambda_j$ that start with the powers in (\ref{lamasymp1}).
		
		We conclude by noting that as $\xi \to 1$, the metric function, $\Omega_0$, and the scalars, $\mu_j$  must be finite, and so it follows form (\ref{FHdefns}) that one as $\xi \to 1$ one has:
		\begin{equation}
			H ~\sim~ (1-\xi)^{-2}   \,.
			\label{Hinflim}
		\end{equation}
		Comparing this with  (\ref{integral1}), and noting the symmetric relationship between $\lambda_1$ and $\lambda_2$, it follows that, as $\xi \to 1$, one has:
		\begin{equation}
			\lambda_1\,, \lambda_2 ~\sim~ (1-\xi)   \,.
			\label{laminflim}
		\end{equation}
		As we will see, this is consistent with the conformal dimension of the operators dual to the $\lambda_j$. 
		
		\subsubsection{The ``pure NS'' superstratum}
		\label{ss:pureNS2}
		
		We conclude this section by  summarizing a particular superstratum solution that we will need in Section~\ref{ss:wkb_approximation}, where we will construct microstrata by making non-BPS perturbations of it.   This solution corresponds to solving the BPS equations with $\nu_1 \equiv \nu_2 \equiv \lambda_2 \equiv 0$ and was originally described in \cite{Ganchev:2021iwy,Ganchev:2022exf}.  It is a ``pure NS'' superstratum in that its S-dual in the NS5-F1 frame only involves fields in the NS sector of the IIB string.  
		
		The fact that $\lambda_2 \equiv 0$ means that the integral of the motion (\ref{integral1}) becomes trivial, with $c_0 =0$, and one must solve the system {\it ab initio}, as was done in  \cite{Ganchev:2022exf}.
		This smooth solution is given by\footnote{Our parametrization of time and normalisations of scalars differ from that of \cite{Ganchev:2021iwy,Ganchev:2022exf}. In particular, notice the presence of an additional factor of $2$ in \eqref{newparams}, reflecting the same factor in the definition of $\lambda_1 \equiv 2\mu_0$.}:
		\begin{equation}
			\begin{aligned}
				\nu_1\equiv\nu_2\equiv\lambda_2\equiv m_5\equiv m_6~\equiv~ 0   \,, \qquad &e^{\lambda_1} ~=~   \frac{ (1-  \lambda^2 \, \xi^{2(2\,n_1 +1)}) ~+~ \lambda\, (2\,n_1 +1)\, \xi^{2\,n_1} \,  (1-  \xi^{2})  }{ (1-  \lambda^2  \, \xi^{2(2\,n_1 +1)}) ~-~ \lambda\, (2\,n_1 +1)\,  \xi^{2\,n_1} \,  (1-  \xi^{2}) }      \,, \\
				e^{\mu_1}   ~=~ 1~-~  \frac{2 \,\lambda^2 \, (1- \xi^{2(2\,n_1 +1)}) }{(1 + \lambda^2) (1- \lambda^2 \, \xi^{2(2\,n_1 +1)})}&\,,  \qquad
				e^{\mu_2}   ~=~ 1~+~  \frac{2 \,\lambda^2 \, (1- \xi^{2(2\,n_1 +1)}) }{ (1 + \lambda^2)(1- \lambda^2 \, \xi^{2(2\,n_1 +1)})}  \,,    
			\end{aligned}
			\label{c3zero-simp}
		\end{equation}
		where
		\begin{equation}
			\beta_1  ~\equiv~  2\lambda \, (2\,n_1 +1) \,,
			\label{newparams}
		\end{equation}
		is an integration constant.  The gauge fields  are given by:
		\begin{equation}
			\begin{aligned}
				\Phi_1 ~=~&\frac{\Omega_1}{2}e^{-\mu_2}  \,, \qquad \Psi_1 ~=~  \frac{1}{2}\, \bigg(F(\xi) ~+~  \frac{k}{ (1-\xi^2) } \, e^{-\mu_2}  \bigg)  \,, \qquad \Phi_3\equiv\Psi_3\equiv0\,,\\
				\Phi_2 ~=~&-\frac{\Omega_1}{2}\,e^{- \mu_1}\,, \qquad\Psi_2 ~=~  \frac{1}{2}\, \bigg(G(\xi) ~-~ 1 ~-~  \frac{k}{ (1-\xi^2) } \, e^{-\mu_1}  \bigg)  \,, \qquad \Phi_4\equiv\Psi_4\equiv0\,,\\
				F(\xi)   ~=~&  \frac{1}{2} \,\bigg[(2\,n_1 +1) \frac{1+ \lambda^2 \, \xi^{2(2\,n_1 +1)}}{1- \lambda^2 \, \xi^{2(2\,n_1 +1)}} ~-~   \frac{1+  \, \xi^{2 }}{1-  \, \xi^{2 }} \bigg]    \,, \qquad G(\xi)   ~=~  F(\xi) ~+~   \frac{1+  \, \xi^{2}}{1-  \, \xi^{2 }}    \,.
			\end{aligned}
			\label{GaugeFields}
		\end{equation}
		and the metric functions are: 
		\begin{equation}
			\begin{aligned}
				\Omega_0^2   ~=~ &
				\frac{ (1 - \lambda^2)(1+ \lambda^2 \, \xi^{2(2\,n_1 +1)}) }{(1 + \lambda^2)(1- \lambda^2 \, \xi^{2(2\,n_1 +1)})}  & \\
				& \,\, \times \bigg(1 \,+\,  \frac{2 \,\lambda^2 \, (1- \xi^{2(2\,n_1 +1)}) }{ (1 + \lambda^2)(1- \lambda^2 \, \xi^{2(2\,n_1 +1)})} \bigg)  \bigg(1 \,-\, \lambda^2 \, (2\,n_1 +1)^2\, \xi^{4\,n_1  }\,\frac{ (1- \xi^2)^2 }{(1- \lambda^2 \, \xi^{2(2\,n_1 +1)})^2} \bigg)   \,, \\
				\frac{k}{1-\xi^2}   ~=~& \Omega_1^{-1}\bigg[\frac{\xi^2}{1-\xi^2}+\frac{(2\,n_1 +1)\,\lambda^2\,\xi^{2\,(2\,n_1 +1)}\Big(1-\lambda^2\big(3-\xi^{2\,(2\,n_1 +1)}(1+\lambda^2)\big)\Big)}{(1+\lambda^2)(1-\lambda^2\,\xi^{2\,(2\,n_1 +1)})^2}\bigg]\,.
			\end{aligned}
			\label{Om0-c3zero}
		\end{equation}
		There are several integration constants arising in the solution of the BPS system, but they are fixed by smoothness and coordinate re-definitions.
		
		The CTC bound (\ref{CTC-bound}) is satisfied if   
		\begin{equation}
			\lambda^2   ~\le~  \frac{1}{(4\,n_1 +1)} \,.
			\label{CTC2}
		\end{equation}
		As before,  if the bound is strictly satisfied then the radius of $\psi$-circle diverges as $(1-\xi)^{-1} \sim \rho^2$, and the metric is asymptotic to AdS$_3$.  If (\ref{CTC2}) is an equality, then the radius of the $\psi$-circle limits to $n_1^{-1} \sqrt{16 n_1^3 + 14 n_1^2+ 7  n_1+1} R_y  $, and the metric is asymptotic to AdS$_2$ $\times S^1$.

		\subsection{The ``special locus''}
		\label{ss:speciallocus}

		The {\it special locus} will play an essential role in the construction of smooth non-BPS solutions, and in Section~\ref{sec:CFT} we will  discuss its holographic interpretation.  Here we will introduce it through its rather prosaic description in  supergravity.  This locus was first identified as having special properties in \cite{Ganchev:2021pgs} and its essential role in finding non-BPS solutions was found in \cite{Ganchev:2021ewa}.  Further investigation of its properties for supersymmetric theories have been carried out in \cite{Ganchev:2022exf,Ganchev:2021iwy}.
		
		In supergravity, the special locus is defined by a very specific algebraic relationship between the scalars $\chi_I$, the scalars $m_{IJ}$ and the Maxwell fields, $F_{\mu \nu}{}^{IJ}$.  On the special locus, $\chi_I$  is an eigenvector of $m_{IJ}$, and the corresponding eigenvalue, $m_0$, is the only non-trivial eigenvalue of $m_{IJ}$, with 
		\begin{equation}
			m_0=1-\frac{1}{2}\,\chi_I \chi_I \,.
			\label{eigenvalue}
		\end{equation}
		All other eigenvalues of $m_{IJ}$ are equal to $1$, and thus, it has an $SO(3)$ invariance in the directions orthogonal to $\chi_I$. The matrix, $m_{IJ}$, encodes the deformations of the $S^3$ on which the parent supergravity theory has been reduced from six to three dimensions. However, the $SO(3)$ is typically  broken by the gauge fields, $A_{\mu}^{IJ}$.
		
		On the CFT side, $m_{IJ}$ corresponds to a matrix of dimension 2 operators and its form on the \textit{special locus} implies that many of their VEVs vanish. This results in a very satisfying picture physically from the field theory perspective - namely the CFT state on the \textit{special locus} can be naturally interpreted as a multi-particle operator defined by the standard OPE product between its single-particle constituents. This will be described in detail in  Sections \ref{sec:heavystates} and \ref{ssech:preshol}.
		
		Furthermore, on the special locus we find that $\chi_I$ is a null vector of the dual field strengths.  
		\begin{equation}
			\widetilde F_{\mu \nu}{}^{IJ} \,  \chi_J ~=~ 0 \,.
			\label{nullvec}
		\end{equation}

		We should stress that these are ``empirical'' properties of the special locus that were found through the construction of smooth non-BPS solutions. While its meaning in the CFT is described in sections \ref{sec:heavystates} and \ref{ssech:preshol}, the independent significance of the special locus in supergravity remains somewhat mysterious.
		
		\subsubsection{BPS ``special locus'', single sector solution}
		\label{ss:specialBPSsol}
		
		The \textit{special locus} also admits a single-sector, BPS solution, given in Section 2.8 of \cite{Ganchev:2022exf}. Since it is single-sector, it takes the same form in both diagonal and axial gauges. We summarize it here, as it will be needed later on:
		\begin{align}\label{eq:singleSectorBPSspec}
			\eta_\tau & = -\frac{1}{2} \Omega_1\,, \quad \nu_1=\alpha_1\,\xi^{n_1}\,\frac{8\,(2\,n_1+1)\,\sqrt{64\,(2\,n_1+1)^2-\alpha_1^4\,\xi^{4\,n_1+2}}}{64\,(2\,n_1+1)^2-\alpha_1^4\,\xi^{4\,n_1+2}+8\,(2\,n_1+1)^2\,\alpha_1^2\,\xi^{2\,n_1}\,(1-\xi^2)},\quad\nu_2=0\,,\notag\\
			\mu_1&=\lambda_1=-2\,\arctanh\bigg[\frac{8\,(2\,n_1+1)^2\,\alpha_1^2\,\xi^{2\,n_1}\,(1-\xi^2)}{64\,(2\,n_1+1)^2-\alpha_1^4\,\xi^{4\,n_1+2}}\bigg],\quad\mu_2=\lambda_2=m_5=m_6=0,\notag\\
			\Phi_1&=\frac{1}{2}\,\Omega_1  \,,\qquad \Phi_2=\Omega_1   \, \bigg(1-\frac{\xi^{-2\,n_1}\big(64\,(2\,n_1+1)^2-\alpha_1^4\,\xi^{4\,n_1+2}\big)}{8\,\alpha_1^2\,(2\,n_1+1)^2\,(1-\xi^2)}\bigg)^{-1},\notag\\
			\Psi_1&=\frac{n_1}{2},\quad\Psi_2=\frac{\xi^2\,\big(8\,(2\,n_1+1)-\alpha_1^2\,\xi^{2\,n_1}\big)}{8\,(2\,n_1+1)\,(1-\xi^2)}\bigg(1-\frac{\xi^{-2\,n_1}\big(64\,(2\,n_1+1)^2-\alpha_1^4\,\xi^{4\,n_1+2}\big)}{8\,\alpha_1^2\,(2\,n_1+1)^2\,(1-\xi^2)}\bigg)^{-1},\notag\\
			\Phi_3&=\Phi_4=\Psi_3=\Psi_4=\Psi_5=\Psi_6=0,\notag\\
			\Omega_0&=1-\frac{8\,(2\,n_1+1)^2\,\alpha_1^2\,\xi^{2\,n_1}\,(1-\xi^2)}{64\,(2\,n_1+1)^2-\alpha_1^4\,\xi^{4\,n_1+2}},\quad k=\frac{\xi^2}{\Omega_1}\,\bigg(1-\frac{(2\,n_1+1)\,\alpha_1^4\,\xi^{4\,n_1}\,(1-\xi^2)}{64\,(2\,n_1+1)^2-\alpha_1^4\,\xi^{4\,n_1+2}}\bigg).
		\end{align}
		This is the $(1,0,n_1)$ sector solution. Exchange of indices $1\leftrightarrow2$ produces the solution in the $(1,1,n_2)$ sector. 
		
		One can verify that the CTC bound (\ref{CTC-bound}) becomes   
		\begin{equation} 
			\frac{8\,(2\,n_1+1)^2\, (8 - \alpha_1^2\,\xi^{2\,n_1})^2 ~-~ 4\,n_1^2\,\alpha_1^4\, \xi^{4\,n_1 +2}}{(1-\xi^2)\,(64\,(2\,n_1+1)^2-\alpha_1^4\,\xi^{4\,n_1+2})} ~\ge~  0 \,, 
			\label{CTC3a}
		\end{equation}
		which is always satisfied for 
		\begin{equation}
			\alpha_1^2   ~\le~  \frac{8\,(2\,n_1+1)}{(4\,n_1+1)} \,.
			\label{CTC3b}
		\end{equation}
		Again,  if the bound is strictly satisfied then the radius of $\psi$-circle diverges as $(1-\xi)^{-1} \sim \rho^2$, and the metric is asymptotic to AdS$_3$.  Similarly, if  (\ref{CTC3b}) is an equality, then the radius of the $\psi$-circle limits to $ \sqrt{n_1} R_y$, and the metric is asymptotic to AdS$_2$ $\times S^1$.

		One should note that because there are no fields that are minimally coupled to $\Phi_2$, one can shift it by a gauge transformation, and, in the conventions of  \cite{Ganchev:2021iwy} and  \cite{Ganchev:2022exf}, this solution has $\eta_\tau = \frac{1}{2}  \Omega_1$ and $\Phi_2(1) =0$.  One can always use a constant gauge to take this to $\eta_\tau = 0$ and $\Phi_2(1) =- \frac{1}{2}  \Omega_1$.  See Section \ref{ss:boundaryconds} for further discussion of this issue.	
		
		\section{Non-BPS solutions}
		\label{sec:nonBPS}

		We now turn to the construction of non-BPS states.  Our starting point will be the 19 functions \eqref{functionlist2}  in the axial gauge of Section \ref{sub:qball}.  The  equations of motion yield 19 second order ODEs for these functions plus three first-order constraints.   The constraints come from the  $R_{\tau\psi}$ Einstein equation and from the  radial components of the Maxwell equations in the internal $(1,3)$ and $(2,4)$ directions.  One can easily verify that these constraints are simply integrals for combinations of the 19 second-order field equations. 
		
		We will show that the system has two classes of smooth ``solitonic'' solutions, which we refer to as $\alpha$-class and $\beta$-class. Each solution is defined by a locus with  parameters: $(n_1,n_2, \hat \omega_1,  \hat \omega_2,\contparama,\contparamb)$ where $n_1,n_2 \in \ZZ$ are ``magnetic quantum numbers,'' $\hat \omega_1,  \hat \omega_2$ are discrete frequencies  of normal modes of oscillation and  $\contparama$ and $\contparamb$ are continuous parameters that determine the magnitudes of various component fields.  As one would expect, these parameters emerge from the choice of boundary conditions that result in smooth solutions, and so our task will be to isolate the locus of smooth solutions from the plethora of possible boundary conditions for our 19 fields.  There are two steps to achieving this.  The first pass will be to determine as many of the boundary conditions as we can without resorting to any approximate methods. In particular, this will fix  $n_1,n_2 \in \ZZ$.  The second pass will pin down the solution locus exactly using perturbation theory.  We will then return to the possible characterizations of the parameters $\contparama,$ and $\contparamb$ and finally, in Section \ref{sec:Holography}, we will show how one extracts the energies, $\hat \omega_1,  \hat \omega_2$, from the boundary data.

		The simplest class of perturbative solutions is to start with global AdS$_3$, which corresponds to  the CFT vacuum (we work in the NS sector\footnote{One can move from the NS to the R sector by the usual spectral flow trasformation, which acts on the gauge fields as $\Phi_1\to \Phi_1+\frac{1}{2}$, $\Phi_2\to \Phi_2-\frac{1}{2}$, $\Psi_2\to \Psi_2+\frac{1}{2}$.}, as explained in Section \ref{sec:CFT}), and we follow the  procedure used in  \cite{Ganchev:2021pgs,Ganchev:2021ewa}. We will subsequently use WKB methods to construct near-extremal  perturbations around a  deep supersymmetric microstate geometry. This uses the two somewhat decoupled $(1,0,n_1)$ and $(1,1,n_2)$ sectors to great advantage by setting up the deep supersymmetric solution in one sector and adding non-supersymmetric perturbations in the other. Finally, in Section~\ref{sec:Nums}, we will construct numerical solutions to the full non-linear system of equations in a further-symmetry-restricted subclass and thus explore a deep non-BPS regime of our theory.

		\subsection{General boundary conditions} 
		\label{sub:gen-bcs}
		
		First and foremost, we fix the geometry to be asymptotic to AdS$_3$.  This means that the leading divergences in  (\ref{genmet1}) must cancel as $\xi \to 1$, and this implies: 
		\begin{equation}
			\Omega_1(1) ~=~ \frac{\Omega_0(1)}{k(1)} \,.
			\label{AdScond1} 
		\end{equation}
		The metric now has divergences proportional to $(1-\xi)^{-1} \sim \rho^2$ that multiply the metric in the  $(d \psi, d\tau)$-directions.   This describes some, possibly boosted, asymptotically AdS geometry.  
		
		It is convenient to use the coordinate $\sigma$, defined in \eqref{xidef}. The components of (\ref{genmet1}) then satisfy:
		\begin{equation}
			g_{\sigma\sigma}~+~ g_{\tau\tau} ~-~ 2\,g_{\sigma\tau}  ~=~ \Omega_1^2 \,,
			\label{niceidentity}
		\end{equation}
		for all $\xi$.
		The fixed canonical periodicity of $\sigma$ prevents it from being rescaled.  On the other hand, we have not yet fixed  the rescaling of the time coordinate, $\tau \to c \, \tau$ (along with the concomitant rescaling of the electric gauge fields and metric functions\footnote{Precisely, the change of coordinates $\tau\to c\, \tau$ leads to the redefinitions $\Omega_1 \to c\, \Omega_1, \ \Phi_{1,2} \to c\, \Phi_{1,2}, \ k \to c^{-1} k$.}).	 In previous work \cite{Ganchev:2021pgs}, this gauge was fixed by requiring $\Omega_1(\xi = 1) = 1$, however, as was pointed out in \cite{Ganchev:2021ewa}, holography suggests a more natural choice.  We want the asymptotic geometry to be sectioned by canonically normalized (cylindrical) Minkowski slices with  
		\begin{equation}
			ds_2^2 ~\sim~  \rho^2 (d \tau^2 - d\sigma^2)  \,.
			\label{MinkSliced}
		\end{equation}
		We therefore fix the scaling of  $\tau$ by requiring:
		\begin{equation}
			\lim_{\xi \to 1}\frac{g_{\sigma\sigma}}{g_{\tau\tau}} ~=~ - 1 \,.
			\label{goodmetbc}
		\end{equation}
		However, combining this with (\ref{niceidentity}), one sees that this condition also removes the $\rho^2$ divergent term in $g_{\sigma\tau}$, and so (\ref{goodmetbc}) sets the asymptotic metric to the {\it unboosted}, conically-normalized, Minkowski-sliced AdS$_3$ with (\ref{MinkSliced}).  
		
		We therefore use the boundary conditions (\ref{AdScond1}) and (\ref{goodmetbc}) throughout this paper.

		From (\ref{genmet1}) we see that we can also absorb a scale of $\Omega_0$ into the definition of $R_{AdS}$, and thus we can, without loss of generality 		take\footnote{The overall scale, $R_{AdS}$, is fixed by the equations of motion.  Since we will actually set $R_{AdS} = g_0^{-1}$, the condition $\Omega_0(1) ~=~1$ will then emerge from the equations of motion.}:
		\begin{equation}
			\Omega_0(1) ~=~1 \,.
			\label{scaleRAdS} 
		\end{equation}

		In order for the solution to limit to the  supersymmetric vacuum at infinity, the scalar fields must limit to the supersymmetric critical point of the potential. 		
		This means that  $\chi_I$ and $m_{IJ}$ must vanish at infinity, which in terms of the fields of the ansatz translates to:
		\begin{equation} \label{eq:boundary_scalars}
			\nu_{1,2}(1) ~=~ \mathcal{O}(1) \,, \qquad \mu_{1,2}(1) ~=~ \lambda_{1,2}(1) ~=~ 0 \,, \qquad m_{5,6}(1) ~=~ 0 \,.
		\end{equation}

		We are, of course, going to require that all the fields in \eqref{functionlist2} are smooth functions of $\xi$ on the whole interval $\left[0,1\right]$, including at the boundaries.    In particular, this imposes some conditions at the origin.  Since we are working with $\tau$ and $\psi$ independent solutions, the value at the origin of the potentials $\Psi_1$ and $\Psi_2$ fixes the ``modes,'' or ``magnetic quantum numbers,'' of our excitations. The other magnetic potentials must limit to $0$ at the origin\footnote{We work here in the axial gauge. In the diagonal gauge, one also needs to impose $\Psi_5(0) = \Psi_6(0) = 0$.}:
		\begin{equation}
			\Psi_{1}(0) ~=~  \frac{1}{2} \, n_{1} \,, \qquad 	\Psi_{2}(0) ~=~ \frac{1}{2} \, n_2  \,, \qquad \Psi_{3}(0) ~=~ \Psi_{4}(0) ~=~ 0 \,.
			\label{gaugeliminf}
		\end{equation}
		The periodicity of the $\psi$-circle and smoothness requires\footnote{Combining (\ref{covderiv}) and  (\ref{gauge_ansatz}), one sees that the minimal couplings of the gauge potentials inherit a factor of $2$, and hence all the factors of $\frac{1}{2}$ in (\ref{gaugeliminf}).}  that $n_j \in \ZZ$.
		
		Finally, ensuring the regularity of the metric at the origin also requires that  $\Omega_0$ and $\Omega_1$ remain finite at $\xi =0$ and that
		\begin{equation}
			k(\xi) = \mathcal{O}(\xi^2) \qq{as} \xi\to 0\,.
			\label{korigin}
		\end{equation}
		%
		
		\subsection{Perturbative solutions} 
		\label{sub:perSols}
		
		To make further progress, we need to employ perturbation theory, and we start with the global AdS$_3$ solution, which has:
		\begin{gather}
			\label{XAdS}
			\Omega_0^{\rm AdS}=1,\qquad \Omega_1^{\rm AdS}=1,\qquad k^{\rm AdS}=\xi^2, \notag\\
			\nu_{1,2}^{\rm AdS}=0,\qquad \mu_{1,2}^{\rm AdS}=0,\qquad \lambda_{1,2}^{\rm AdS}=0,\qquad m_{5,6}^{\rm AdS}=0,\notag\\
			\Psi_{1,2}^{\rm AdS}=\coeff{1}{2}\, n_{1,2},\qquad \Psi_{3,4}^{\rm AdS}=0, \qquad \Phi_{1,2}^{\rm AdS}=\coeff{1}{2}\, \omega_{1,2} ,\qquad \Phi_{3,4}^{\rm AdS}=0 \,,
		\end{gather}
		for some parameters $\omega_{1,2}$.
		
		The only regular and normalisable perturbations around this vacuum involve $\nu_{1,2}$, $\lambda_{1,2}$ and $m_{\pm}=m_5\pm m_6$, and can be written in terms of hypergeometric functions whose parameters are combinations of $\omega_{1,2}$ and $n_{1,2}$.  Given that $n_{1,2} \in \ZZ$, these  hypergeometric functions can only be smooth for $\omega_{1,2} \in 2\ZZ  - 1$ for $\nu_{1,2}$, or $\omega_{1,2} \in \ZZ$ for $\lambda_{1,2}$. (Smoothness also excludes finite  ranges for  $\omega_{1,2}$, depending on $n_{1,2}$.)  
		Here we focus on the solutions with $n_{1,2} \ge 0$ and $\omega_{1,2} > 0$, for which these linearised solutions have the form:
		\begin{align}\label{eq:nusLambdas}
			&\nu_{1,2}=\alpha_{1,2}\,\xi^{n_{1,2}}\,_2F_1\big((1-\omega_{1,2})/2,\,n_{1,2}+(1+\omega_{1,2})/2,\,1+n_{1,2},\,\xi^2\big),\notag\\
			&\lambda_{1,2}=\beta_{1,2}\,\xi^{2\,n_{1,2}}\,_2F_1\big(-\omega_{1,2},\,2\,n_{1,2}+\omega_{1,2},\,1+2\,n_{1,2},\,\xi^2\big),\notag\\
			&\hspace*{-1.6mm}\begin{array}{rcl}
				m_+=
				\left\{\begin{array}{@{}l@{}}
					\beta_+\,\xi^{ n_1-n_2}\,_2F_1\big(\frac{\omega_2-\omega_1}{2},n_1-n_2+\frac{\omega_1-\omega_2}{2},\,1+n_1-n_2,\,\xi^2\big),\quad n_1\geq n_2,\vspace{1mm}\\[\jot]
					\beta_+\,\xi^{ n_2-n_1}\,_2F_1\big(\frac{\omega_1-\omega_2}{2},n_2-n_1-\frac{\omega_1-\omega_2}{2},\,1+n_2-n_1,\,\xi^2\big),\quad n_1<n_2,\vspace{1mm}
					\vspace{-1mm}\end{array}\right.
			\end{array},\notag\\ &m_-=\beta_-\,\xi^{n_1+n_2}\,_2F_1\big(-(\omega_1+\omega_2)/2,\,n_1+n_2+(\omega_1+\omega_2)/2,\,1+n_1+n_2,\,\xi^2\big)\,.
		\end{align}
		These  are smooth for   $n_{1,2} \ge 0$ and $\omega_{1,2} \in 2\ZZ_+  -1$ for $\nu_{1,2}$, or $\omega_{1,2} \in \ZZ_+$ for $\lambda_{1,2}$.		
		
		Equations \eqref{eq:nusLambdas} provide the perturbative beginning of two classes of solutions, which we will refer to as $\alpha$ and $\beta$ class in reference to the coefficients in (\ref{eq:nusLambdas}). Each complete solution, at non-linear order, will be parametrized by $(n_1,n_2, \hat \omega_1,  \hat \omega_2,\contparama,\contparamb)$.  The $\zeta_j$ represent the ``amplitude'' of the solution, subsuming the linearized amplitudes, $\alpha$ and $\beta$. The magnetic quantum numbers $n_1,n_2$ have already been fixed.  At first order, the frequencies $\hat \omega_1,  \hat \omega_2$ coincide with $\omega_1,  \omega_2$, but the former will be corrected at higher orders in perturbation theory. More precisely, higher orders in perturbation theory, when combined with the smoothness requirement, will require a modification of the boundary conditions for the gauge fields, $\Phi_j(1)$, which will be necessary to cancel the non-normalisable contributions to some of the fields, and for the 3D metric, $\Omega_1(1)$ and $k(1)$ subject to (\ref{AdScond1}), which will ensure a properly normalised time coordinate, according to \eqref{goodmetbc}. Both type of corrections will reflect on the physical frequencies, and we will then have to disentangle the $\hat \omega_1,  \hat \omega_2$ from all the perturbatively corrected boundary data.  Nevertheless, the frequencies $\hat \omega_1,  \hat \omega_2$ will be discrete, inheriting this property from the discreteness of the perturbative normal modes. We write:
		\begin{equation}
			\hat{\omega}_{1,2}=\omega_{1,2}+\delta\omega_{1,2},
			\label{hatomdefn}
		\end{equation}
		where $\delta\omega_{1,2}$ is the correction, for which, in Section~\ref{sec:Holography}, we will give an expansion in powers of $N_{1,2}/N$ using perturbation theory, whereas in Section~\ref{sec:Nums}, we compute the exact value using numerics, for a certain subclass of geometries. 
		 
		The $\alpha$-class of solutions is defined by turning on one, or both, of $\nu_{1,2}$ at linear order (with $\lambda_{1,2}=0$ {\it at linear order}), while the $\beta$-class is defined as the solutions in which either one or both of $\lambda_{1,2}$ are switched on at linear order (and $\nu_{1,2}=0$ {\it at linear order}).  These were already investigated in detail in \cite{Ganchev:2021ewa} for the $(1,0,n)$ solutions and their features in the $(1,1,n)$ sector are exactly the same, so we will only briefly go over them below. At linear order the scalar perturbations, $m_{\pm}$ are global $SO(4)$ rotations of the $\beta$-class perturbations.  However, at higher orders, they could lead to new branches of solutions. We choose not to investigate this here and we will focus entirely on the $\alpha$- and $\beta$-classes of solutions. 
		
		The identification of the bulk fields in (\ref{eq:nusLambdas}) with CFT operators was done in \cite{Ganchev:2021ewa} and the complementary field theory interpretation of this gravitational story is in Section \ref{sec:CFT}, where a more detailed take on it is presented - in particular, the operator map is given in \eqref{eq:holoMap}. The main difference with the approach in \cite{Ganchev:2021pgs,Ganchev:2021ewa} is that here we will, with one or two exceptions, take\footnote{This is equivalent to the choice $\omega_0=0$ in the notation and conventions of \cite{Ganchev:2021pgs}, and $m=0$ in the notation of \cite{Ganchev:2021ewa}} $\omega_{1,2}=1$.  A generic combination of such fields will  break all the supersymmetry.  However, if only one of the fields is turned on, then the result will be supersymmetric:   For $\omega_{1}=1$, the supersymmetry follows from the analysis of Section \ref{sub:BPSsystem}, while for $\omega_{2}=1$, the supersymmetry involves a flip in parity in the analysis of Section \ref{sub:BPSsystem}.   Thus the individual constituents preserve particular supersymmetries, but the combined excitation preserves none.
		
		One can see this rather directly from the perturbative solutions (\ref{eq:nusLambdas}).  For $\omega_{1,2}=1$, one has $\nu_{1,2}=\alpha_{1,2}\,\xi^{n_{1,2}}$ and the solutions will, naively  seem to become double superstrata as in (\ref{nu_dblss0}).  However there is a critical difference in the boundary conditions on  $\Psi_j(0)$ in (\ref{nu_dblss0}) and (\ref{gaugeliminf}), and the latter breaks the supersymmetry.  If one takes $\alpha_2 =0$, then the result is a superstratum that can be written in  terms of the holomorphic variable, $\xi e^{i \psi}$ \cite{Heidmann:2019xrd}, and if $\alpha_1 =0$ the  result is a ``anti-superstratum'' that can be written in  terms of the anti-holomorphic (complex conjugate) variable, $ \xi e^{-i \psi}$. (These excitations can be related by a flip in $(\tau,\,\psi) \to -(\tau,\,\psi)$.) However, when both are excited together, there is no way to write the solution in terms of a single holomorphic coordinate, and so supersymmetry is broken \cite{Heidmann:2019xrd}.
		
		There is, however, an interesting exception to this discussion.   When $n_1=n_2$,  $\alpha_{1,2}\neq0$ and $\beta_{1,2}=0$, the resulting solution can be generated via a finite $SU(2)_R$ gauge rotation\footnote{One can  think of this solution as holomorphic with respect to a different complex structure.} of the single sector  superstratum with $\alpha_2=\beta_1=\beta_2=0$.  Such a rotation is inconsistent with the projectors (\ref{basic-proj}), and so falls outside the analysis of Section \ref{sub:BPSsystem}.   One should note that the orbit of such global transformations will only access particular corners of the moduli space of our general solutions and so such supersymmetry may be viewed as ``accidental.''  This exception belongs to the alpha class of solutions, presented in Section \ref{ss:alphaClass}, and we  discuss the details  in the paragraph after \eqref{eq:N1N2a1a2}.
		
As we will discuss in Section \ref{sec:heavystates}, generic values of $\omega$ lead to microstrata that carry both left-moving and right-moving momentum waves.  However, taking  $\omega_1 = \omega_2=1$ means that the microstratum is rather special in that it  carries purely left-moving momentum.  The supersymmetry breaking then arises from the incompatibility  of the supersymmetry of such a left-moving wave with the supersymmetry of the supertube ground state.

		\subsubsection{Perturbation theory}
		\label{ss:ptBC}
		
		To construct the perturbative solutions, we expand each field as:
		\begin{equation}
			X(\xi)=X^{\rm AdS}+\sum_{k=1}^{\infty}\epsilon^k\,\delta^{(k)}X,
		\end{equation}
		where $\epsilon$ is a bookkeeping parameter and the expressions for $X^{\rm AdS}$ are given in (\ref{XAdS}). For the alpha class, the exponent, $k$, at each order is the sum of powers of $\alpha_1$ and $\alpha_2$. Similarly, for the beta class, at each order, $k$ equals the sum of the powers of $\beta_1$ and $\beta_2$.	As already mentioned, we will work with $\omega_{1,2}=1$ and $n_{1,2}$ are identified with the two sectors $(1,0,n_1)$ and $(1,1,n_2)$ respectively. 
		
		The equations at each order split in five sub-sectors: 1)  the three metric functions $\Omega_{0,1}$ and $k$; 2)  the $\chi^I$ scalars $\nu_{1,2}$; 3)  the $m_{IJ}$ scalars $\mu_{1,2}$, $\lambda_{1,2}$ and $m_{5,6}$; 4)  the four gauge fields $\Phi_{1,2}$ and $\Psi_{1,2}$; and 5)  the other four gauge fields $\Phi_{3,4}$ and $\Psi_{3,4}$. These are easier to solve using the redefinitions:
		\begin{align}
			\mu_{\pm}=\mu_1\pm\mu_2,\quad m_{\pm}=&m_5\pm m_6,\quad \Phi^{(1,2)}_{\pm}=\Phi_1\pm\Phi_2,\quad \Psi^{(1,2)}_{\pm}=\Psi_1\pm\Psi_2,\notag\\
			\Phi^{(3,4)}_{\pm}=&\Phi_3\pm\Phi_4,\quad \Psi^{(3,4)}_{\pm}=\Psi_3\pm\Psi_4.
		\end{align}
		We implement the regularity and boundary conditions described in Section \ref{sub:gen-bcs}: we impose (\ref{gaugeliminf})  on the magnetic gauge fields and require the metric functions to obey (\ref{AdScond1}), (\ref{goodmetbc}),  (\ref{scaleRAdS})  and  (\ref{korigin}). The scalar boundary conditions, (\ref{eq:boundary_scalars}), are imposed as:
		\begin{equation}
			\mu_{1,2},\lambda_{1,2},m_{5,6}=\mathcal{O}\big(1-\xi^2\big),\quad \nu_{1,2}=\mathcal{O}(1) \qq{as} \xi \to 1 \,.
		\end{equation}
		This means that the scalars decay at infinity as normalizable perturbations.
		
		At infinity, $\xi\rightarrow1$, all the gauge fields are required to approach finite values, which are fixed by regularity.
		
		\subsubsection{The  $\alpha$-class and special locus}
		\label{ss:alphaClass}
		
		Taking $\alpha_i$, $i\in\{1,2\}$, to be non-zero and setting $\beta_{1,2}=0$ in \eqref{eq:nusLambdas} at linear order gives us the alpha class of solutions. It turns out that at second order in perturbation theory it is necessary to turn on the corresponding $\lambda_i$ field in order to have a normalisable solution to the equations of motion. This is not an independent mode  because at fourth order regularity requires:
		\begin{equation}\label{eq:specialLoc}
			\beta_i=-\frac{\alpha_i^2}{4}.
		\end{equation}
		This puts the solution on the \textit{special locus}\footnote{In our previous papers, \cite{Ganchev:2021pgs,Ganchev:2021ewa,Ganchev:2021iwy}, \eqref{eq:specialLoc} has 8 in the denominator because of the slightly different gauge we use there. Specifically, $\lambda_1$ here is equivalent to $2\,\mu_0$ in the previous works.} introduced in Section \ref{ss:speciallocus} with all its specific properties being clearly verifiable within the perturbative expansion. Of course, we can turn on both alphas at the same time and we get the expected result: $\lambda_1$ and $\lambda_2$ have to be excited and their amplitudes locked onto $\alpha_1$ and $\alpha_2$ via \eqref{eq:specialLoc}. For the $\alpha$-class solutions we may parametrize them by identifying $\contparam{1,2}$ for instance with $\alpha_{1,2}$. The CFT representation of the resulting non-BPS state is given in \eqref{eq:alpha1alpha2}.

		The properties of $m_{IJ}$ on the \textit{special locus} (see Section \ref{ss:speciallocus}) allow us to easily solve for the scalar fields in the perturbative expansion. In fact, the perturbative solution suggests certain relations between some of the scalar fields:
		\begin{align}
			\mu_{1,2}(\xi)&=\lambda_{1,2}(\xi)=\frac{1}{2}\log\Big(1-\frac{1-\xi^2}{2}\nu_{1,2}(\xi)^2\Big),\notag\\
			m_5(\xi)&=-\frac{1}{2}(1-\xi^2)\,\nu_1(\xi)\,\nu_2(\xi),\quad m_6(\xi)=0,
		\end{align}
		with the only non-trivial eigenvalue of the $m_{IJ}$ given by \eqref{eigenvalue}. Remarkably, one can check that these indeed hold in the full non-linear system of equations (on the \textit{special locus}) by substituting them in and then reducing the equations to a consistent set. What is more,  the gauge fields, as well as the metric function $k(\xi)$, satisfy first-order, linear ordinary differential equations. One can, furthermore, derive a first-order equation for a linear combination of $\Omega_0(\xi)$ and $\Omega_1(\xi)$, in addition to a ``Wronskian-like''  constraint that relates the first derivatives of $\nu_1(\xi)$ and $\nu_2(\xi)$. The equations are given in Appendix \ref{app:spLocusEqs}. We have checked that solutions to this reduced system of equations also verify all the second-order, non-linear equations of motion. We will make some comments about this in Section \ref{ss:AlmostBPS}.

		Therefore, we only need to determine $\nu_{1,2}$, $\Phi_i$, $\Psi_i$ and the metric functions $\Omega_{0,1}$ and $k$ on the \textit{special locus} in order to satisfy the equations of motion. Here we will  give the first three orders of the solution for general $n_1, n_2$ and provide the $4^{\rm th}$ order expressions (the highest we have for arbitrary $n_{1,2}$) in a \textit{Mathematica} notebook as ArXiv auxiliary file.
				
		At first order, only the $\nu$ fields are non-zero:
		\begin{gather}
			\delta^{(1)}\nu_1=\alpha_1\,\xi^{n_1},\quad \delta^{(1)}\nu_2=\alpha_2\,\xi^{n_2},\notag\\
			\delta^{(1)}\Phi_i=\delta^{(1)}\Psi_i=\delta^{(1)}\Omega_{0,1}=\delta^{(1)}k=0.
		\end{gather}
		
		At second order  $\nu_{1,2}$ receive vanishing contributions, while all the others are excited:
		\begin{align}\label{eq:alphaGen2ndO}
			\delta^{(2)}\nu_{1,2}=&0,\notag\\
			\delta^{(2)}\Phi_1=&\frac{1}{8}\Bigg[-\alpha_1^2+\alpha_2^2\bigg(-1+\frac{2\,(n_1-n_2)^2}{(n_1+n_2)\,(n_1+n_2+1)\,(n_1+n_2+2)}-\xi^{2\,n_2}(1-\xi^2)\bigg)\Bigg],\notag\\
			\delta^{(2)}\Phi_2=&\frac{1}{8}\Bigg[-\alpha_2^2+\alpha_1^2\bigg(-1+\frac{2\,(n_1-n_2)^2}{(n_1+n_2)\,(n_1+n_2+1)\,(n_1+n_2+2)}-\xi^{2\,n_1}(1-\xi^2)\bigg)\Bigg],\notag\\
			\delta^{(2)}\Phi_3=&\frac{1}{4}\,\alpha_1\,\alpha_2\,\xi^{n_1+n_2}\bigg(\frac{\xi^2\,(n_2+1)}{n_1+n_2+2}-\frac{n_2}{n_1+n_2}\bigg),\notag\\
			\delta^{(2)}\Phi_4=&-\frac{1}{4}\,\alpha_1\,\alpha_2\,\xi^{n_1+n_2}\bigg(\frac{\xi^2\,(n_1+1)}{n_1+n_2+2}-\frac{n_1}{n_1+n_2}\bigg),\notag\\
			\delta^{(2)}\Psi_1=&-\frac{1}{8}\,\alpha_2^2\,\xi^{2+2\,n_2},\quad\delta^{(2)}\Psi_2=-\frac{1}{8}\,\alpha_1^2\,\xi^{2+2\,n_1},\notag\\
			\delta^{(2)}\Psi_3=&-\frac{\alpha_1\,\alpha_2\,(n_2+1)\,\xi^{n_1+n_2+2}}{4\,(n_1+n_2+2)},\quad\delta^{(2)}\Psi_4=\frac{\alpha_1\,\alpha_2\,(n_1+1)\,\xi^{n_1+n_2+2}}{4\,(n_1+n_2+2)},\notag\\
			\delta^{(2)}\Omega_0=&-\frac{1}{8}\,(1-\xi^2)\,\big(\alpha_1^2\,\xi^{2\,n_1}+\alpha_2^2\,\xi^{2\,n_2}\big),\quad \delta^{(2)}\Omega_1=-\frac{1}{4}\big(\alpha_1^2+\alpha_2^2\big),\quad \delta^{(2)}k=\frac{\xi^2}{4}\big(\alpha_1^2+\alpha_2^2\big).
		\end{align}
		
		Then at third order again only the $\nu$ fields are non-trivial:
		\begin{align}
			\delta^{(3)}\,\nu_{1,2}=\frac{\alpha_{1,2}\,\xi^{n_{1,2}}}{8}\Bigg[-\alpha_{1,2}^2\,\xi^{2\,n_{1,2}}\,(1-\xi^2)+2\,\alpha_{2,1}^2\Bigg(\xi^{2\,n_{2,1}}\bigg(-\frac{n_{1,2}}{n_1+n_2}+\frac{(1+n_{1,2})\,\xi^2}{2+n_1+n_2}\bigg)\notag\\
			+\frac{(n_1-n_2)^2\big(-\xi^{2\,n_{2,1}}+n_{2,1}\,\xi^{2\,n_{2,1}}\,\Phi(\xi^2,\,1,\,n_{2,1})+n_{2,1}\,\log(1-\xi^2)\big)}{n_{2,1}\,(n_1+n_2)\,(1+n_1+n_2)\,(2+n_1+n_2)}\Bigg)\notag\\
			+\frac{2\,(n_1-n_2)^2\,\alpha_{2,1}^2\Big(n_1+n_2+\frac{n_{1,2}}{n_{2,1}}+(n_1-n_2)\,\big(\gamma+\psi^{(0)}(n_{2,1})\big)\Big)}{(n_1+n_2)\,(1+n_1+n_2)\,(2+n_1+n_2)}\Bigg],
		\end{align}
		\begin{equation}
			\delta^{(3)}\Phi_i=\delta^{(3)}\Psi_i=\delta^{(3)}\Omega_{0,1}=\delta^{(3)}k=0,
		\end{equation}
		where $\Phi(z,\,s,\,a)$ is the Lerch transcendent, $\gamma$ is the Euler-Mascheroni constant and $\psi^{(m)}(z)$ is the polygamma function of order $m$. The function $\psi^{(0)}(z)$ is also known as the digamma function.
		
		The solution is smooth. In  particular, the log gets cancelled with a log part of the Lerch transcendent, $\Phi(\xi^2,\,1,\,n_{2,1})$, and the entire expression is a polynomial in $\xi$. Moreover, the coefficients are all rational despite the appearance of $\gamma$ and $\psi^{(0)}(z)$. We see no issues, in principle, to continuing this to arbitrarily high orders. Indeed, we have have obtained the solution to $8^{\rm th}$ order for specific values of $n_1$ and $n_2$ and have encountered no problems.
		
		In constructing these solutions we find that the asymptotic values at infinity of the metric coefficient, $\Omega_1$, and the electromagnetic potentials, $\Phi_j$ and $\Psi_j$, get corrected in terms of $\alpha_{1,2}$ at higher orders in perturbation theory.  As we will discuss in Section~\ref{sec:Holography}, this results in an energy shift of the normal modes away from their AdS values, $\omega_{1,2}$, to new, lower energies represented by $\hat \omega_{1,2}$.  For this reason, a more physical parametrization of our new solutions would be to identify $\contparam{1,2}$ with $\hat{\omega}_{1,2} -\omega_{1,2}=\delta\omega_{1,2}$.

		\subsubsection{Almost-BPS solutions}
		\label{ss:AlmostBPS}
		
		Before describing more non-BPS solutions to our three-dimensional system, it is worth stepping back and commenting on the somewhat surprising result summarized above and in Appendix  \ref{app:spLocusEqs}: some of the fields are algebraically related to others and some of the equations of motion  can be reduced to first-order, linear ordinary differential equations. 
		
		A similar phenomenon has already played a significant role in the construction of special families of non-BPS solutions in higher 	dimensions.  It was observed in \cite{Goldstein:2008fq,Bena:2009ev,Bena:2009en,Bobev:2009kn,Bena:2009qv,Bena:2009fi,Bena:2011ca,Vasilakis:2011ki}  that one could construct non-supersymmetric solutions by assembling ``locally supersymmetric components'' whose individual supersymmetries are incompatible with one another and so the solution preserves no  supersymmetry when considered globally. At a computational level, this could be implemented  by solving some algebraic and first-order BPS equations in backgrounds that satisfied incompatible BPS conditions, or backgrounds that break supersymmetry altogether. For example, one can find non-BPS solutions by solving  BPS equations for electromagnetic fields in a gravitational background that has non-trivial, supersymmetry-breaking holonomy.   The non-trivial result of such an approach is to show that, despite the mismatched supersymmetry conditions, the background one creates is still a solution to the complete set of equations of motion. 
		
		Thus the construction of such ``Almost-BPS solutions'' typically involves solving a combination of algebraic and first-order BPS-like equations and second-order equations of motion.  The results of Section \ref{ss:alphaClass} are highly reminiscent of this process, especially because one or two of the equations in Appendix  \ref{app:spLocusEqs} display a close parallel to some of the BPS equations. 
		
		We will not pursue this issue here, but it would be very interesting to revisit the Almost-BPS approach in three dimensions and see, to what extent, some of our  solutions might be characterized as being Almost-BPS.   Intuitively, the success of the Almost-BPS procedure requires a ``decoupling'' between the distinct,  but incompatible, supersymmetric elements. While we do not expect our general non-BPS solutions to behave in such a manner,   the way we broke the supersymmetry by combining incompatible supersymmetric elements from the $(1,0,n_1)$ and $(1,1,n_2)$ sectors is very suggestive.   It is also possible that the special locus might be characterized through Almost-BPS behaviour in supergravity.

		\subsubsection{The $\beta$-class solutions}
		\label{ss:betaClass}
		
		This sector starts, at linear order, with $\alpha_{1,2}=0$ in \eqref{eq:nusLambdas}  and  non-zero $\beta_i$, $i\in\{1,2\}$. There is no need of a \textit{special locus}, and one finds $\nu_{1,2}=0$ at all orders in perturbation theory.  (This is a simple consequence of a discrete parity invariance $\chi_I \to - \chi_I$ in the three-dimensional action.) One also finds that  $m_{5,6},\,\Phi_{3,4},\,\Psi_{3,4}$ vanish to all orders we have checked. When both $\beta_{1,2}$ are turned on, the CFT representation of the resulting non-BPS state is given in \eqref{eq:beta1beta2}.  
		
		As with the $\alpha$-class, the individual solutions with $\beta_1 =0$ or $\beta_2=0$  are separately supersymmetric, but the combined excitation breaks all the supersymmetry.  Unlike the $\alpha$-class, there is no accidental supersymmetry for $n_1 =n_2$: global $SO(4)$ rotations of $\alpha_{1}=\alpha_{2}=\beta_2=0$ create additional background fields that move outside our Ansatz. We have computed the perturbative solution for general $\beta_{1,2}$ and $n_{1,2}$ up to third order:
		\begin{gather}
			\delta^{(1)}\lambda_1=\beta_1\,\xi^{2\,n_1}\,(1-\xi^2),\quad \delta^{(1)}\lambda_2=\beta_2\,\xi^{2\,n_2}\,(1-\xi^2),\notag\\
			\delta^{(1)}\mu_{1,2}=\delta^{(1)}\Phi_i=\delta^{(1)}\Psi_i=\delta^{(1)}\Omega_{0,1}=\delta^{(1)}k=0.
		\end{gather}
		\begin{align}
			\delta^{(2)}\lambda_{1,2}=&\,0,\notag\\
			\delta^{(2)}\mu_1=&\,\beta_2^2\,\frac{1-\xi^{2\,(1+2\,n_2)}}{2\,(1+2\,n_2)^2}-\beta_1^2\,\frac{1-\xi^{2\,(1+2\,n_1)}}{2\,(1+2\,n_1)^2},\notag\\
			\delta^{(2)}\mu_2=&\beta_1^2\,\frac{1-\xi^{2\,(1+2\,n_1)}}{2\,(1+2\,n_1)^2}-\,\beta_2^2\,\frac{1-\xi^{2\,(1+2\,n_2)}}{2\,(1+2\,n_2)^2},\notag\\
			\delta^{(2)}\Phi_1=&\,\beta_1^2\,\frac{\xi^{2\,(1+2\,n_1)}-2\,(1+2\,n_1)}{4\,(1+2\,n_1)^2}+\beta_2^2\,\frac{(1+n_1+n_2)\,\xi^{2\,(1+2n_2)}-2\,(n_1+n_1\,n_2+n_2^2)}{4\,(1+n_1+n_2)\,(1+2\,n_2)^2},\notag\\
			\delta^{(2)}\Phi_2=&\,\beta_2^2\,\frac{\xi^{2\,(1+2\,n_2)}-2\,(1+2\,n_2)}{4\,(1+2\,n_2)^2}+\beta_1^2\,\frac{(1+n_1+n_2)\,\xi^{2\,(1+2n_1)}-2\,(n_2+n_1\,n_2+n_1^2)}{4\,(1+n_1+n_2)\,(1+2\,n_1)^2},\notag\\
			\delta^{(2)}\Psi_1=&\,-\beta_2^2\,\frac{\xi^2\,\big(1-\xi^{2\,(1+2\,n_2)}\big)}{4\,(1+2\,n_2)^2\,(1-\xi^2)}-\beta_1^2\,\frac{\xi^2\,\Big(1-\xi^{4\,n_1}\,\big(1+2\,n_1\,(1-\xi^2)\big)\Big)}{4\,(1+2\,n_1)^2\,(1-\xi^2)},\notag\\
			\delta^{(2)}\Psi_2=&\,-\beta_1^2\,\frac{\xi^2\,\big(1-\xi^{2\,(1+2\,n_1)}\big)}{4\,(1+2\,n_1)^2\,(1-\xi^2)}-\beta_2^2\,\frac{\xi^2\,\Big(1-\xi^{4\,n_2}\,\big(1+2\,n_2\,(1-\xi^2)\big)\Big)}{4\,(1+2\,n_2)^2\,(1-\xi^2)},\notag\\
			\delta^{(2)}\Omega_0=&-\frac{(1-\xi^2)^2}{8}\,\big(\beta_1^2\,\xi^{4\,n_1}+\beta_2^2\,\xi^{4\,n_2}\big),\quad\delta^{(2)}\Omega_1=-\frac{1}{2}\bigg[\frac{\beta_1^2}{1+2\,n_1}+\frac{\beta_2^2}{1+2\,n_2}\bigg],\notag\\
			\delta^{(2)}k=&\frac{\xi^2}{4}\bigg[\beta_1^2\,\frac{\big(2+\xi^{4\,n_1}(1-\xi^2)\big)}{1+2\,n_1}+\beta_2^2\,\frac{\big(2+\xi^{4\,n_2}(1-\xi^2)\big)}{1+2\,n_2}\bigg].
		\end{align}
		\begin{equation}
			\delta^{(3)}\mu_{1,2}=\delta^{(3)}\Phi_i=\delta^{(3)}\Psi_i=\delta^{(3)}\Omega_{0,1}=\delta^{(3)}k=0,
		\end{equation}
		\begin{align}
			\delta^{(3)}\lambda_{1,2}=&\frac{\beta_{1,2}\,\beta_{2,1}^2\,\xi^{2\,n_{1,2}}}{4\,(1+n_1+n_2)\,(1+2\,n_{2,1})^2}\bigg[4\,(1-\xi ^2)\,(1+2\,n_{2,1})\,\big(\log(4)+2\,H_{4\,n_{2,1}-1}-H_{2\,n_{2,1}-\frac{1}{2}}\big)\notag\\
			&+\frac{1}{n_{2,1}\,(1+4\,n_{2,1})}\Big[2\,(1-\xi ^2)\,\xi^{4\,n_{2,1}}\,(8\,n_{2,1}^2+6\,n_{2,1}+1)\, _2F_1\big(1,\,4\,n_{2,1};\,1+4\,n_{2,1};\,\xi\big)\notag\\
			&-8\,(1-\xi^2)\,\xi^{1+4\,n_{2,1}}\,n_{2,1}\,(1+2\,n_{2,1})\, _2F_1\Big(1,\,\frac{1}{2}+2\,n_{2,1};\,\frac{3}{2}+2\,n_{2,1};\,\xi^2\Big)\notag\\
			&+(1+4\,n_{2,1})\,\Big(n_{2,1}\, \big(5+n_{1,2}-\xi^2(1-n_{1,2})-\big(4+\xi^2\,(1-\xi^2)(1+n_{1,2})\big) \,\xi^{4\,n_{2,1}}\big)\notag\\
			&-n_{2,1}^2\,(1-\xi^2)(\xi^{2\,(1+2\,n_2)}+7)+2\,(1-\xi^2)\,(1-\xi^{4\,n_{2,1}})\Big)\Big]\notag\\
			&+4\,(1-\xi^2)(1+2\,n_{2,1})\,\log(1-\xi^2)\bigg]\notag\\
			&+\frac{\beta_{1,2}^3\,(1-\xi^2)\,\xi^{2\,n_{1,2}}}{12\,(1+2\,n_{1,2})^2}\,\Big[(1+2\,n_{1,2})^2\,\xi^{4\,n_{1,2}}+\big(1-8\,n_{1,2}\,(1+n_{1,2})\big)\xi^{2\,(1+2\,n_{1,2})}\notag\\
			&+(1+2\,n_{1,2})^2\,\xi^{4\,(1+n_{1,2})}-3\Big],
		\end{align}
		where $H_n$ is a harmonic number. We again find a smooth solution, and as for the $\alpha$-class, despite appearances, this is a polynomial in $\xi$ with rational coefficients. Hence, there should be no issues, in principle, to continuing this to arbitrarily high orders. As before, we also have examples for specific values of $n_1$ and $n_2$ that we have taken to $8^{\rm th}$ order with no complications. We can therefore parametrize this class of solution by identifying $\contparam{1,2}$ with $\beta_{1,2}$. 
		
		Like the $\alpha$-class solutions,  the $\beta$-class  solutions also develop non-trivial perturbative corrections to the asymptotic values, at infinity, of the metric coefficient, $\Omega_1$, and the electromagnetic potentials, $\Phi_j$ and $\Psi_j$. This also results in an energy shift of the normal modes away from their AdS values, $\omega_{1,2}$, to new, lower energies represented by $\hat \omega_{1,2}$.  Again, one could also use a more physical parametrization of our new solutions by identifying $\contparam{1,2}$ with $\hat{\omega}_{1,2}-\omega_{1,2}=\delta\omega_{1,2}$.
		
		\subsection{Perturbation theory comparison with numerical results}
		\label{sub:trophies}
		
To examine the accuracy and convergence of our perturbative solutions, it is invaluable to compare them to  accurate numerical solutions of the full non-linear system of equations. 
 Here we will make one such comparison, and defer the discussion of our  numerical integration methods to Section~\ref{sec:Nums}.
 
We consider a simplified form of the $\beta$-class of solutions of Section~\ref{ss:betaClass}:   we impose an additional $\mathbb{Z}_2$ symmetry between the two sectors,  taking $\beta_1=\beta_2=\beta$ and $n_1=n_2=n$. That implies $\lambda_1=\lambda_2$, $\mu_1=\mu_2$, $\Phi_1=\Phi_2$, $\Psi_1=\Psi_2$, in addition to all the vanishing fields already mentioned in Section~\ref{ss:betaClass}. The reasons for this choice of simplification  are explained in the chapter on numerics. Importantly, for $n\neq0$ the geometries are still non-BPS.
		
\begin{figure}[!ht]
			\centering
			\includegraphics[width=\textwidth]{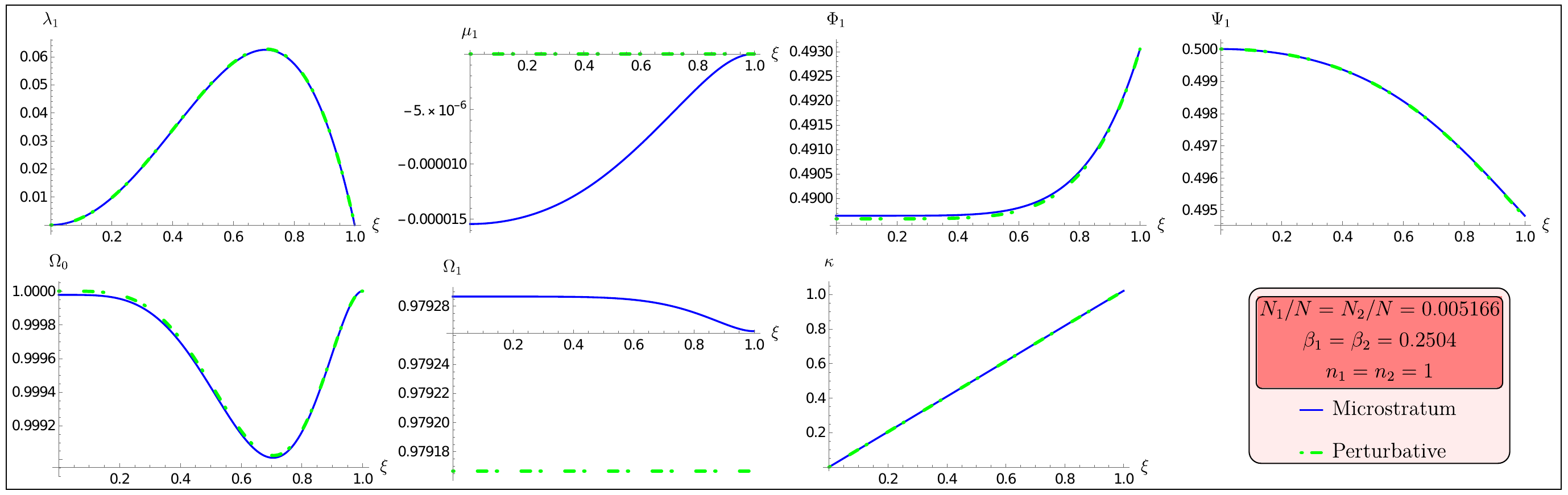}
			\caption{The non-zero fields in the $\mathbb{Z}_2$ symmetric beta class of solutions, where $\beta_1=\beta_2 ~\approx 1/4$, $n_1=n_2$ and $\lambda_1=\lambda_2$, $\mu_1=\mu_2$, $\Phi_1=\Phi_2$, $\Psi_1=\Psi_2$. We compare the full numerical results to the perturbative solution with $\omega_1=\omega_2=1$.}
			\label{fig:BetaZ2SymFields1}
		\end{figure}

		It turns out that for small values of $\beta$ the perturbative expansion to third order\footnote{We use the results from Section~\ref{ss:betaClass}, as we obtain numerical solutions for a few different values of $n_1=n_2$.} gives remarkably good approximation to the full solution. This can be seen in Fig.~\ref{fig:BetaZ2SymFields1}, where we show a comparison for all the relevant fields  with $\beta\sim\frac{1}{4}$, $n_1=n_2=n=1$ and $\omega_1=\omega_2=1$. Note that the plots show $\kappa$ instead of $k$, defined via $k=\xi\,\kappa$. We also include the quantities $N_1/N=N_2/N$ for later reference. They are defined on the CFT side and can be extracted from the bulk using holographic renormalization, as detailed in Section~\ref{sec:Nums}.
		
\begin{figure}[!ht]
			\centering
			\includegraphics[width=\textwidth]{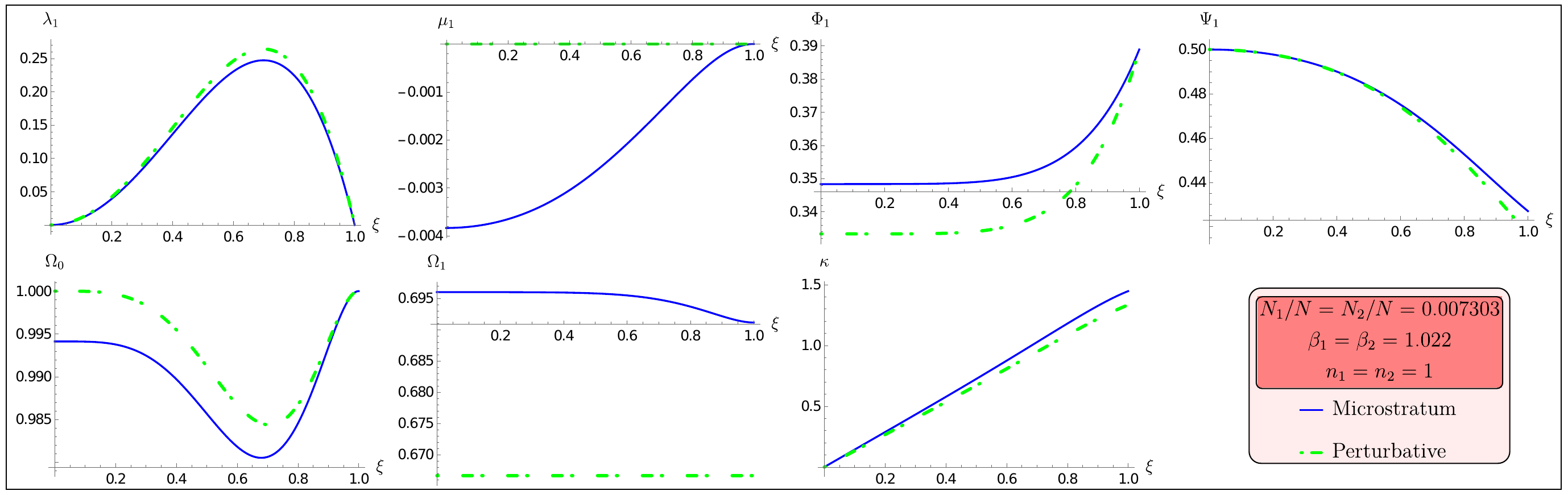}
			\caption{The non-zero fields in the $\mathbb{Z}_2$ symmetric beta class of solutions, where $\beta_1=\beta_2~\approx 1$, $n_1=n_2$ and $\lambda_1=\lambda_2$, $\mu_1=\mu_2$, $\Phi_1=\Phi_2$, $\Psi_1=\Psi_2$. We compare the full numerical results to the perturbative solution with $\omega_1=\omega_2=1$.}
			\label{fig:BetaZ2SymFields2}
		\end{figure}

\begin{figure}[!ht]
			\centering
			\includegraphics[width=\textwidth]{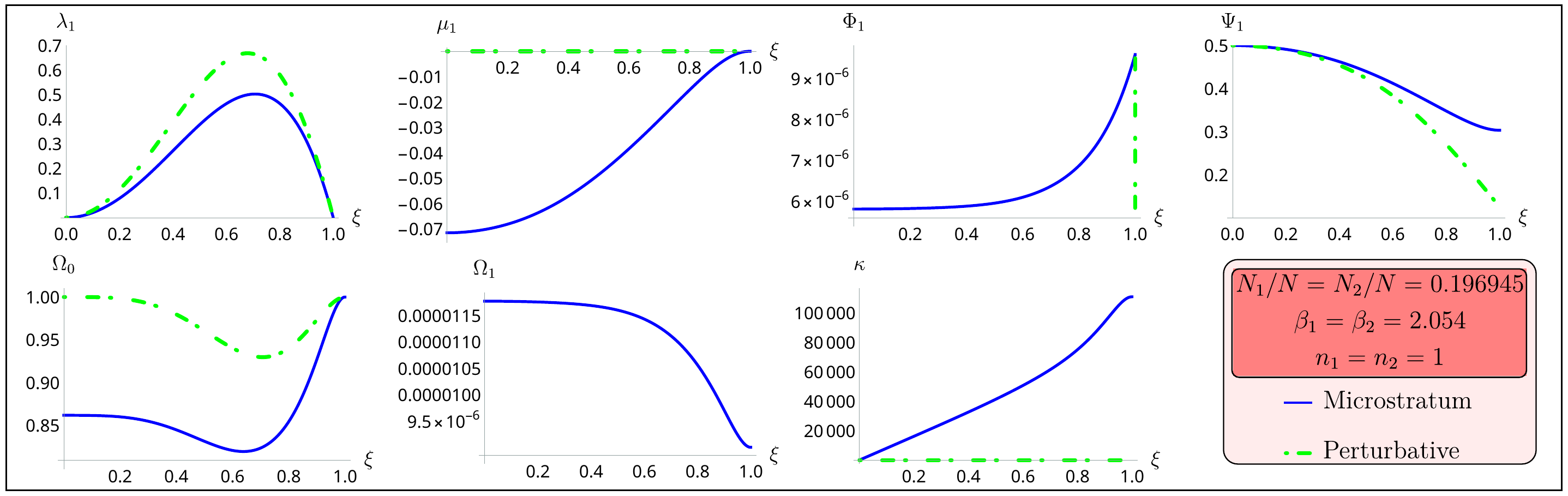}
			\caption{The non-zero fields in the $\mathbb{Z}_2$ symmetric beta class of solutions, where $\beta_1=\beta_2 \approx 2.05$, $n_1=n_2$ and $\lambda_1=\lambda_2$, $\mu_1=\mu_2$, $\Phi_1=\Phi_2$, $\Psi_1=\Psi_2$. We compare the full numerical results to the perturbative solution with $\omega_1=\omega_2=1$.}
			\label{fig:BetaZ2SymFields3}
		\end{figure}

		Figures~\ref{fig:BetaZ2SymFields2} and \ref{fig:BetaZ2SymFields3} are analogues of Figure~\ref{fig:BetaZ2SymFields1}, also with $n_1=n_2=n=1$ and $\omega_1=\omega_2=1$, but with larger values of $\beta$. They are meant to illustrate how the increasing amount of ``non-BPSness'' deforms the solution, all the way to the CTC bound, \eqref{CTC-bound}. The fields that change the most are $\Phi_1$, $\Omega_1$ and $\kappa$. In particular, $\kappa$ increases massively in value at infinity, as one gets closer to the maximum value of $\beta$, which together with \eqref{AdScond1} and \eqref{scaleRAdS} implies that $\Omega_1$ decreases quickly towards $0$ as $\xi\to1$.		
		
		For $n=1$, the CTC bound sets a maximum value of $\beta$ at $\sim2.05$, and so while $\beta\sim1/4$ is not infinitesimal, it is also not very large and the AdS throat will still be fairly shallow. Given that we are using only third order perturbation theory, the accuracy is surprisingly good, except for $\mu_1$. The latter receives its first non-trivial correction at fourth order\footnote{We have checked this for explicit values of $n_1=n_2$.} in $\beta$, hence the visible disagreement on our plots. Moreover, at third order $\Omega_1$ is just a constant, nevertheless, the horizontal scale shows that the fully back-reacted solution is very close to that value.
		
		Fig.~\ref{fig:BetaZ2SymFields3} displays the closest we have managed to get numerically to the CTC bound in between all the families of geometries we have explored. Perturbation theory is clearly not a good guide in this regime. At infinity ($\xi=1$), $\kappa$  has grown by 5 orders of magnitude to $\mathcal{O}\big(10^{5}\big)$, with $\Omega_1$ proportionally small by \eqref{AdScond1} and \eqref{scaleRAdS}. Thus, were we to also show the perturbative curve for $\Omega_1$, it will make the numerical result appear as identically zero. Similarly, for $\Phi_1$, hence we zoom these two graphs on the numerical results for clarity. As discussed in Section~\ref{sec:Nums}, the sign of the left-hand side of \eqref{CTC-bound} at infinity is dictated by the sign of $\Omega_1$ there, indicating that we are clearly close to the CTC threshold.
		
		We have similar plots for all the families of geometries we have constructed numerically, but these are looked at in more detail in Section~\ref{sec:Nums}. Here we just wanted to give the reader an indication of the accuracy of the the perturbation theory, as well as show that in certain regimes we have fully back-reacted geometries, providing  confidence in our results.
		
		\subsection{Perturbation theory around the pure-NS superstrata solutions}
		\label{sub:perturbation_pureNS}

		In the previous sections, we explored the construction of microstrata using perturbation theory around the vacuum. However, given that we have analytic solutions for BPS microstate geometries for any amplitude, it is natural to ask whether one can build new microstrata as perturbations on top of these BPS geometries.
		
		This kind of analysis is, in general, difficult to achieve because the BPS geometries are complicated and perturbations theory involves systems of very complicated, coupled differential equations. However, this work is facilitated in many ways by our ansatz. A first, trivial reason is that the ansatz reduces the problem to solving equations involving functions of only one variable. This alone is not sufficient, as the resulting equations can be arbitrarily complex and highly-coupled. Another simplification comes from a remarkable property of the ansatz:  the near-decoupling of two sets of mutually incompatible, BPS sectors. This lets one build an analytic BPS solution in one sector, and add supersymmetry-breaking excitations through the second sector. 
		
		The first step is thus to choose the BPS geometry that will be used as a base for the perturbation. Because it is very simple and already well-studied, it would be tempting to use the standard single-sector superstratum geometry, found by setting $\alpha_2 = 0$ in the double-sector solution \eqref{algBPS1}. However, as we have noted, the construction of non-BPS, $\alpha$-class solutions require  excitations  along the special locus.  Thus the proper starting point for such an analysis  is  the ``special locus'' solution in \eqref{eq:singleSectorBPSspec}.  This can then be used as a base to construct non-BPS linear excitations of the scalar $\nu_2$, and using the set of reduced equations on the special locus in Appendix~\ref{app:spLocusEqs}, one finds that at first order, the problem reduces to a system of first-order differential equations for the fields $\nu_2, \Phi_{3,4}, \Psi_{3,4}$ (see Appendix~\ref{app:linear_special_locus}).

		It is, however, simpler   to use the ``pure-NS'' solution of Section~\ref{ss:pureNS2} as a base. Indeed, at first order in the $\beta_2$-perturbation, one finds that only one field receives a contribution: the scalar $\lambda_2$. We consider the solution given in section~\ref{ss:pureNS2}, where we fix $\Omega_1$ by requiring~\eqref{goodmetbc}. There is still a symmetry that shifts the fields $\Phi_2$ and $\Psi_2$. In accordance with the discussion in section~\ref{ss:boundaryconds}, we first need to introduce the mode numbers of the $\beta_2$-perturbation by applying shifts to the gauge fields\footnote{We first subtract $\Omega_1/2$ from $\Phi_2$ so that the shifted field satisfy $\Phi_2(\xi=1) = c_{\phi_2}$.}:
		\begin{equation}
			\Phi_2 \to \Phi_2 - \frac{\Omega_1}2 + c_{\phi_2} \,,\qquad \Psi_2 \to \Psi_2 + \frac{1}{2}\, n_2 \,. 
		\end{equation}
		where, as will be explained in Section \ref{sec:Holography}, holography can be used to unravel the relation between the constant $c_{\phi_2}$ and the frequency:
		 \begin{equation}
		     c_{\phi_2} ~=~  \frac{(2n_1 + 1) \beta_1^2}{4(2n_1 + 1)^2 + \beta_1^2} + \frac{\hat{\omega}_2}2 \,;
		 \end{equation}
		this relation is valid at first order in $\beta_2$ but is exact in $\beta_1$.
		
		At first-order, the problem reduces to solving a single linear differential equation:
		\begin{equation} \label{eq:first_eq_wkb}
			\xi \partial_\xi \qty(\xi \partial_\xi \lambda_2) + V_0(\xi) \lambda_2 ~=~ 0
		\end{equation}
		where the function $V_0$ is defined in terms of the fields of the ``pure-NS'' BPS solution of Section \ref{ss:pureNS2}. Using the notation introduced in the Section \ref{ss:bps_eqs},
		\begin{equation} \label{eq:pot0wkb}
			V_0(\xi) ~=~ 16 (F_2^2 + H_0 \Phi_2^2)- 4 H_0 \Omega_1^2 \qty(e^{-2\mu_1} - e^{-\mu_1-\mu_2} \cosh\lambda_1)  \,.
		\end{equation}

		We wish to find bound states of the scalar $\lambda_2$, where the excitation lives entirely in the throat of the geometry, and compute the normal modes of these excitations. These normal modes, once fully back-reacted, translate to shifts in the energy of the microstratum. Thus, solving the problem defined by the equation \eqref{eq:first_eq_wkb} with suitable boundary conditions  (which will be discussed below) on the scalar $\lambda_2$, one can have access to the first-order corrections to the energy of these non-BPS microstates, and, on the CFT side, to the anomalous dimension of the constituent operators of the microstratum.
		
		\subsubsection{The WKB approximation} 
		\label{ss:wkb_approximation}
		
		One cannot solve equation \eqref{eq:first_eq_wkb} exactly, so one has to resort to approximations. We will use the WKB method, which is well-suited to dispersive or dissipative problems with large separation of scales: the potential must vary slowly with respect to the oscillations of the scalar. This method has often been used to compute normal or quasi-normal modes, and has recently been applied successfully to the computation of the quasi-normal modes of superstrata in \cite{Bena:2019azk,Bena:2020yii}. Here we will review the basics of the approximation. Its application to the problem at hand will be explained in the next section.
		
		 Since the WKB approximation is typically only accurate for large frequencies, we will have   to take $\omega_2 \gg 1$.  In most of the rest of this paper we have taken $\omega_2 = 1$ and so our  results here will probe a different extreme of the microstratum solution space.
		
		The WKB method has been originally developed for equations in the Schrödinger form:
		\begin{equation} \label{eq:schrodinger}
			\dv[2]{f}{x} - V(x) f ~=~ 0 \,.
		\end{equation}
		One can easily transform any second-order linear equation, such as \eqref{eq:first_eq_wkb}, into the Schrödinger form, by making a change of variables of the form
		\begin{equation} \label{eq:change_variable_schrodinger}
			\xi \,\to \, x(\xi) \,,\qquad \lambda_2(\xi) \,\to \, \Lambda(x(\xi))\, f(x(\xi)) \,,
		\end{equation}
		for suitably chosen functions $x$ and $\Lambda$.
		
		The WKB method consists in approximating the solutions of \eqref{eq:schrodinger} by 
		\begin{equation} \label{eq:wkb_solutions}
			f_{\text{WKB}}^{\pm}(x) ~=~ C_{\pm} \, \abs{V(x)}^{-1/4} \exp( \pm \int^x V(y)^{1/2} dy)
		\end{equation}
		where $C_{\pm}$ are integration constants. Formal estimations of the quality of this approximation can be done (see, for example,  \cite{bender_advanced_1999} for a detailed review). For the approximation to be close to the exact solution, the potential must obey
		\begin{equation} \label{eq:condition_1_wkb}
			\abs{V(x)^{-3/2} V'(x)} ~\ll~ 1 \,,
		\end{equation}
		which ensures the separation of scales between the oscillations of the solution and the variation of the potential.
		
		Notice that if $V(x) >0$, the solutions \eqref{eq:wkb_solutions} are exponentially growing or decaying, while if $V(x) < 0$, they are oscillating. The solutions break down at the \emph{turning points} of the classical motion, that is, when $V(x) = 0$. It is also easy to see that the conditions \eqref{eq:condition_1_wkb} cannot be satisfied in the neighbourhood of the turning points. It does not mean however that one must give up on using the WKB approximation: one can use a different approximation to construct the solutions in the problematic regions, then match with the WKB solution at the edges of these domains.
		
		Assuming that the first derivative of the potential does not vanish, one can approximate the potential around a turning point $x_\star$ by a linear function $V(x) \sim V'(x_\star) (x - x_\star)$. The solutions to the Schrödinger equation for a linear potential are well known and can be expressed in terms of the Airy functions
		\begin{equation} \label{eq:wkb_aroung_turning}
			f_\star(x) ~=~ C_1 \, \mathrm{Ai}\qty(V'(x_\star)^{1/3} (x - x_\star)) \,+\, C_2 \, \mathrm{Bi} \qty(V'(x_\star)^{1/3} (x - x_\star)) \,.
		\end{equation}
		
		One can then construct the full solutions to the problem in patches. Away from the turning points, the solution is well-described by \eqref{eq:wkb_solutions}. Around each turning point, one uses the solution \eqref{eq:wkb_aroung_turning}. It is then necessary to ensure that the solutions are regular, by relating the integration constants of neighbouring patches to each other. These conditions are called \emph{connection formulas}.
		
		For any matching to be possible, it is of course necessary that the domains of validity of the solutions \eqref{eq:wkb_solutions} and \eqref{eq:wkb_aroung_turning} overlap. This leads to another condition on the slope and curvature of the potential near the turning points:
		\begin{equation} \label{eq:condition_2_wkb}
			V'(x_\star)^{-4/3} V''(x_\star) ~\lesssim~ 1 \,.
		\end{equation}
		%
		
		\subsubsection{Quantization condition} 
		\label{ss:quantization}
		
		To obtain WKB approximations to  $\lambda_2$, and find the possible  frequencies, $\omega_2$,  one starts by rewriting (\ref{eq:first_eq_wkb}) in the Schrödinger form. There are many possible choices for \eqref{eq:change_variable_schrodinger} and this  choice of variable $x(\xi)$ can affect the precision of the WKB method as well as the simplicity of the treatment of boundary conditions. We make the same choice as in \cite{Bena:2019azk,Bena:2020yii}, as it makes the boundary conditions relatively simple:
		\begin{equation} \label{eq:new_var_x}
			x(\xi) ~\equiv~ \log(\frac{\xi}{\sqrt{1-\xi^2}}) ~=~ \log\rho \,. 
		\end{equation}

\begin{figure}[ht]
			\centering
			\includegraphics[width=.8\textwidth]{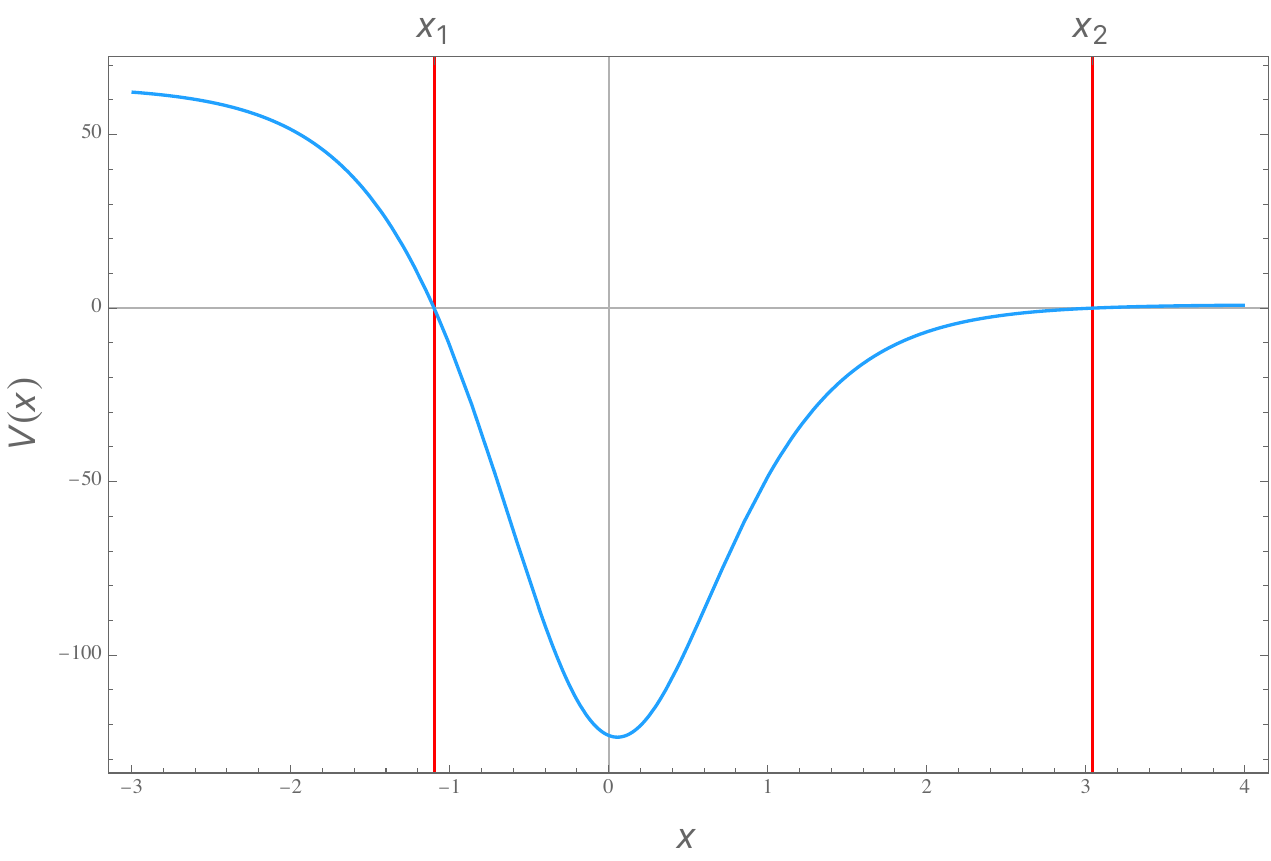}
			\caption{Shape of the potential for $\lambda_2$, written in terms of the variable $x$ introduced in \ref{eq:new_var_x}. The specific curve plotted here corresponds to $n_1 = 2$, $n_2 = 4$, $\beta_1 = 2$, and we have chosen to fix arbitrarily $\hat\omega_2 = 5$.}
			\label{fig:potential_shape}
		\end{figure}

		After this change of variable, the function $f$ defined by
		\begin{equation}
			\lambda_2(\xi) ~\equiv~ \sqrt{1-\xi^2} \, f(x(\xi))
		\end{equation}
		satisfies the Schrödinger equation \eqref{eq:schrodinger} with potential
		\begin{equation}
			V(x) ~=~ 1 - \frac{1}{ (1 + e^{2x})^{2} } (1 + V_0(x))
		\end{equation}
		where $V_0$ was defined in \eqref{eq:pot0wkb}. This potential depends on several parameters: the initial BPS superstrata parameters $\beta_1, n_1$, and the mode numbers for the excitation $n_2$, and $\omega_2$ through $c_{\phi_2}$. Figure \ref{fig:potential_shape} depicts the typical shape of the potential. It has two turning points noted $x_1$ and $x_2$ ; it limits to $1$ at positive infinity, and to $4 \, n_2^2$ at negative infinity.

		One can see numerically that the potential satisfies the conditions \eqref{eq:condition_1_wkb} and \eqref{eq:condition_2_wkb} around each turning point when the frequency of the scalar, $\omega_2$, is large. In practice, we find that $\omega_2 \sim \mathcal{O}(10-100)$ is sufficient to obtain a good approximation. 
		
		The solutions away from the turning points are given by
		\begin{equation}
			f(x) ~=~ \left\{ 
			\begin{aligned}
				& V(x)^{1/4} \qty( C^{(1)}_+  \exp(\int_x^{x_1} \sqrt{V(y)} dy) + C^{(1)}_- \exp(- \int_x^{x_1} \sqrt{V(y)} dy)) \,, \,\, x < x_1 \,,
				\\
				& (-V(x))^{1/4} \qty( C^{(2)}_+  \exp(i \int_{x_1}^{x} \sqrt{- V(y)} dy) + C^{(2)}_-  \exp(- i \int_{x_1}^{x} \sqrt{V(y)} dy))  \,, \,\, x_1 < x < x_2 \,,
				\\
				& V(x)^{1/4} \qty( C^{(3)}_+  \exp(\int_{x_2}^{x} \sqrt{V(y)} dy) + C^{(3)}_- \exp(- \int_{x_2}^{x} \sqrt{V(y)} dy)) \,, \,\, x_2 < x \,,
			\end{aligned}
			\right.
		\end{equation}
		where $C_{\pm}^{(I)}$ are constants of integration. All these constant are of course not independent: the solution to a second-order differential equation is determined by only two constants. Looking at the neighbourhood of each turning point, the solution is well described in terms of Airy functions \eqref{eq:wkb_aroung_turning}, and imposing that the solution is smooth leads to relations between the integration constants on both sides of the turning point. The details of the computation can be found in \cite{bender_advanced_1999}, one finds that the connection formulas are given by
		\begin{equation}
			\begin{aligned}
				C_+^{(2)} ~&=~ \frac 12 e^{i \pi/4} C_+^{(1)} + e^{-i \pi /4} C_-^{(1)} \,, \qquad& C_+^{(3)} ~&=~ - \sin\Theta \, C_+^{(1)} + 2 \cos\Theta \, C_-^{(1)} \,,
				\\
				C_-^{(2)} ~&=~ \frac 12 e^{-i \pi/4} C_+^{(1)} + e^{i \pi /4} C_-^{(1)} \,, \qquad& C_-^{(3)} ~&=~ \frac 12 \cos\Theta \, C_+^{(1)} + \sin\Theta \, C_-^{(1)} \,,
			\end{aligned}
		\end{equation}
		where
		\begin{equation}
			\Theta ~\equiv~ \int_{x_1}^{x_2} \sqrt{\abs{V(y)}} dy \,.
		\end{equation}
		
		We must now impose the boundary conditions. First, the excitation must be localized in the throat, and have a finite energy. As a consequence, the growing mode at infinity must be cancelled $C_+^{(3)} \,=\, 0$. Moreover, contrary to excitations over a black hole background, we must impose a smoothness condition at the origin, as the objective is to construct smooth microstate geometries. This translates to $C_+^{(1)} \,=\, 0$. One integration constant, $C_-^{(1)}$, is not fixed. This constant is the strength of the perturbation, and must be infinitesimally small compared to $\beta_1$.
		
		Combining these boundary conditions with the connection formulas leads to a quantization condition:
		\begin{equation} \label{eq:wkb_quantization}
			\cos\Theta = 0 \quad \Longleftrightarrow \quad \frac 1\pi \int_{x_1}^{x_2} \sqrt{\abs{V(y)}} dy ~=~ 2m_2 + \frac{1}{2} \,, \quad 2m_2 \in \mathbb{N} \,,
		\end{equation}
		where $m_2$ has been introduced so as to match the conventions of  \eqref{eq:beta1beta2-2}. This equation has to be understood as an implicit quantization condition for the frequency $\hat\omega_2$, which in turn determines the first-order correction to the energy of the microstratum. We are working in an approximation that is valid in the limit $m_2 \gg 1$.

		\subsubsection{Normal modes of $\lambda_2$} 
		\label{ssub:normal_modes_of_lambda_2_}
		
		We first look at a simple application of the quantization condition:  set the background to be empty AdS by putting the amplitude of the superstratum to zero, $\beta_1 = 0$. The perturbation theory around this background has been studied in Section~\ref{ss:betaClass}, we know that the frequencies $\hat\omega_2 \equiv \omega_2$ of the normal modes of $\lambda_2$ must be positive integers. If we additionally impose $n_2 = 0$, the integral in \eqref{eq:wkb_quantization} can be computed explicitly, and one finds:
		\begin{equation}
			2m_2 + \frac12 = \frac 1\pi \int_{x_1}^{x_2} \sqrt{\abs{V(y)}} dy ~=~ \sqrt{\omega_2^2 - \frac14} \,-\, \frac12 ~\sim~ \omega_2 - \frac12 - \frac{1}{8 \, \omega_2} + \dots \,.
		\end{equation}
		From that relation and the quantization condition, we find that $2m_2 = \omega_2 - 1$, and that the error of the WKB approximation is of order $1/ (16 m_2)$.
		
		Returning to the general case, $\beta_1 \neq 0$, we can now solve \eqref{eq:wkb_quantization} numerically, and find the quantization conditions on $\hat\omega_2$. For the purposes of the numerics, we choose as an example $n_1=2$, $n_2=4$. We are mostly interested in the limit where the geometry develop a very long throat, mimicking the geometry of a black hole. This is obtained when 
		\begin{equation}
		    \beta_1 \to \beta_1^{(0)} = \frac{4 n_1 + 2}{\sqrt{4 n_1 + 1}} ~=~ \frac{10}3 \,.
		\end{equation}
		The value $\beta_1^{(0)}$ also corresponds to the CTC bound: all the geometries with $\beta_1 > \beta_1^{(0)}$ have closed timelike curves at the boundary.
		
		\begin{figure}[t]
			\centering
			\includegraphics[width=0.7\textwidth]{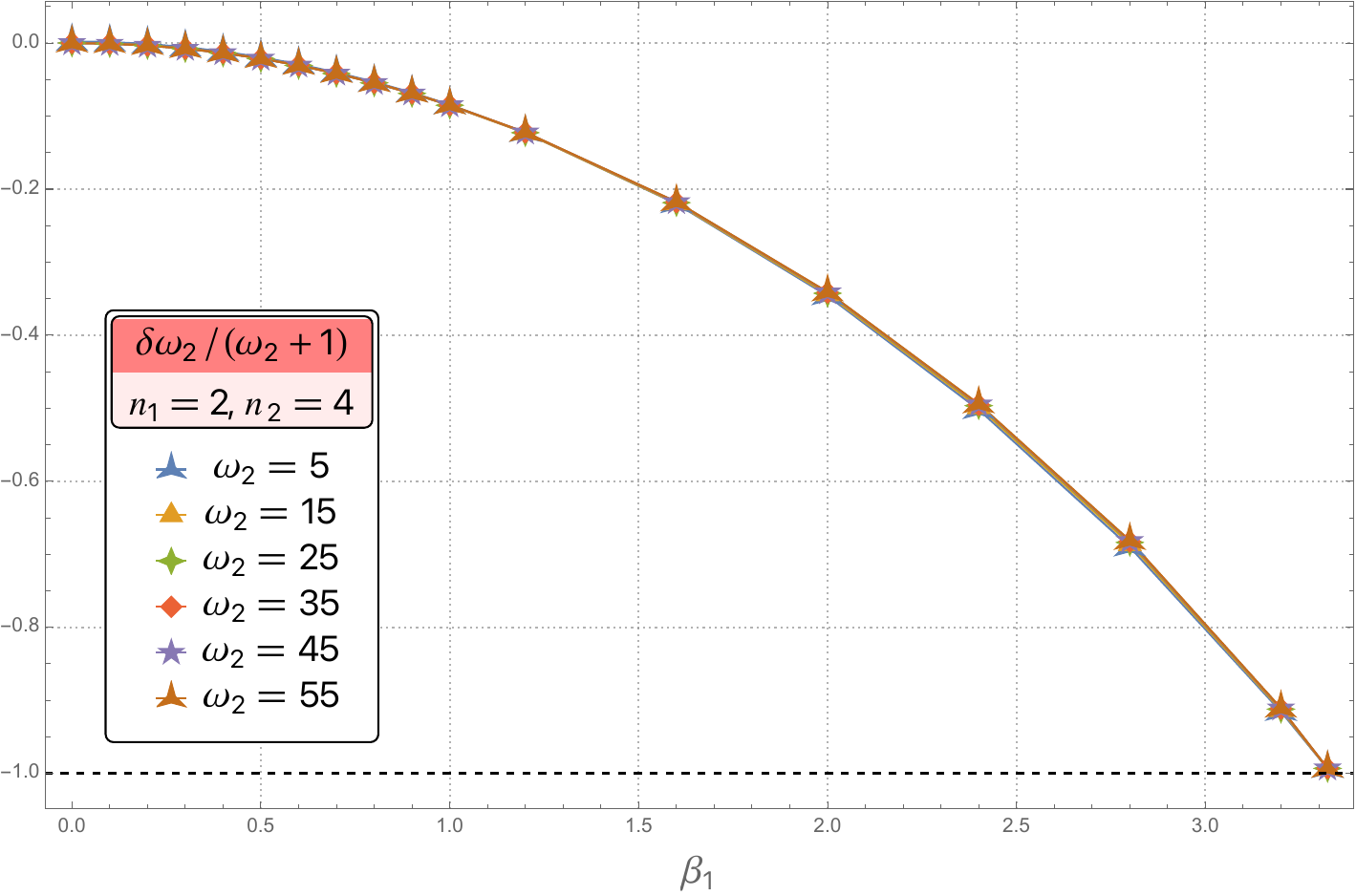}
			\caption{The ratio between the shift in the frequency $\delta\omega_2$ and $1+\omega_2$, as a function of $\beta_1$, for $n_1 = 2$, $n_2 = 4$, and for several values of $\omega_2$. Up to the precision of the WKB approximation, the curves superimpose.}
			\label{fig:resultsWKB}
		\end{figure}

		Figure \ref{fig:resultsWKB} shows the evolution of 
		\begin{equation}
		\label{eq:WKBshifts}
		    \frac{\delta\omega_2}{\omega_2 + 1} ~\equiv~ \frac{\hat\omega_2 - \omega_2}{\omega_2 + 1}
		\end{equation}
		as a function of $\beta_1$, for different values of $\omega_2$. We first note that $\delta\omega_2 \leq 0$, which, as we will see in Section~\ref{sec:CFT}, is expected from the fact that this quantity is interpreted in the CFT as a binding energy. We also note that $\frac{\delta\omega_2}{\omega_2 + 1}$ seems to be independent of $\omega_2$.
		
\begin{figure}[t]
			\centering
			\includegraphics[width=0.48\textwidth]{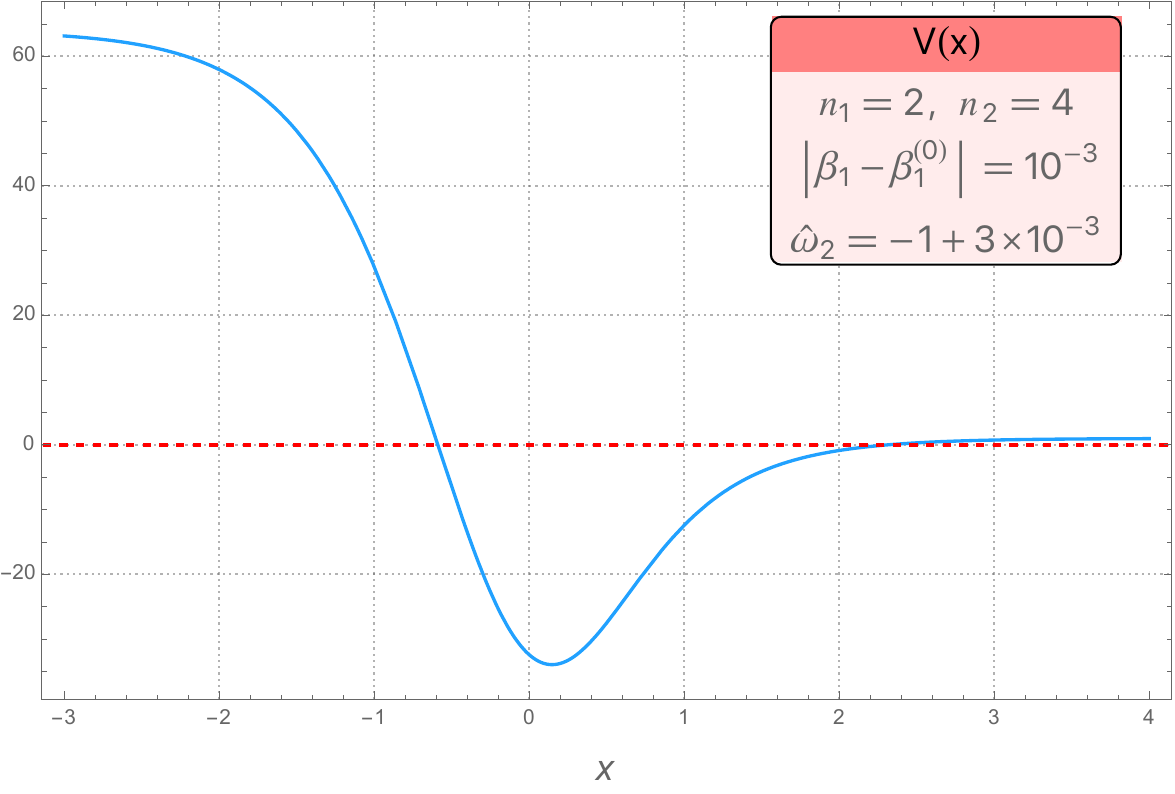}
			\includegraphics[width=0.48\textwidth]{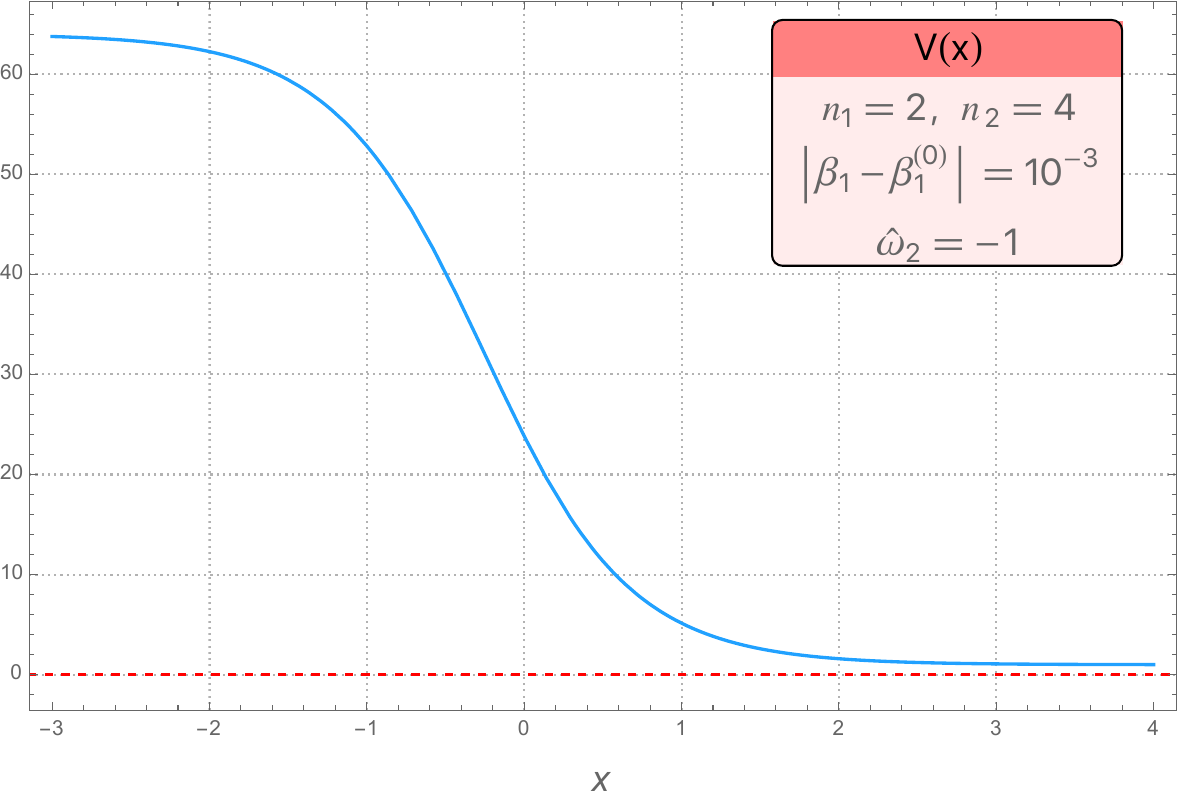}
			\caption{The potential $V$ of the scalar $\lambda_2$, close to the CTC bound, $\beta_1 = \beta_1^{(0)} - 10^{-3}$, and with $n_1=2, n_2=4$. To the left, a typical plot of the potential when the parameter $\hat\omega_2$ is chosen very close but above the limit $-1$, here at $\hat\omega_2 = -1 + 3 \times 10^{-3}$. It is possible to form bound states in this range of parameters. To the right, at $\hat\omega_2 = -1$, the potential is everywhere positive, there is no bound state.}
			\label{fig:potential_limit}
		\end{figure}

		At the CTC bound, the WKB results suggest that $\hat\omega_2 \to -1$. This can be understood without resorting to the WKB approximation: the potential $V$ of the scalar field $\lambda_2$ undergoes a transition when $\hat\omega_2 \to -1$, that is visible on Figure~\ref{fig:potential_limit}. Indeed, when $\hat\omega_2 > -1$, and sufficiently close to the CTC bound, there is always a region where the potential is negative, it is therefore possible to form a bound state. That is no longer true when $\hat\omega_2 = -1$, because the potential reduces to
		\begin{equation}
		\begin{aligned}
		  V(x) ~=~ 1 +   \frac{1}{(1 + e^{2x})^2}  \bigg( & (3+4n_1-2n_2)(1+4n_1-2n_2-4e^{2x})   \\ 
		& -   \frac{4(1+2n_1)(1+2n_1-2n_2-2e^{2x})}{1-(1+e^{-2x})^{-(2n_1+1)} (\frac{\beta_1}{4n_1+2})^2 }\bigg)  \,,
		\end{aligned}
		\end{equation}
		which is everywhere positive.  We will come back to lower bounds on $\hat\omega_2$  in Section~\ref{sub:momexcit} after discussing the CFT interpretation of the dual states and then in Section~\ref{ss:NumResults} as part of our numerical analysis.

		In the perturbative regime, we can make a further check on these numerical results, by comparing them to the perturbative solution around the vacuum. Indeed, the terms proportional to $\beta_1^2$ and $\beta_2^2$ in the shift $\delta\omega_2$ can be accessed by the perturbation theory around the vacuum: We can reproduce the computations of Section~\ref{ss:betaClass}, that were done for $\omega_2 = 1$, for different values of $\omega_2$. The general analysis is complicated, but we can restrict ourselves to the case $n_1 = 2$, $n_2 = 4$.

		\begin{table}[ht] \label{fig:table_comparison}
			\begin{equation*}
				\begin{array}{c|c|c}
					\text{$\omega_2$} & \text{From WKB} & \text{From vacuum perturbation theory} \\[.2em]
					\hline & & \\[-1em]
					1 & -0.173513 & - \dfrac{6}{35} \ \approx -0.17143 \\[.8em]
					\hline & & \\[-.8em]
					5 & -0.520276 & -\dfrac{14844427}{28601650} \ \approx -0.51901 \\[.8em]
					\hline & & \\[-.8em]
					9 & -0.862479 & -\dfrac{102879754}{119409675} \ \approx -0.86157 \\[.8em]
					\hline & & \\[-.8em]
					13 & -1.20262 & -\dfrac{154958717}{128927388} \ \approx -1.20191 \\[.8em]
					\hline & & \\[-.8em]
					17 & -1.54174 & -\dfrac{2066874954282}{1341119288509} \ \approx -1.54116
				\end{array}
			\end{equation*}
			\caption{Comparison between the computations of the frequency shift from the WKB analysis, and from the perturbation theory around the vacuum, for a few values of $\omega_2$. The numbers shown correspond to the term proportional to $\beta_1^2$ in the expansion of $\delta\omega_2$.}
		\end{table}

		\begin{figure}[!ht]
			\centering
			\includegraphics[width=0.48\textwidth]{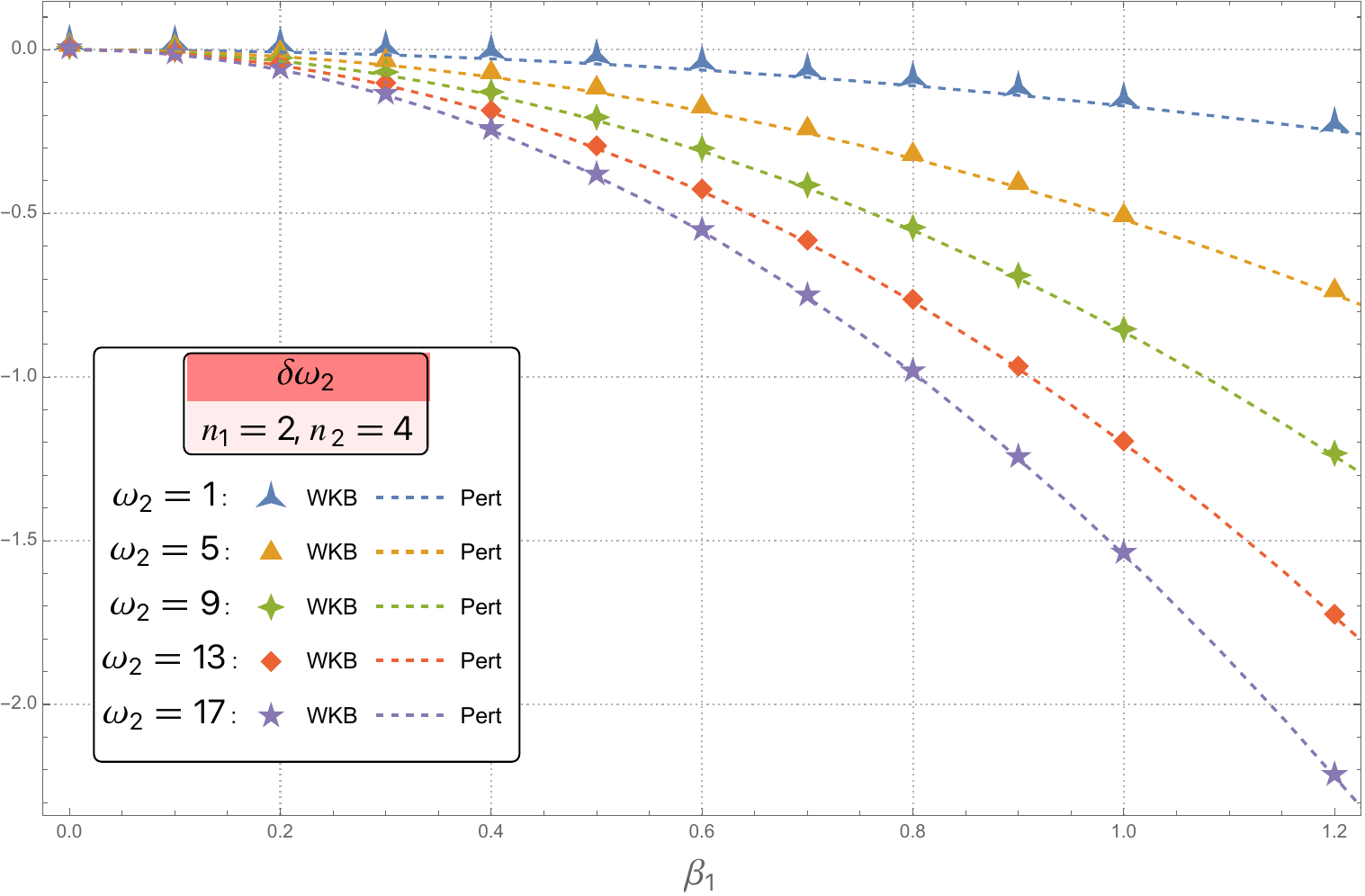}
			\includegraphics[width=0.48\textwidth]{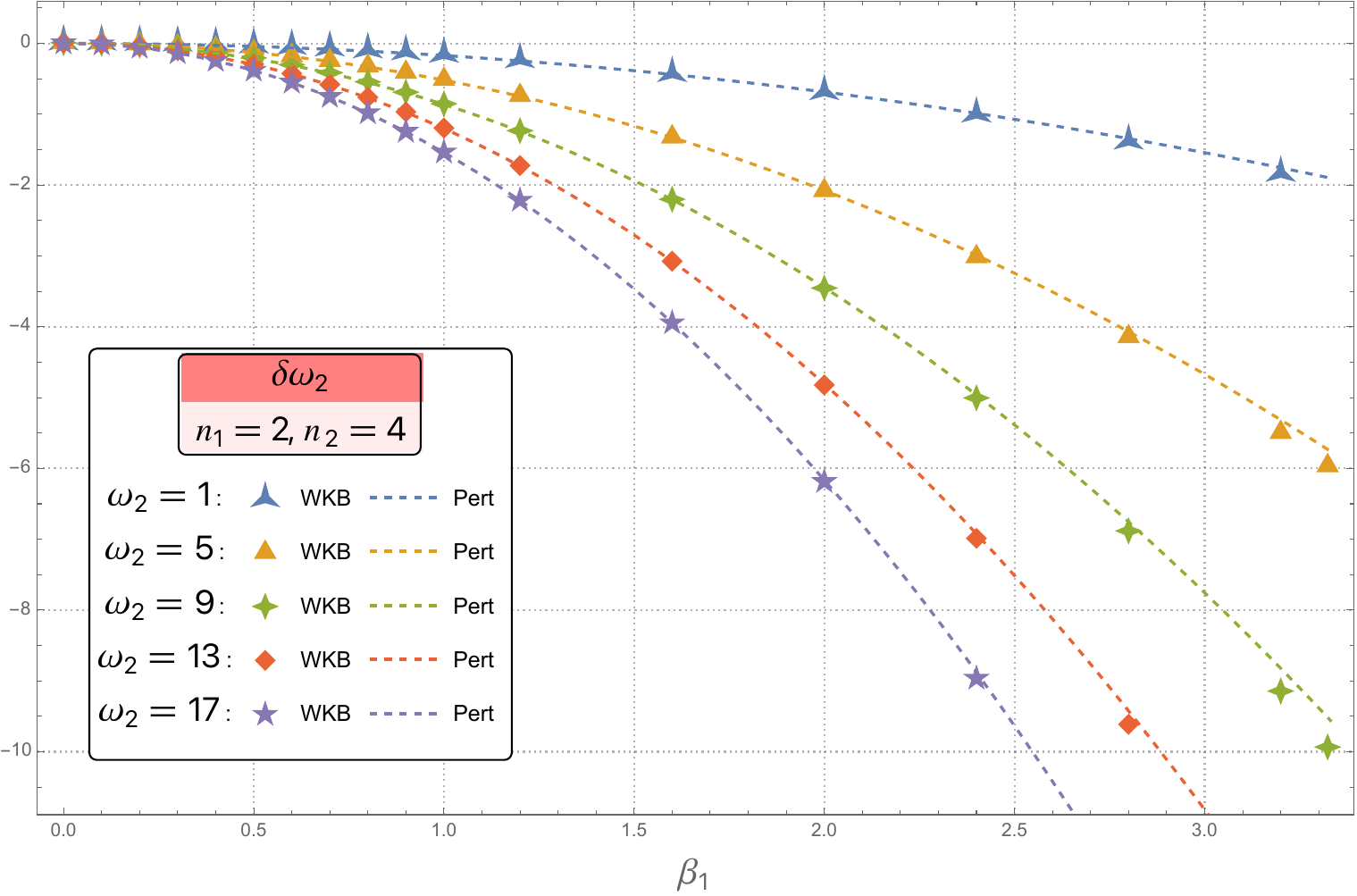}
			\caption{The shift in the frequency, $\delta\omega_2$, as a function of $\beta_1$, for several values of $\omega_2$. The points are the numerical results from the WKB approximation, while the dashed lines are the perturbative results, exact up to order $\beta_1^2$. The left figure is a zoom of the right figure in the perturbative region. Despite the smallness of the values for $\omega_2$, the WKB approximation is accurate.}
			\label{fig:compar_wkb_pert}
		\end{figure}		

		The main interest of the WKB approximation is that its results are valid for finite values of $\beta_1$, while the perturbation theory is only giving access to expansions around the vacuum. Nevertheless, we can check the validity of the WKB approximation in the perturbative regime using perturbation theory. It can also teach us about the perturbation theory in the regime $\omega_2 \gg 1$, that is hard to attain with perturbatively.

		Figure~\ref{fig:compar_wkb_pert} shows the agreement between the WKB results and the perturbation theory: we plot simultaneously the values of $\delta\omega_2$ computed using the WKB method, and its perturbative expression\footnote{This computation is detailed in a few examples in Section~\ref{sec:Holography}.} at order $\beta_1^2$. In Table~\ref{fig:table_comparison}, we compare the exact coefficient of order $\beta_1^2$ appearing in $\delta\omega_2$ from perturbation theory, with the ones obtained by fitting a polynomial in $\beta_1$ through the WKB results: we see that the error is reduced when $\omega_2$ is large.

		We can also extract perturbative information from the WKB results. For example, on Figure~\ref{fig:resultsWKB}, we have seen that the quantity $\delta\omega_2 / (1 + \omega_2)$ is independent of $\omega_2$. By fitting the results obtained at large $\omega_2$ and for $n_1 = 2, n_2 = 4$ with a quadratic function, we find
		\begin{equation}
		    \delta\omega_2 ~\approx~ -0.0843 (1 + \omega_2)\beta_1^2 + \dots\,.
		\end{equation}
		This coefficient, along with the functional dependence of $\delta\omega_2$ in $\omega_2$, constitute predictions that a more advanced perturbative expansion, in which $\omega_2$ is kept free, could check.
		\section{The CFT interpretation}
		\label{sec:CFT}
		
		The three-dimensional supergravity that underlies our work can be embedded in Type IIB string theory compactified on a 4-dimensional Calabi-Yau $\mathcal{M}$ times a circle $S^1$. The four-dimensional compact space will play no role in our analysis and can be taken to be either $\mathbb{T}^4$ or K3. The remaining six-dimensional space-time has AdS$_3\times S^3$ asymptotics and thus our solutions are holographically dual to states of the D1-D5 CFT. In this section we review the operator content of this CFT and describe the heavy states that are dual to the geometries discussed in this work. As we will see it is sometimes convenient to refer to the CFT free locus, where the theory reduces to the orbifold $\mathcal{M}^N/S_N$ ($N=n_1 n_5$ is the product of the numbers of D1 and D5 branes and $S_N$ is the permutation group). The characterisation in terms of orbifold is useful because the elementary constituents of our heavy states, even the non-BPS ones, are by themselves BPS and thus are expected to be stable across the moduli space of the CFT. Convenient observables to compare the gravity and the CFT free locus pictures are the expectation values of chiral primary operators in the states dual to the geometries: when the states are supersymmetric, these expectation values do not depend on the moduli. We will adopt the common nomenclature and refer to these expectation values as vacuum expectation values (VEV), but it should be kept in mind that they are not computed in the NS vacuum but in non-trivial states; equivalently these VEVs can be thought as 3-point correlators and they will be the focus of Section~\ref{ssech:preshol}.
		
		\subsection{Supergravity operators of the D1-D5 CFT}
		
		The six-dimensional supergravity obtained by compactifying type IIB on $\mathcal{M}$ contains a gravity multiplet, $\sigma$, and $N_f$ tensor multiplets, $s^{(f)}$,  with $N_f=5$ or $21$ if $\cal{M}=$$\,\mathbb{T}^4$ or K3. The scalar chiral primary operators (CPO) of the D1-D5 CFT are dual to the $S^3$-harmonics of these fields, and are denoted respectively as $\sigma_n$ and $s^{(f)}_n$: they have conformal dimension $(h,\bar h)=(\frac{n}{2},\frac{n}{2})$, with $n$ a positive integer starting from 1 for the tensor fields $s^{(f)}_n$, and from 2 for the gravity fields $\sigma_n$.
		
		The three-dimensional truncation we use does not allow all these operators to have freely tunable VEVs: only the operators dual to the fields that admit homogeneous and normalisable solutions of the three-dimensional supergravity equations can take arbitrary expectation values. This does not imply that the other operators necessarily have vanishing VEVs, but these VEVs will be determined in terms of the freely tunable ones by the non-linearity of the three-dimensional supergravity equations. Thus only a subset of the D1-D5 states can fit in our truncation.  However, as we have seen, this subset is large enough to encompass non-trivial non-BPS states, including non-supersymmetric deformations of BPS states with long AdS$_2$ throats. 
		
		For the restricted $U(1)^2$ truncation considered in \cite{Mayerson:2020tcl,Ganchev:2021pgs}, the possible homogeneous deformations and the dual operators where identified in \cite{Ganchev:2021ewa}. In that restricted truncation, only two fields, $\nu_1$ and $\lambda_1$, have normalisable homogeneous modes. The dual of $\nu_1$ is the component of spin $(\frac{1}{2},\frac{1}{2})$ of one of the tensor multiplet CPOs\footnote{For the supergravity fields, we follow the notation of~\cite{Rawash:2021pik} and references therein, which is motivated by the analysis of the six-dimensional supergravity sketched in Appendix~\ref{app:uplift}.}, $s^{(7)}_1$. At the orbifold locus, $s^{(7)}_1$ is described by a bilinear of free fermions (see for instance \cite{Avery:2010qw,Hampton:2018ygz} for a summary of the orbifold CFT construction) that we schematically indicate as 		
		\begin{equation} \label{eq:1/2CPO}
			s^{(7)(\frac{1}{2},\frac{1}{2})}_1 \sim \; \sum_{r=1}^N \,\Bigl |\frac{1}{2},\frac{1}{2}\Bigr \rangle^{(r)} + \ldots \;,
		\end{equation}
		where the sum is over the $N$ copies of the sigma-model on $\mathcal{M}$ and the dots stand for possible corrections where the elementary fields act on different copies\footnote{In the K3 case such corrections are absent for the low dimension field in~\eqref{eq:1/2CPO}, while for the dimension 2 states in~\eqref{eq:11CPO} they are spelled out explicitly in~\cite{Rawash:2021pik}.}. Analogously, we denote by $O^{(1,1)}_2$ the state of spin $(1,1)$ dual to $\lambda_1$, which is a linear combination of the $(1,1)$ components of the operators $s_2^{(6)}$ and $\sigma_2$:
		\begin{equation}\label{eq:11CPO}
			O^{(1,1)}_2 \equiv \sqrt{3} \,s_2^{(6)(1,1)}-\sigma_2^{(1,1)} \sim \; \sum \, |1,1\rangle^{(r_1 r_2 r_3)} + \ldots \;.
		\end{equation}
		In the orbifold CFT picture, $s_2^{(6)(1,1)}$ and $\sigma_2^{(1,1)}$ are two orthogonal linear combinations of the supersymmetric twist field of order 3 and of an operator that is quadrilinear in the elementary fermions.  The state in~\eqref{eq:11CPO} is such that the single-trace terms with four fermions cancels. Thus, the leading term on the right-hand side of~\eqref{eq:11CPO} is build by gluing three copies and the sum is over the possible 3-cycles $(r_1 r_2 r_3)$. 
		
		The more general truncation we use in this paper contains a larger set of homogeneous modes, including also the fields $\nu_2$, $\lambda_2$ and $m_5\pm m_6$, in addition to $\nu_1$ and $\lambda_1$. The states dual to these new modes are different spin components of the states introduced above. Indeed $SO(4)$ rotations, dual to the $SU(2)_L\times SU(2)_R$ R-symmetry of the CFT, permute the fields $\nu_1$ and $\nu_2$ and the fields $\lambda_1$, $m_5\pm m_6$ and $\lambda_2$. The full holographic map\footnote{The map depends on the choice of phase for the fields: for example the state $|\frac{1}{2},-\frac{1}{2}\rangle$ turns into $|-\frac{1}{2},\frac{1}{2}\rangle$ if one conjugates the field $\nu_2$. The map given here corresponds to the choice of phase used in the non-BPS solutions of this article.} is summarised below:
		\begin{subequations}\label{eq:holoMap}
			\begin{equation}
				\nu_1 \to s^{(7)(\frac{1}{2},\frac{1}{2})}_1 \;,\quad \nu_2\to s^{(7)(\frac{1}{2},-\frac{1}{2})}_1\;,
			\end{equation}
			\begin{equation}
				\lambda_1 \to O^{(1,1)}_2 \,, \quad m_5+m_6\to O^{(0,1)}_2 \;,\quad  m_5-m_6 \to O^{(1,0)}_2\;,\quad \lambda_2\to O^{(1,-1)}_2\,,
			\end{equation}
		\end{subequations}
		where as before the superscript $(j,\bar j)$ denotes the spin of each operator with respect to the R-symmetry group $SU(2)_L\times SU(2)_R$.
		
		\subsection{Heavy states}
		\label{sec:heavystates}
		The states described above are ``light", having a conformal dimension that does not scale with the CFT central charge $c=6N$. We will also refer to such light operators as single-particle (or single-trace) operators, as they are dual to elementary supergravity fields. Supergravity solutions like the ones described in Section~\ref{sec:3D-Sugr}, that are large deformations of AdS$_3\times S^3$, are dual to ``heavy" states whose conformal dimension grows like $c$ in the large central charge limit.
		
		Heavy states can be realised as multi-particle (or multi-trace) operators made by a large (of order $c$) number of single-trace constituents. For supersymmetric states, a precise characterisation of the multi-particle operators dual to the supergravity solutions can be given in the language of the $\mathcal{M}^N/S_N$ orbifold sigma model \cite{Skenderis:2006ah,Kanitscheider:2006zf,Kanitscheider:2007wq,Giusto:2015dfa}. We first introduce the basic formalism for a simple geometry, the single two-charge superstratum corresponding to the solution of Section~\ref{sec:doublestratum} with $\alpha_2=0$ and $n_i=0$, and then generalise to three-charge and non-BPS solutions. 
	
		\subsubsection{Two-charge states}
		\label{sub:2chg}
			
		The single superstratum geometry depends on one continuous parameter $\alpha_1$ and, at linear order in $\alpha_1$, the only field excited is $\nu_1$ which, as we have seen, is dual to the state in~\eqref{eq:1/2CPO}. This indicates that the heavy state dual to the geometry for finite values of $\alpha_1$ is a multi-particle state made by many copies of a basic constituent of spin $(\frac{1}{2},\frac{1}{2})$ that we indicate as $|\frac{1}{2},\frac{1}{2}\rangle$. A more precise definition is given by the following coherent-like state~\cite{Skenderis:2006ah,Kanitscheider:2006zf,Kanitscheider:2007wq}
		\begin{equation}
			\label{eq:coherentalpha1}
			|\alpha_1\rangle \equiv \sum_{p=0}^N A_1^p \,A_0^{N-p} \,\Bigl (\Bigl |\frac{1}{2},\frac{1}{2}\Bigr\rangle\Bigr)_*^p \,(|0\rangle)^{N-p}\quad \mathrm{with}\quad A_1 = \sqrt{N}\,\frac{\alpha_1}{2}\,,\, A_0 = \sqrt{N}\,\left(1-\frac{\alpha_1^2}{4}\right)^{1/2}\,,
		\end{equation}
		where $|0\rangle$ is the NS-sector\footnote{We represent the states in the NS sector in this article. Of course an alternative representation can be given by spectrally flowing to the R sector. As far as one works with asymptotically AdS solutions the two representations are equivalent, but one should keep in mind that only the R-sector states can be extended to asymptotically-flat solutions.} vacuum and $(|\frac{1}{2},\frac{1}{2}\rangle)_*^p$ denotes a multi-particle state made by $p$ copies of the state $|\frac{1}{2},\frac{1}{2}\rangle$, which will be more explicitly defined below. In the large $N$ limit, the sum over $p$ in \eqref{eq:coherentalpha1} is peaked over the average value 
		\begin{equation}
			{\bar p} = A_1^2 = N\,\frac{\alpha_1^2}{4}\equiv N_1 \,.
		\end{equation}
		This can be checked by calculating the norm of the summands in the state \eqref{eq:coherentalpha1}, see for instance~\cite{Giusto:2015dfa}. For many purposes it suffices to approximate the sum with its average:
		\begin{equation}\label{eq:leadingalpha1}
		|\alpha_1\rangle \sim \Bigl(\Bigl |\frac{1}{2},\frac{1}{2}\Bigr \rangle\Bigr )_*^{N_1}\quad \mathrm{with} \quad \frac{N_1}{N}=\frac{\alpha_1^2}{4}\,;
		\end{equation}
		the precise form of the state \eqref{eq:coherentalpha1} is however needed to compute expectation values of operators that mix the constituents $|\frac{1}{2},\frac{1}{2}\rangle$ and $|0\rangle$ in the heavy state. 
		
		A crucial point for us is the definition of the product $(|\frac{1}{2},\frac{1}{2}\rangle)_*^p$ and, for this purpose, we must recall a few basic facts about the orbifold $\mathcal{M}^N/S_N$. 
		All operators must be invariant under the exchange of the copies of the sigma-model on each $\mathcal{M}$: 
		a simple way to achieve this for the heavy state is to define each term in~\eqref{eq:coherentalpha1} as
		\begin{equation}\label{eq:multistar}
		\Bigl(\Bigl |\frac{1}{2},\frac{1}{2}\Bigr \rangle\Bigr)_*^p \equiv \sum_{\genfrac{}{}{0pt}{}{r_1,r_2,\ldots,r_p}{r_1\not=r_2\not\ldots r_p}} \Bigl |\frac{1}{2},\frac{1}{2}\Bigr \rangle^{(r_1)} \Bigl |\frac{1}{2},\frac{1}{2}\Bigr \rangle^{(r_2)}\ldots  \Bigl |\frac{1}{2},\frac{1}{2}\Bigr \rangle^{(r_p)}\,,
		\end{equation}
		so that any of the $N$ copies is either occupied by {\it one} CPO $|\frac{1}{2},\frac{1}{2}\rangle^{(r)}$ or by the vacuum. Another approach, more natural at a generic point in the CFT moduli space, is to consider the (non-singular) OPE product of $p$ single-particle operators
		\begin{equation}\label{eq:multi}
		\Bigl(\Bigl|\frac{1}{2},\frac{1}{2}\Bigr \rangle\Bigr )^p \equiv \lim_{z_i\to 0} s^{(7)(\frac{1}{2},\frac{1}{2})}_1(z_1) \ldots s^{(7)(\frac{1}{2},\frac{1}{2})}_1(z_p) \,|0\rangle\,.
		\end{equation}
		This is {\it not} the definition of the multi-particle operator in \eqref{eq:multistar}: to obtain the latter one must remove the copies where more than one operator $|\frac{1}{2},\frac{1}{2}\rangle^{(r)}$ acts simultaneously and also the multi-trace corrections mentioned in~\eqref{eq:1/2CPO}.   For example for $N=p=2$ we have
		\begin{equation}\label{eq:multistarex}
		\Bigl(\Bigl |\frac{1}{2},\frac{1}{2}\Bigr \rangle\Bigr)_*^2 \equiv  2\, \Bigl |\frac{1}{2},\frac{1}{2}\Bigr \rangle^{(1)} \Bigl |\frac{1}{2},\frac{1}{2}\Bigr \rangle^{(2)}\,,
		\end{equation}
		while the definition through the OPE limit yields extra terms where two operators $s^{(7)}_1$ act on the same copy
		\begin{equation}\label{eq:multiex}
		\Bigl(\Bigl|\frac{1}{2},\frac{1}{2}\Bigr \rangle\Bigr )^2 \equiv 2\, \Bigl |\frac{1}{2},\frac{1}{2}\Bigr \rangle^{(1)} \Bigl |\frac{1}{2},\frac{1}{2}\Bigr \rangle^{(2)} + |{1},{1}\rangle^{(1)}  |0\rangle^{(2)}+  |0\rangle^{(1)} |{1},{1}\rangle^{(2)}\,.
		\end{equation}
		The fact that the superstratum geometry is dual to the state \eqref{eq:coherentalpha1} with the definition \eqref{eq:multistar} has been confirmed by several precision-holography checks \cite{Giusto:2015dfa,Giusto:2019qig,Rawash:2021pik,Ceplak:2021wzz}, where the difference between \eqref{eq:multistar} and \eqref{eq:multi} plays a crucial role. 
		
		This raises a very interesting question: what is the gravity dual of a heavy state defined in terms of the multi-particle excitations \eqref{eq:multi} that is more natural from the OPE point of view. As discussed in Section~\ref{sec:nonBPS}, in the three-dimensional truncation there is a family of supersymmetric solutions parametrised by the two parameters $\alpha_1,\beta_1$: at one extreme is the geometry with $\beta_1=0$, whose dual is the state $|\alpha_1\rangle$ in \eqref{eq:coherentalpha1}; at the other extreme is the so-called ``special-locus" solution, with $\beta_1=-\frac{\alpha_1^2}{4}$, that exhibits a remarkable simplicity and, in particular, its scalar matrix $m$ has a single non-trivial eigenvalue and thus it enjoys an $SO(3)$ symmetry. As we will explain in Section \ref{ssech:preshol}, this structure of the matrix $m$ is naturally explained by assuming that the state dual to the ``special-locus" geometry has the leading order form\footnote{It is easier to derive the relation between the supergravity parameter $\alpha_1$ and $N_1$ by looking at 3-charge case~\eqref{eq:alp1} and~\eqref{eq:alp1b}, where it is possible to use~\eqref{eq:hhbar} to match the momentum of the gravity solution with that of the corresponding CFT state. The 2-charge result is obtained simply by setting $n_1=0$.}
		\begin{equation}\label{eq:slstate}
		|\alpha_1\rangle_{s.l.}\sim \Bigl (\Bigl |\frac{1}{2},\frac{1}{2}\Bigr \rangle\Bigr )^{N_1}\quad \mathrm{with} \quad \frac{N_1}{N}=\frac{2\alpha_1^2}{8+\alpha_1^2}\,,
		\end{equation}
where the multi-particle $(|\frac{1}{2},\frac{1}{2}\rangle)^{N_1}$ is defined as in \eqref{eq:multi}. This identification of the holographic dual of the special locus is one of the central results of this paper.  
		 
		The precise form of the state $|\alpha_1\rangle_{s.l.}$, analogous to \eqref{eq:coherentalpha1}, has not yet been determined, but it is important to keep in mind that \eqref{eq:slstate} represents only the leading order term of the state and, like in \eqref{eq:coherentalpha1}, the actual state also contains terms with $(|\frac{1}{2},\frac{1}{2}\rangle)^{p}$, where $p$ is spread around the average value $N_1$; these terms are relevant when computing generic 1-point functions of CPOs in the state $ |\alpha_1\rangle_{s.l.}$.

		\subsubsection{Three-charge states}
		\label{sub:3chg}
		
		It is straightforward to generalise the CFT description of the states to the 3-charge BPS geometries and then to the non-BPS solutions constructed in the previous section. For simplicity, and since it will be sufficient for the scope of this paper, we will only give the leading order form of the state, in the spirit of \eqref{eq:leadingalpha1} and~\eqref{eq:slstate}. Adding the momentum charge $n$ amounts to replacing the CPO $|\frac{1}{2},\frac{1}{2}\rangle$ by its descendant $L_{-1}^n|\frac{1}{2},\frac{1}{2}\rangle$. Thus the three-charge superstratum, \eqref{algBPS1} with $\alpha_2=0$, and the special locus, \eqref{eq:singleSectorBPSspec}, solutions are dual to
		\begin{subequations}\label{eq:alp1}
		\begin{equation}\label{eq:alp1a}
			|\alpha_1,n\rangle  \sim \Bigl (L_{-1}^n\Bigl |\frac{1}{2},\frac{1}{2}\Bigr \rangle\Bigr )_*^{N_1}\quad \mathrm{with} \quad \frac{N_1}{N}=\frac{\alpha_1^2}{4}\,,
		\end{equation}
		\begin{equation}\label{eq:alp1b}
			|\alpha_1,n\rangle_{s.l.}  \sim \Bigl(L_{-1}^n \Bigl |\frac{1}{2},\frac{1}{2}\Bigr \rangle\Bigr)^{N_1}\quad \mathrm{with} \quad \frac{N_1}{N}=\frac{2(2n+1)\alpha_1^2}{8(2n+1)+\alpha_1^2}\,,
		\end{equation}
		\end{subequations}
		respectively, and
		 the multi-particle operators are defined as in \eqref{eq:multistar} and \eqref{eq:multi}. One also has the ``pure NS" superstratum of Section~\ref{ss:pureNS2}, which depends on the single parameter $\beta_1$ associated with the CPO defined in \eqref{eq:11CPO}. We thus expect the heavy state dual to the pure NS superstratum to be of the form
		\begin{equation}\label{eq:bet1}
		|\beta_1,n_1\rangle \sim (L_{-1}^{2n_1} |1,1\rangle)^{N_1}\quad \mathrm{with} \quad \frac{N_1}{N}=\frac{(2n_1+1)\beta_1^2}{4\,(2n_1+1)^2+\beta_1^2}\,.
		\end{equation}
We conjecture that the multi-particle constituents appearing in this state are defined as in  \eqref{eq:multi}. For instance, for $n_1=0$ the coherent-like sum~\eqref{eq:coherentalpha1} should run over the states 
		\begin{equation}\label{eq:multi2}
		\Bigl(|1,1 \rangle\Bigr )^p \equiv \lim_{z_i\to 0} O^{(1,1)}_2(z_1) \ldots O^{(1,1)}_2(z_p) \,|0\rangle\,.
		\end{equation}
		This is to be contrasted with the geometries defined through the Lunin-Mathur profile~\cite{Kanitscheider:2007wq} which involve linear combinations between single and multi-trace operator that are different from the one needed to define $O^{(1,1)}_2$, see \cite{Giusto:2019qig,Rawash:2021pik}. We do not present, however, any direct check of this conjecture, since this would require computing the VEVs of dimension-four CPOs, a task that is unfeasible with our current holographic tools.
		
		It is interesting to compare the CFT identification of the states in \eqref{eq:alp1} and \eqref{eq:bet1} with the CTC bounds, \eqref{CTC1}, \eqref{CTC2} and \eqref{CTC3b}, derived from the dual geometries. One can see that for the ``$\alpha$-class"  solutions \eqref{eq:alp1} the maximum value of $\frac{N_1}{N}$ allowed by the bound is
		\begin{equation}
		\frac{N_1}{N}\Bigl |_\mathrm{max} = 1\qquad (\alpha\mathrm{-class)}\,,
		\end{equation}
		and for the ``$\beta$-class" \eqref{eq:bet1}
		\begin{equation}
		\frac{N_1}{N}\Bigl |_\mathrm{max} = \frac{1}{2}\qquad(\beta\mathrm{-class)}\,.
		\end{equation}
		In two-dimensional theories with an $\mathcal{N}\ge 2$ superconformal algebra the ``stringy exclusion principle" of \cite{Lerche:1989uy,Maldacena:1998bw} bounds the conformal dimension of operators: for chiral primary operators with $h=j$ the bound is
		\begin{equation}
		h=j\le \frac{c}{6}=N\,,
		\end{equation}
		which applies to both the left-moving and the right-moving sectors of the CFT. It is immediate to see from \eqref{eq:alp1} and \eqref{eq:bet1} that the stringy exclusion bound implies
		\begin{equation}
		\frac{N_1}{N}\le w \quad \mathrm{with}\quad w =  \begin{cases} 2 & \alpha \mathrm{-class}\,,\\ 1 & \beta\mathrm{-class} \,.\end{cases}
		\end{equation}
		The supergravity solutions given in Section~\ref{sec:3D-Sugr} thus cover only half the allowed range of conformal dimensions, for both the $\alpha$- and the $\beta$-  class of supersymmetric states. One wonders what are the supergravity duals of the remaining half of the allowed states. These are easily derived from the solutions of Section~\ref{sec:3D-Sugr} by the following chain of transformations: i) start from the solutions of Section~\ref{sec:3D-Sugr} written in the NS sector, which have $0\le j_{NS}\le \frac{N}{2}$; ii) flow to the R-sector, obtaining states with $-\frac{N}{2}\le j_{R}\le 0$; iii) flip the sign of all the gauge fields, $\Phi_i\to -\Phi_i$, $\Psi_i\to -\Psi_i$, thus reversing the sign of the angular momenta: $0 \le j_{R}\le \frac{N}{2}$; iv) flowing back to the NS sector gives states with $\frac{N}{2} \le j_{NS}\le N$, which provide precisely the missing half of the states. 
		
		We would also like to point out a qualitative difference between the solutions with the maximum value of $N_1$ allowed by the supergravity CTC bound of Section~\ref{sec:3D-Sugr}, for the $(1,0,n)$ superstratum, \eqref{eq:alp1a}, and for the special locus or the purely NS superstratum, \eqref{eq:alp1b} and \eqref{eq:bet1}. Due to the definition of the ``starred" multi-particle operator, \eqref{eq:multistar}, when $N_1=N$ the $(1,0,n)$ superstratum state \eqref{eq:alp1a} is made by $N$ identical strands of type $\Bigl |\frac{1}{2},\frac{1}{2}\Bigr \rangle$: this is reflected by the vanishing of the VEV of the operator $s_1^{(7)}$, which, as we will see in \eqref{eq:ssd1}, goes to zero when $\alpha_1^2=4 \Leftrightarrow N_1=N$. By contrast, in the special locus state, the operator $s_1^{(7)}$ is allowed to act once or twice on each strand, and thus even when $N_1=N$ the state contains strands of different type -- some in the vacuum, some in the state $\Bigl |\frac{1}{2},\frac{1}{2}\Bigr \rangle$ and some in the state $J^+_{-1} |0\rangle \sim (s_1^{(7)(\frac{1}{2},\frac{1}{2})})^2 |0\rangle$; as a consequence the VEV of $s_1^{(7)}$, computed in \eqref{eq:sld1}, does not vanish when $\alpha_1^2=8\Leftrightarrow N_1=N$. A similar phenomenon is expected to hold for the purely NS superstratum with $N_1=\frac{N}{2}$: this could be verified by looking at the VEV of $O_2^{(1,1)}$.
		
		The non-BPS microstates we construct in this paper are obtained by starting from one of the BPS solutions described in Section \ref{sub:BPSsystem} and  adding single-particle constituents that do not preserve the same supercharges of the original constituents. Hence, though each single-particle component is by itself BPS, and as such it is dual to a certain supergravity mode, the supersymmetry of the full state is broken by the non-trivial interaction between the mutually non-BPS constituents. For example the states $|\frac{1}{2},\frac{1}{2}\rangle$ and $|\frac{1}{2},-\frac{1}{2}\rangle$ preserve different supercharges on the right sector but the same on the left sector, and thus a multi-particle state containing both states is still BPS; to break the remaining supersymmetries one can excite the left sector by acting, for instance, with some power of the Virasoro generator $L_{-1}$. One is thus led to consider the state\footnote{In the Ramond sector, the CPOs, such as $|\frac{1}{2},\frac{1}{2}\rangle$, are mapped to ground states, while the super-descendants become states with a non-trivial momentum: for instance $|\frac{1}{2},-\frac{1}{2}\rangle$ is mapped to a state with an extra $\tilde{J}^-_{-1}$ insertion. Thus it is clear that in the Ramond sector the state~\eqref{eq:alpha1alpha2} carries both left-moving and right-moving momentum due to the $L_{-1}$ and $\tilde{J}^-_{-1}$ insertions respectively. Note also that, while the OPE between the states $|\frac{1}{2},\frac{1}{2}\rangle$ and $|\frac{1}{2},-\frac{1}{2}\rangle$ is now singular, the multi-particle state is defined by keeping only the leading regular term of the OPE. In the OPE between mutually non-BPS states, such as the constituents in~\eqref{eq:alpha1alpha2}, there are generically no regular terms and the multi-particle state relevant for our supergravity solutions appears in the singular term $\sim\epsilon^{\delta u_{12}}$, where $\epsilon$ is the distance between the two colliding operators and $\delta u_{12}$ should match the supergravity result~\eqref{eq:deltau} in the strong coupling regime of the CFT.}
		\begin{equation}\label{eq:alpha1alpha2}
		|\alpha_1,\alpha_2,n_1,n_2\rangle \sim \Bigl (L_{-1}^{n_1}\Bigl |\frac{1}{2},\frac{1}{2}\Bigr \rangle\Bigr )^{N_1} \Bigl (L_{-1}^{n_2} \Bigl |\frac{1}{2},-\frac{1}{2}\Bigr \rangle\Bigr )^{N_2} \,,
		\end{equation}
		where the relation between the microscopic quantities, $N_1$, $N_2$, and the macroscopic ones, $\alpha_1$, $\alpha_2$, should have the general form
		\begin{equation}\label{eq:N1N2a1a2}
		\frac{N_1}{N}=\frac{2(2n_1+1)\alpha_1^2}{8(2n_1+1)+\alpha_1^2}+O(\alpha_1^2\alpha_2^2)\quad ,\quad \frac{N_2}{N}=\frac{2(2n_2+1)\alpha_2^2}{8(2n_2+1)+\alpha_2^2}+O(\alpha_1^2\alpha_2^2)\,;
		\end{equation}
		we will provide a derivation of the first non-trivial correction terms of order $O(\alpha_1^2,\alpha_2^2)$ in the next section. While the state $|\alpha_1,\alpha_2,n_1,n_2\rangle$ is non-BPS for generic values of $n_1$ and $n_2$, it is BPS for $n_1=n_2$, since it can be obtained by acting with a finite $SU(2)_R$ rotation generated by $\tilde J^-_0$ on the supersymmetric state $|\alpha_1,n_1\rangle_{s.l.}$ (this follows from $\tilde J^-_0 |\frac{1}{2},\frac{1}{2}\rangle=|\frac{1}{2},-\frac{1}{2}\rangle$ and $(\tilde J^-_0)^k |\frac{1}{2},\frac{1}{2}\rangle=0$ for $k>1$). It is important to clarify why, in \eqref{eq:alpha1alpha2},   we have used the multi-particle definition \eqref{eq:multi} rather than the one in \eqref{eq:multistar}. Both options lead to perfectly acceptable heavy states which should have a dual representation in supergravity. It is however a non-trivial outcome of the supergravity analysis of Section~\ref{ss:alphaClass} that only the state in \eqref{eq:alpha1alpha2}, which is a non-BPS deformation of the special locus geometry, fits in the Q-ball Ansatz. A rough explanation of this finding rests on two facts: first, the Q-ball Ansatz requires the phase associated with the $\lambda_1$ mode to be exactly twice the one of the $\nu_1$ mode (and the same for $\lambda_2$ and $\nu_2$) and, second, the scalar matrix of the special locus geometry enjoys an $SO(3)$ symmetry. This symmetry, as we will explain in Section~\ref{ssech:preshol}, implies that the VEVs of all dimension two operators vanish and thus it guarantees that the supergravity fields satisfy the particular relations needed for this vanishing. Hence, it is somehow this symmetry that protects the ratio of the phases of the scalar modes against the non-BPS corrections and locks it to the value required by the Q-ball Ansatz. 		
		
		Analogously one can consider the ``pure NS" non-BPS state
		\begin{equation}\label{eq:beta1beta2}
		|\beta_1,\beta_2,n_1,n_2\rangle \sim (L_{-1}^{2\,n_1}|1,1\rangle)^{N_1} (L_{-1}^{2\,n_2}|1,-1\rangle)^{N_2} \,,
		\end{equation}
		where
		\begin{equation}\label{eq:N1N2b1b2}
		\frac{N_1}{N}=\frac{(2n_1+1)\beta_1^2}{4\,(2n_1+1)^2+\beta_1^2}+O(\beta_1^2\beta_2^2)\quad ,\quad \frac{N_2}{N}=\frac{(2n_2+1)\beta_2^2}{4\,(2n_2+1)^2+\beta_2^2}+O(\beta_1^2\beta_2^2)\,.
		\end{equation}
		This state is non-BPS for the same reason explained before \eqref{eq:alpha1alpha2} but, contrary to the state $|\alpha_1,\alpha_2,n_1,n_2\rangle$, it remains non-BPS even for $n_1=n_2\not=0$. Indeed, since $\tilde J^-_0 |1,1\rangle=|1,0\rangle$ and $(\tilde J^-_0)^2 |1,1\rangle=|1,-1\rangle$, acting with a finite rotation generated by $\tilde J^-_0$ on the BPS state $|\beta_1,n_1\rangle$ produces both copies of type $|1,0\rangle$ and $|1,-1\rangle$ and does not yield the state $|\beta_1,\beta_2,n_1,n_1\rangle$.

		\subsubsection{Frequencies and  momentum excitations}
		\label{sub:momexcit}
		
The major part of this paper will be concerned with the CFT states (\ref{eq:alpha1alpha2}) and (\ref{eq:beta1beta2}), and their geometric duals. These states have purely left-moving momentum excitations and are obtained by choosing the perturbative frequencies to be $\omega_1 = \omega_2 =1$.

More generally, it follows from the results of   \cite{Ganchev:2021ewa} that the perturbative  states of the form:
\begin{equation}\label{eq:alpha1alpha2-2}
\Big  (L_{-1}^{n_1+m_1} \tilde L_{-1}^{m_1}\Big |\frac{1}{2},\frac{1}{2}\Big \rangle\Big)^{N_1} \Big  (L_{-1}^{n_2+m_2} \tilde L_{-1}^{m_2}\Bigl |\frac{1}{2},-\frac{1}{2}\Big  \rangle\Bigr )^{N_2} \,,
\end{equation}
should be identified with the $\alpha$-class solutions for $\omega_1 = 2m_1+ 1$ and $\omega_2 = 2m_2+ 1$.
Similarly, the perturbative  states of the form:
\begin{equation}\label{eq:beta1beta2-2}
\Big (L_{-1}^{2(n_1+m_1)} \tilde L_{-1}^{2m_1}  \big |1,1\big\rangle\Big )^{N_1}  \Big (L_{-1}^{2(n_2+m_2)} \tilde L_{-1}^{2m_2}  \big |1,-1\big\rangle\Big )^{N_2}  \,,
\end{equation}
should be identified with the $\beta$-class solutions for $\omega_1 = 2m_1+ 1$ and $\omega_2 = 2m_2+ 1$.  Most of our results will focus on $m_1 =m_2 =0$, but it is sometimes straightforward to generalize to different values of $m_j$, as we have done in Section  \ref{sub:perturbation_pureNS}, and will do in Section \ref{ss:NumResults}. In the WKB analysis of Section~\ref{ssub:normal_modes_of_lambda_2_}, we considered the state~\eqref{eq:beta1beta2-2} with $n_1=2$, $n_2=4$, $m_1=0$ and $m_2 \gg 1$. The right moving dimension of such a state is $\bar{h}= N_1+\hat{\omega}_2 N_2$, where we already included the effect of the interactions between the mutually non-BPS constituents since we used $\hat{\omega}_2$ at the place of $\omega_2$.  In the regime $N_1 \gg N_2$, which applies directly to the WKB analysis, $\bar{h}$ is always positive.

		\subsection{The holographic charges}\label{sec:holochar}
		\label{ss:holcharges}
		
		We start by briefly reviewing the relevant holographic formulas (following \cite{Kanitscheider:2006zf,Ganchev:2021ewa}) needed to extract the angular momenta, $J^a$, $\tilde J^a$, and the conformal dimensions, $h$, $\bar h$, from the geometry. To compute these quantities one only needs the three-dimensional Einstein metric $ds^2_3$ \eqref{genmet1} and the gauge fields $\tilde A^{ij}$ \eqref{gauge_ansatz}. We will replace our radial coordinate, $\xi$, with the more usual  AdS$_3$ radial coordinate $z$.  The coordinates $x^\mu\equiv \{\tau,\sigma\}$ on the boundary, where $\tau$ coincides with the coordinate used in~\eqref{xidef} up to a rescaling, while the normalisation of $\sigma$ cannot be changed since we need to preserve the periodicity $2\pi$. The relation between $\xi$ and $z$ is chosen in such a way that the three-dimensional metric~\eqref{genmet1} has the asymptotic form as $z\to 0$ \cite{AST_1985__S131__95_0,deHaro:2000xn}
		\		\begin{equation}\label{eq:asympt3D}
		- ds^2_3 = \frac{dz^2}{z^2} +\frac{-d\tau^2+d\sigma^2}{z^2}+ g^{(2)}_{\mu\nu} dx^\mu dx^\nu +O(z^2)\,.
		\end{equation}
		The $z$ (or $\xi$) component of the gauge fields should vanish at $z=0$ ($\xi=1$) and the asymptotic values of the $\tau$ and $\psi$ components are denoted by $\Phi_i^{(\infty)}$, $\Psi_i^{(\infty)}$ ($i=1,2,3,4$). In terms of these asymptotic values, the $SU(2)_L\times SU(2)_R$ angular momenta $J^a$, $\tilde J^a$ ($a=3,\pm$) are
		\begin{subequations}\label{eq:angmomhol}
		\begin{equation}
			J^3=-\frac{N}{2}\left(\Phi_1^{(\infty)} + \Phi_2^{(\infty)}+2(\Psi_1^{(\infty)} + \Psi_2^{(\infty)})\right)\,\,,\,\, J^\pm=\frac{N}{2}\left(\Phi_3^{(\infty)} + \Phi_4^{(\infty)}+2(\Psi_3^{(\infty)} + \Psi_4^{(\infty)}) \right)\,,
		\end{equation}
		\begin{equation}
			\tilde J^3=-\frac{N}{2}\left(\Phi_1^{(\infty)} - \Phi_2^{(\infty)}\right)\,\,,\,\, \tilde J^\pm=\frac{N}{2}\left(\Phi_3^{(\infty)} - \Phi_4^{(\infty)} \right)\,.
		\end{equation}
		\end{subequations}
		The values of the left and right conformal dimensions, $h$ and $\bar h$, are computed from
		\begin{subequations}\label{eq:hhbar}
		\begin{equation}
			h=\frac{N}{4}\left(1+g^{(2)}_{\tau\tau} +2 g^{(2)}_{\tau\sigma} + g^{(2)}_{\sigma\sigma} \right)+\frac{(J^3)^2 + J^+ J^-}{N}\,,
		\end{equation}
		\begin{equation}
			\bar h=\frac{N}{4}\left(1+g^{(2)}_{\tau\tau} -2 g^{(2)}_{\tau\sigma} + g^{(2)}_{\sigma\sigma} \right)+\frac{(\tilde J^3)^2 + \tilde J^+ \tilde J^-}{N}\,.
		\end{equation}
		\end{subequations}
		Note that the formulas \eqref{eq:angmomhol} assume the gauge given in \eqref{gauge_ansatz}, but the result in any other gauge follows simply by transforming \eqref{eq:angmomhol} under the appropriate $SO(4)$ rotation. Under a generic transformation $\mathcal{U}$ the matrix of the gauge fields, $\tilde A$, transforms as 
		\begin{equation}
		\tilde A\to \mathcal{U} \tilde A \mathcal{U}^{-1} + \frac{1}{2} d\mathcal{U} \mathcal{U}^{-1}\,.
		\label{eq:NAGT-1a}
		\end{equation}
		In particular, for a block-diagonal phase rotation of the form
		\begin{equation}\label{eq:phaserot}
		\mathcal{U} =\begin{pmatrix} \mathcal{U}_1&0\\ 0 & \mathcal{U}_2 \end{pmatrix}\quad \mathrm{with}\quad \mathcal{U}_i = \begin{pmatrix}\cos\delta_i &-\sin\delta_i\\ \sin\delta_i & \cos\delta_i \end{pmatrix}\quad \mathrm{and}\quad \delta_i = {\delta}^\tau_i \,\tau + {\delta}^\psi_i\, \psi\,, 
		\end{equation}
		the angular momenta transform as 
		\begin{equation}
		\begin{aligned}
			&J^3\to J^3+\frac{N}{4}({\delta}^\tau_1+ {\delta}^\tau_2+2({\delta}^\psi_1+{\delta}^\psi_2))\quad,\quad J^\pm \to J^\pm e^{\pm i (\delta_1+\delta_2)}\,,\\
			&\tilde J^3\to \tilde J^3+\frac{N}{4}({\delta}^\tau_1- {\delta}^\tau_2)\quad,\quad \tilde J^\pm \to \tilde J^\pm e^{\pm i (\delta_1-\delta_2)}\,.
		\end{aligned}
		\end{equation}
		
		One should recall that in Section \ref{sub:qball} we imposed a gauge choice that made all our fields independent of $\tau$ and $\psi$ and encoding the mode numbers into the electromagnetic potentials.  The gauge transformations defined by (\ref{eq:NAGT-1a})  and (\ref{eq:phaserot}) will undo our original gauge choice and make some of the scalars depend explicitly on  $\tau$ and $\psi$.  Since this gauge transformation is based on holographic identification of states, we will view the resulting mode numbers and frequencies of the fields as being the ``physical quantities.''  In particular, this will lead to the physical frequencies, $\hat \omega_i$, of the fields.
		
		\subsection{Precision holography tests for supersymmetric states}
		\label{ssech:preshol}

In this section we provide some explicit checks supporting the identification between the supergravity solutions discussed in Section~\ref{sec:nonBPS} and the heavy states introduced in the Section \ref{sec:heavystates}. In order to have a precise, quantitative check, we focus on {\em protected} physical observables, such as discrete quantum numbers and supersymmetric $3$-point correlators~\cite{Baggio:2012rr}. In particular for the $3$-point correlators we follow the approach initiated by~\cite{Skenderis:2006ah,Kanitscheider:2006zf,Kanitscheider:2007wq}: we extract the expectation values of the BPS operators dual to single particle states from the asymptotic expansion of the supergravity solutions at the AdS boundary and then match these results against those obtained in the CFT description. The main goal of this section is to show how the different definitions of the multi-particle heavy states are reflected in the corresponding supergravity solutions, thus providing support for our proposed identification.
		
In Section~\ref{sub:onequat} we focus on the solutions that can be studied in the gauged-fixed approach summarised in Appendix D of~\cite{Rawash:2021pik}, which requires to write the geometry as a fibration over a 4D base metric in the coordinate system where it is explicitly the flat Euclidean space (see  [C.1] of~\cite{Rawash:2021pik}). This is always possible in the two-charge case ($n_1=n=0$), while for the three-charge solutions (with $n_1=n\not=0$) it has to be checked on a case by case basis\footnote{In this section we focus on single-mode configurations, so we set $\alpha_2$ and $n_2$ to zero when relevant.}. For instance, the gauged-fixed approach can be used for the three-charge superstratum of Section~\ref{sec:doublestratum}, but not for the special locus solution of Section~\ref{ss:speciallocus}. However it is possible to see the main feature we wish to highlight, and one of the primary results of this paper, already in the two-charge case: the special locus constraint $\beta_1=-\frac{\alpha_1^2}{4}$ is directly related to the standard definition of multi-particle states~\eqref{eq:multi}. In contrast, the more familiar solutions with $\beta_1=0$ are dual to the states such as~\eqref{eq:coherentalpha1} and~\eqref{eq:alp1a} as discussed in~\cite{Kanitscheider:2007wq,Bena:2017xbt}. This information is derived from the expectation values of the operators of dimension $2$~\cite{Giusto:2019qig,Rawash:2021pik}. Of course the expectation values of dimension $2$ operators are related to subleading terms at the AdS boundary with respect to the contributions relevant for dimension $1$ operators and so one needs to perform a careful analysis of the relevant supergravity solutions.

In Section~\ref{sub:oneeig} we show that this pattern holds also for the geometries with $n_1\not=0$ where again the VEVs of BPS operators are protected thanks to the fact that a chiral part of the supersymmetry is preserved. However, for the three-charge special locus solution~\eqref{eq:singleSectorBPSspec} we cannot use the gauged-fixed approach and we need to revert to the full gauge-invariant discussion of~\cite{Kanitscheider:2006zf,Kanitscheider:2007wq}. So we will briefly review the main ingredients needed to calculate the expectation values of the operators of dimension $1$ and $2$ in a gauged invariant way. The main result is that, in direct contrast to the standard superstratum geometry~\eqref{algBPS1}, for the special locus geometry~\eqref{eq:singleSectorBPSspec}  {\it all operators of dimension $2$ have a vanishing expectation value} for any choice of $n_1$. This is consistent with the dual state~\eqref{eq:slstate} since these vanishing expectation values are proportional to extremal $3$-point correlators or equivalently $2$-point correlators between a single and a double-particle state. These quantities are known to be sensitive to the mixing between multi-particle operators~\cite{DHoker:1999jke} and they vanish when using the natural holographic dictionary~\cite{Arutyunov:2000ima,Aprile:2020uxk} derived from the standard OPE product~\eqref{eq:multi}. This is to be contrasted with what happens for the state~\eqref{eq:coherentalpha1}, where one can use the free orbifold description to check that there are non-zero expectation values for operators of dimensions $2$. Thanks to the non-renormalisation properties of these observables~\cite{Baggio:2012rr}, we can match the CFT results with those obtained from the solution with $\beta_1=0$ and check that there is a non-trivial agreement as done in~\cite{Giusto:2019qig,Rawash:2021pik}. Thus the solutions with $\beta_1=0$ and $\beta_1=-\frac{\alpha_1^2}{4}$ are dual to~\eqref{eq:alp1a} and~\eqref{eq:alp1b} respectively for any value of $n$ showing that the dressing with the $L_{-1}^n$ insertions that carry momentum do not alter the relation between the two products~\eqref{eq:multi} and~\eqref{eq:multistar} and the two extreme values of $\beta_1$. For non-BPS heavy states, such as~\eqref{eq:alpha1alpha2} and~\eqref{eq:beta1beta2}, these expectation values are not protected and so we cannot perform the same quantitative analysis, but we expect the same pattern to hold as the heavy states are still constructed in the same way as the BPS states but just with different momentum carrying excitations.

		\subsubsection{A gauged fixed approach to the holographic VEV}
		\label{sub:onequat}
		
		In this section we use [D.2] and~[D.4] of~\cite{Rawash:2021pik} to read the holographic expectation values. We start from the six-dimensional uplift of Appendix~\ref{app:uplift} and use, for instance,  [C.1] of~\cite{Rawash:2021pik} to read the functions $Z_{1,2,4}$.  Note that to compare with solutions of~\cite{Rawash:2021pik} we need to work in the Ramond sector. From the scalars parametrised as in~\eqref{eq:c4v4} and~\eqref{eq:asxD} we have
		\begin{equation}
		\label{eq:z4z1}
		\frac{Z_4}{Z_2} = C_4 = \sqrt{\frac{Q_1}{2 Q_5}} X\;,\quad \frac{Z_1}{Z_2} = V_4^2 =  \frac{Q_1}{2 Q_5} \left(X^2+2\Delta\right) \;,\quad {\cal P}\equiv Z_1 Z_2 -Z_4^2 = \frac{Q_1}{Q_5} \Delta\, Z_2^2 \;.
		\end{equation}
		Looking now at the six-dimensional Einstein metric $ds_6^2$~\eqref{eq6dem}, we should rewrite it in the form of  [C.1] of~\cite{Rawash:2021pik}. In order to apply the gauged-fixed approach of that reference, we first need to fix the constant $\Omega_1$ and perform an appropriate gauge transformation~\eqref{eq:phaserot} on the three-dimensional gauge fields so as to reproduce the values expected for the CFT in the Ramond sector. Then we should choose coordinates such that the four-dimensional  metric in~[C.1] takes the explicitly flat form $d\rho^2+\rho^2 \left(d\theta^2+\sin^2\theta d\varphi_1^2 + \cos^2\theta d\varphi_2^2 \right)$, where the relation between $\xi$ and $\rho$ is given in~\eqref{xidef}. In general this requires a change of coordinates that we can derive perturbatively around the AdS boundary
		\begin{equation}
		\label{eq:rhorhop}
		\rho \to c_\rho \left(\hat\rho + \frac{f_\rho(\hat\theta,\hat\varphi_i)}{\hat\rho}+\ldots\right)\;,\quad
		\theta \to \hat\theta + \frac{f_\theta(\hat\theta,\hat\varphi_i)}{\hat\rho^2}+\ldots\;,\quad
		\varphi_j \to \hat\varphi_j + \frac{f_{\varphi_j}(\hat\theta,\hat\varphi_i)}{\hat\rho^2}+\ldots\;,
		\end{equation}
		where the dots stand for subleading terms in the large $\rho$ expansion.
		
		We start with the configuration  \eqref{algBPS1} for $\alpha_2=n_2=0$. In order to satisfy~\eqref{goodmetbc} we need to perform a rescaling of $\tau$ which amounts to choosing
		\begin{equation}
		\label{eq:sso1}
		\Omega_1 = 1-\frac{\alpha_1^2}{4}\;,
		\end{equation}
		while to reproduce the Ramond-sector angular momenta from~\eqref{eq:angmomhol} we need to perform the $SO(4)$ rotation~\eqref{eq:phaserot} with parameters\footnote{In the Neveu-Schwarz sector we have $\delta^\psi_2 = 0$ and the potentials $\Psi_{1,2}$ vanish at $\xi=0$.}
		\begin{equation}
		\label{eq:ssgt1}
		\delta^\tau_i = 0\;,\quad \delta^\psi_1 =n_1\;,\quad \delta^\psi_2 = -1\;.
		\end{equation}                
		Finally, the required change of variables~\eqref{eq:rhorhop}  is:
		\begin{equation}
		\label{eq:covss}
		c_\rho= \sqrt{1-\frac{\alpha_1^2}{4}}\;,\quad f_\rho(\hat\theta,\hat\varphi_i) = \frac{1}{8}\,(1+2\cos(2\theta))\;,\quad f_\theta(\hat\theta,\hat\varphi_i) =-\frac{1}{4} \left(1-\frac{\alpha_1^2}{4}\right)\sin(2\theta)\;,
		\end{equation}
		while $f_{\varphi_j}=0$. This change of coordinates becomes singular when $\alpha_1^2 = 4$, which is exactly the CTC bound~\eqref{CTC1} (remember, we have set   $\alpha_2=0$). 
		
		For the \nBPS{4} solution with $n_i=0$, this analysis was first performed in~\cite{Kanitscheider:2007wq} where the dual interpretation in terms of the state~\eqref{eq:coherentalpha1} was also introduced. For this geometry, we see that $Z_2$ does not depend on the $S^3$ coordinates. Using~\eqref{eq:Xd}, and the explicit expressions in~\eqref{algBPS1}, in the equations above, one can expand $\frac{Z_4}{Z_2}$ in spherical harmonics\footnote{We also follow~\cite{Rawash:2021pik} for the conventions of the spherical harmonics, see their Appendix~A.} and read from~(D.4) of~\cite{Rawash:2021pik} the leading order result for the expectation value of the operator $s^{(7)}_1$:
		\begin{equation}
		\label{eq:ssd1}
		\langle s^{(7)(\pm\frac{1}{2},\pm\frac{1}{2})}_1 \rangle = \frac{\alpha_1}{\sqrt{2}}\;\lim_{\xi\to 1} \left(\hat\rho\,\sqrt{1-\xi^2} \right) = \frac{\alpha_1}{\sqrt{2}}\,\sqrt{1-\frac{\alpha_1^2}{4}}\;.
		\end{equation}
		Moving to the harmonics of order $2$, one can check that the operator $O_2$ has vanishing expectation value in the solution defined by~\eqref{algBPS1}. The only components that can be potentially non-trivial are $(\pm 1, \pm 1)$, but  there is a cancellation between the two terms in~\eqref{eq:11CPO} that are separately non-zero
		\begin{equation}
		\label{eq:dim1su}
		s_2^{(6)(\pm 1,\pm 1)} = \frac{1}{\sqrt{3}} \sigma_2^{(\pm 1,\pm 1)} = \frac{1}{2\sqrt{2}} \frac{\alpha_1^2}{4}\left(1- \frac{\alpha_1^2}{4}\right)\quad \Rightarrow \quad \langle O^{(\pm 1,\pm 1)}_2 \rangle = 0 \;.
		\end{equation}
		Then the orthogonal combination
		\begin{equation}
		\label{eq:11CPOC}
		C^{(\pm 1,\pm 1)}_2 \equiv \sqrt{3} \,\sigma_2^{(\pm 1,\pm 1)} + s_2^{(6)(\pm 1,\pm 1)} \;
		\end{equation}
		has a non-trivial expectation value in the heavy state dual to~\eqref{algBPS1}:
		\begin{equation}
		\label{eq:evh2}
		\langle C^{(\pm 1,\pm 1)}_2 \rangle = \frac{\alpha_1^2}{2\sqrt{2}} \left(1- \frac{\alpha_1^2}{4}\right)\;.
		\end{equation}
		
		Now consider the \nBPS{4}  version of the ``special locus'' solution discussed in Section~\ref{ss:specialBPSsol}. In order to satisfy~\eqref{goodmetbc}, we have  to use the freedom to rescale $\tau$ so to fix $\Omega_1$ as
		\begin{equation}
		\label{eq:slo1}
		\Omega_1 = 1-\frac{2\,\alpha_1^2\,(2\,n_1+1)}{8+16\,n_1+\alpha_1^2}\;.
		\end{equation}
		The $SO(4)$ gauge transformation needed to obtain the Ramond-sector angular momenta is
		\begin{equation}
		\label{eq:slgt1}
		\delta^\tau_1 = 0\;,\quad \delta^\tau_2= \frac{\alpha_1^2+4 (\alpha_1^2-4 ) n_1-8}{2 \left(\alpha_1^2+16 n_1+8\right)}\;,\quad \delta^\psi_1 =0\;,\quad \delta^\psi_2 = -1\;.
		\end{equation}                
		For the ``special locus" solution with $n_1\not=0$, the four-dimensional  base metric is not flat, and thus the change of coordinates \eqref{eq:rhorhop} that was used above for the superstratum geometry in ~\eqref{algBPS1} cannot be defined; however one can use the asymptotic expansion~\eqref{eq:asympt3D} to define the holographic radial coordinate $\hat\rho=1/z$
		\begin{equation}
		\label{eq:hrhpsl}
		\rho \to c_\rho\,\hat\rho\;, \qquad c_\rho \equiv \left( \frac{8(2n_1+1)+\alpha_1^2}{8(2n_1+1) -(4n_1+1) \alpha_1^2} \right)^{\frac{1}{2}}  \;.
		\end{equation}
		This change of coordinates becomes singular when the supergravity solution develops CTC, since the denominator in~\eqref{eq:hrhpsl} vanishes when the equality in~\eqref{CTC3b} is satisfied. 
		
		For the rest of this section we focus on the \nBPS{4}  ``special locus" geometry given in \eqref{eq:singleSectorBPSspec} with $n_1=0$, which fits in the general ansatz of [C.1] of~\cite{Rawash:2021pik} with a flat  four-dimensional base metric, so we can apply the gauge fixed approach spelled out there. Then~\eqref{eq:hrhpsl} reads
		\begin{equation}
		\label{eq:crhosl}
		\rho \to \sqrt{\frac{8+\alpha_1^2}{8-\alpha_1^2}}\;\hat\rho\;
		\end{equation}
		and the expectation value for the operator~\eqref{eq:1/2CPO} now reads
		\begin{equation}
		\label{eq:sld1}
		\langle s^{(7)(\pm\frac{1}{2},\pm\frac{1}{2})}_1 \rangle = \frac{4 \sqrt{2}}{8+\alpha_1^2}\;.
		\end{equation}
		Moving to the harmonics of order $2$, one can start again from~[D.4] of~\cite{Rawash:2021pik} and evaluate the second equation in~\eqref{eq:z4z1}. The surprise is that it does not depend on the $S^3$ coordinates and so the expectation value of $s_2^{(6)}$ vanishes for the \nBPS{4}  ``special-locus'' solution. The analysis of the expectation values for $\sigma_2$ is more involved. For the $(\pm 1,\pm 1)$ components, the terms proportional to the second harmonics of $Z_{1,2}$ and those proportional to the square of the first harmonics of $Z_4$ are separately non-trivial, but cancel in the combination in~[D.4] of~\cite{Rawash:2021pik}. For the $(0,0)$ component, also the term proportional to the square of the first harmonic of the vectors contributes, but again there is a non-trivial cancellation when all contributions to~[D.4] are combined.

		To summarize, all components of $\sigma_2$ vanish, which implies  $\langle C_2\rangle = \langle O_2\rangle=0$, and so there are no dimension $2$ expectation values for this ``special locus'' solution.
		
		\subsubsection{A gauge-invariant approach to the holographic VEV}
		\label{sub:oneeig}
		
		We start by briefly reviewing some key ingredients needed to read the holographic VEVs in a gauge-invariant approach, see~\cite{Kanitscheider:2006zf,Kanitscheider:2007wq,Rawash:2021pik} for more details. Again we use the six-dimensional uplift of Appendix~\ref{app:uplift} and then decompose the six-dimensional fields in $S^3$ spherical harmonics. In particular, we need the components of the metric, the 3-forms along the $S^3$ and the scalars, whose decomposition reads
		\begin{equation}
		\begin{aligned}
			&h_{(\alpha\beta)}=\sum_{k,I} \rho^{(I)}_k Y^I_{k(\alpha\beta)} + \rho^{(I)}_{k(v)} D_{(\alpha} Y^I_{k\beta)} +\rho^{(I)}_{k(s)} D_{(\alpha}D_{\beta)} Y^I_k\,,\\
			&h^\alpha_\alpha =\sum_{k,I} \pi^{(I)}_k Y^I_k\,,\quad G^{(r)}|_{S^3}=-\sum_{k,I} \Lambda_k U^{(r)(I)}_k Y^I_k\,,\quad \phi^{(mr)} = \sum_{k,I} \phi^{(mr)(I)}_k Y^I_k\,.
		\end{aligned}
		\end{equation}
		Here the subscript $(\alpha\beta)$ denotes the symmetric traceless part and $Y^I_k$, $Y^I_{k\alpha}$, $Y^I_{k(\alpha\beta)}$ are scalar, vector and tensor harmonics of order $k$. The coefficients, $ \rho^{(I)}_k,  \rho^{(I)}_{k(v)}, \rho^{(I)}_{k(s)}, \pi^{(I)}_k, U^{(r)(I)}_k$ and $\phi^{(mr)(I)}_k$,  in this expansion  are functions on AdS$_3$, and their asymptotics at infinity determine the VEVs we are seeking to extract.  The most direct way to extract the coefficient $\rho^{(I)}_{k(s)}$ is to use the transversality of the tensor and vector harmonics:
		\begin{equation}
		D^\alpha D^\beta h_{(\alpha\beta)}=\sum_{k,I} \rho^{(I)}_{k(s)} D^\alpha D^\beta D_{(\alpha}D_{\beta)} Y^I_k=-2 \sum_{k,I} \Lambda_k\left(1-\frac{\Lambda_k}{3}\right)\rho^{(I)}_{k(s)} Y^I_k\,,
		\end{equation}
		where $\Lambda_k=k(k+2)$ and $-\Lambda_k+s$ is the eigenvalue of the Laplacian for the scalar ($s=0$), vector ($s=1$) and tensor ($s=2$) harmonics. The coefficients of the spherical harmonics are functions of the AdS$_3$ coordinates and in particular of the radial coordinate $\hat\rho$: the leading term for large $\hat\rho$ goes like $\hat\rho^{-k}$ for the coefficients of spherical harmonics of order $k$; for any field $\Phi^{(I)}_k$ one denotes this leading term as 
		\begin{equation}
		\left[\Phi^{(I)}_k\right]\equiv \lim_{\hat\rho\to \infty}  \hat\rho^k \Phi^{(I)}_k\,.
		\end{equation}
		Clearly the numbers $\left[\Phi^{(I)}_k\right]$ depend on the normalization of the coordinate $\hat\rho$ and this is fixed, as discussed around~\eqref{eq:hrhpsl}, by requiring that at leading order the AdS$_3$ part of the metric reduces to the canonical form~\eqref{eq:asympt3D} (with $\hat\rho=1/z$). The formalism is invariant under coordinate re-definitions, as far as \eqref{eq:asympt3D} is satisfied. To lighten the notation, we will omit the square brackets in the following and simply denote $\left[\Phi^{(I)}_k\right]$ as $\Phi^{(I)}_k$.
		
		Scalar operators of the CFT come from tensor multiplets, $s^{(r)(I)}_k$ ($k\ge1$), or from the gravity multiplet, $\sigma^{(I)}_k$ ($k\ge2$). The VEVs of tensor multiplet operators are (proportional to)
		\begin{equation}
		s^{(r)(I)}_k=\frac{\sqrt{k}}{\sqrt{k+1}}\left(\phi^{(5r)(I)}_k +2(k+2) U^{(r)(I)}_k  \right)\quad (r=6,7)\,.
		\end{equation}
		For the gravity multiplet the recipe is more complicated: one first defines
		\begin{equation}
		\sigma^{(I)}_k=\frac{\sqrt{k(k-1)}}{3 (k+1)}\left(6(k+2) {\hat U}^{(5)(I)}_k -{\hat \pi}^{(I)}_k \right)\,,
		\end{equation}
		where ${\hat U}^{(5)(I)}_k$ and ${\hat \pi}^{(I)}_k$ are the gauge-invariant combinations, which are given for $k=2$ by
		\begin{equation}
		{\hat U}^{(5)(I)}_2=U^{(5)(I)}_2-\frac{1}{2} \rho^{(I)}_{2(s)}\,,\quad {\hat \pi}^{(I)}_2=\pi^{(I)}_2 + \Lambda_2 \,\rho^{(I)}_{2(s)}\,.
		\end{equation}
		Then the VEV of the single particle gravity operator of dimension $(1,1)$ is (proportional to)~\cite{Rawash-thesis} 
		\begin{equation}
		{\tilde \sigma}^{(I)}_2=\sigma^{(I)}_2+\frac{11}{24 \sqrt{2}}\sum_{r=6,7} s^{(r)(i)}_1 s^{(r)(j)}_1 a_{Iij}\,, 
		\end{equation}
		where $a_{Iij}$ are the triple overlap coefficients of scalar spherical harmonics
		\begin{equation}
		a_{Iij}=\int d\Omega_3 Y^I_2 Y^i_1 Y^j_1 \,,
		\end{equation}
		with $d\Omega_3$ the round $S^3$ measure normalised such that $\int d\Omega_3=1$. The overlaps that are relevant for us are
		\begin{equation}
		a_{(1,1) (-\frac{1}{2},-\frac{1}{2}) (-\frac{1}{2},-\frac{1}{2})}=\frac{2}{\sqrt{3}}\,,\quad a_{(0,0)(\frac{1}{2},\frac{1}{2}) (-\frac{1}{2},-\frac{1}{2})}=a_{(0,0) (-\frac{1}{2},-\frac{1}{2})(\frac{1}{2},\frac{1}{2})}=\frac{1}{\sqrt{3}}\,.
		\end{equation}
		
		Starting from the scalar sector, we use~\eqref{eq:PhQ1}, \eqref{eq:PhQ2}, \eqref{eq:asxD}, \eqref{eq:GS3} evaluated for the solution~\eqref{eq:singleSectorBPSspec} with generic $n_1$ and obtain
		\begin{equation}
		\phi^{(56)}=0\,,\quad G^{(6)}|_{S^3}=0\,,
		\end{equation}
		since 
		\begin{equation}
		\Delta+\frac{X^2}{2}=1\,,\quad \frac{G^{(1)}|_{S^3}}{G^{(2)}|_{S^3}}=\frac{Q_5}{Q_1}
		\end{equation}
		and this implies that the VEVs of  all the tensor multiplet operators $s^{(6)(I)}_k$ vanish (as discussed after~\eqref{eq:sld1} in the gauged-fixed approach)
		\begin{equation}
		s^{(6)(I)}_k=0\quad \forall\,k\,.
		\end{equation}
		The computation of the VEV of the gravity multiplet operator $\tilde\sigma^{(I)}_2$ is a bit more laborious and we summarise below some relevant steps. The properly normalised radial coordinate $\hat\rho$ is related to $\xi$ by~\eqref{xidef} and~\eqref{eq:hrhpsl}: for convenience we recall this relation for the case at hand below
		\begin{equation}
		\xi=\frac{c_\rho \hat\rho}{\sqrt{1+c_\rho^2 \hat\rho^2}}\quad\mathrm{with}\quad c_\rho^2=\frac{8(1+2n_1)+\alpha_1^2}{8(1+2n_1)-(1+4n_1)\alpha_1^2}\,.
		\end{equation}
		The coefficients extracted from the asymptotic expansion of the metric are (in the equations below spherical harmonic index $I$ is either $(\pm1,\pm1)$ or $(0,0)$) 
		\begin{subequations}
		\begin{equation}\begin{aligned}
			\phi^{(57)(\pm\frac{1}{2},\pm\frac{1}{2})}_{1} &=2U^{(7)(\pm\frac{1}{2},\pm\frac{1}{2})}_{1}= \frac{2\alpha_1(2n_1+1)}{c_\rho\sqrt{64(2n_1+1)^2-\alpha_1^4}}\quad \Rightarrow\\  s^{(7)(\pm\frac{1}{2},\pm\frac{1}{2})}_{1}& = \frac{4\sqrt{2}\alpha_1(2n_1+1)}{c_\rho\sqrt{64(2n_1+1)^2-\alpha_1^4}}\;,
		\end{aligned}\end{equation}
		\begin{equation}\begin{aligned}
			\rho^{(I)}_{2(s)} &= \pi^{(I)}_2=-2 U^{(5)(I)}_2= \frac{4\alpha_1^2(2n_1+1)^2}{\sqrt{3}c^2_\rho (64(2n_1+1)^2-\alpha_1^4)}\quad \Rightarrow \\ \hat \pi^{(I)}_2&=-9 \hat U^{(5)(I)}_2=\frac{12\sqrt{3}\alpha_1^2(2n_1+1)^2}{c^2_\rho (64(2n_1+1)^2-\alpha_1^4)}\,,
		\end{aligned}\end{equation}
		\begin{equation}
			\sigma^{(I)}_2 =-\frac{44\sqrt{2}\alpha_1^2(2n_1+1)^2}{3\sqrt{3}c^2_\rho (64(2n_1+1)^2-\alpha_1^4)}\quad\Rightarrow \quad \tilde \sigma^{(I)}_2=0\,.
		\end{equation}
		\end{subequations}
		We conclude that both dimension-two operators $s^{(6)(I)}_2$ and $\tilde\sigma^{(I)}_2$ have vanishing VEVs in the special locus geometry with generic values of $n_1$.

		\section{The Holographic analysis of non-BPS microstates}
		\label{sec:Holography}

                The non-BPS solutions constructed in Section~\ref{sec:nonBPS} are holographically related to the heavy CFT states reviewed in Section~\ref{sec:heavystates}. Supersymmetry is broken because, for generic values of the parameters, these states contain mutually non-BPS constituents. In Section \ref{ssech:preshol} we focused on the regime where supersymmetry is restored and extracted information about the 3-point functions in order to characterize the special locus within the CFT. We now extend the holographic analysis to the generic non-BPS case: the main goal is to extract some {\em dynamical} holographic data in the strongly coupled (supergravity) regime of the CFT. In particular we will determine the binding energies between the mutually non-BPS constituents or, in the CFT language, the (average) anomalous dimension of some multi-particle operators. 
		
		When dealing with non-BPS configurations the set of precise-holography tools that has been used to conjecture and verify with great accuracy the map between asymptotically AdS$_3$ geometries and D1-D5 supersymmetric states \cite{Kanitscheider:2006zf,Kanitscheider:2007wq,Giusto:2015dfa,Giusto:2019qig,Rawash:2021pik} shrinks considerably. This is because three-point functions where some of the operators break all supersymmetries generically depend on the CFT moduli and their values at the gravity point are unknown. The only available link between gravity and CFT is provided by few quantised charges, the momentum charge, $n_p=h-\bar h$, and the $SU(2)_L\times SU(2)_R$ angular momenta, $J^a$ and $\tilde J^a$ ($a=3,\pm$). The gravity computations of Section~\ref{sec:nonBPS} have been performed in a particular gauge where all the dependence on the worldsheet coordinates $\tau$ and $\psi$ has been gauged away. The quantised charges computed in this gauge do not reproduce the values expected from the CFT and thus a first step for a holographic interpretation of our gravity solutions is to perform the appropriate $SO(4)$ rotation that yields the values of the angular momenta predicted by the CFT. 
		
		A further step is finding the relation between the microscopic quantities, $N_1$, $N_2$, and the gravity parameters, $\alpha_1$, $\alpha_2$ or $\beta_1$, $\beta_2$, expressed by the perturbative relations \eqref{eq:N1N2a1a2} and \eqref{eq:N1N2b1b2}. To this purpose we will need to match the momentum charge $h-\bar h$ but also the energy $h +{\bar h}$. This second point requires some clarification: Unlike the quantised momentum, the energy is a non-protected (moduli-dependent) quantity, which receives corrections from the interactions between the mutually non-BPS single-particles constituents of the heavy states \eqref{eq:alpha1alpha2}, \eqref{eq:beta1beta2}. From the CFT perspective, these corrections correspond to the anomalous dimensions of the non-BPS multi-particle operators formed by the interacting single-particles and are interesting dynamical data that are, generically, unknown. These energy corrections can be read off from the geometry in two different ways: from the $\tau$-dependence of fields and from the asymptotic behaviour of $(\tau,\psi)$ component of the metric (the precise formulas will be reviewed below). Thus, though we cannot predict the energy corrections from the CFT, requiring the consistency\footnote{In the solutions of \cite{Ganchev:2021ewa}, the problem was already constrained enough to allow for an independent check of this consistency requirement; in this article, we turn things around and impose by hand the consistency, as a way to infer the relations \eqref{eq:N1N2a1a2} and \eqref{eq:N1N2b1b2}.} between these two ways of extracting them imposes a non-trivial constraint, which allows us to completely determine the holographic dictionary and the energy corrections order by order in perturbation theory. 
		
		A schematic summary of our procedure is as follows:
		
		i) Start from the solutions of Section~\ref{sec:nonBPS} and perform an $SO(4)$ rotation so that: a) the angular momenta of the rotated solution have the values expected from the CFT, b) the $\xi$-components of the gauge fields vanish at infinity, $\xi=1$ (this is the gauge where one usually computes the angular momenta), and c) the $\psi$-components of the gauge fields vanish at the origin, $\xi=0$ (this is needed for regularity, since the $\psi$ coordinate degenerates at the origin);
		
		ii) Extract the conformal dimensions $h$, $\bar h$ from the metric (with the holographic formulas reviewed below) and determine the relation between $N_1$, $N_2$ and the supergravity parameters ($\alpha_1$, $\alpha_2$ or $\beta_1$, $\beta_2$) by matching the momentum $h-\bar h$ with the CFT value and the energy $h+\bar h$ with the value inferred from the frequency of the supergravity fields.
		
		The procedure outlined above cannot actually be carried out exactly for the non-BPS states. The matching of the angular momenta between gravity and CFT, done in point i) to identify the $SO(4)$ rotation, requires the knowledge of the relation between $N_1$, $N_2$ and $\alpha_1$, $\alpha_2$ (or $\beta_1$, $\beta_2$), but this relation is determined only in point ii). One can resort, however, to a perturbative approach, where one uses in i) the relations, \eqref{eq:alp1} or \eqref{eq:bet1}, valid in the supersymmetric limit, and one computes in ii) the first order corrections to these relations in the non-BPS parameter. In principle this procedure could be iterated to any arbitrary perturbative order.

		\subsection{The \texorpdfstring{$\alpha$}{alpha}-class}
		
		We focus here on the gravity solution constructed in Section~\ref{ss:alphaClass} and show that, after an appropriate $SO(4)$ rotation, it is dual to the heavy state described schematically in \eqref{eq:alpha1alpha2}.
		
		Our guides in finding the correct gauge are, as we outlined above, the CFT quantised charges, that can be easily extracted from \eqref{eq:alpha1alpha2}:
		\begin{equation}\label{eq:CFTchargesalpha}
		J^3 = \frac{1}{2}(N_1+N_2)\,,\,\, J^\pm=0\,,\,\, \tilde J^3= \frac{1}{2}(N_1-N_2)\,,\,\, \tilde J^\pm=-\delta_{n_1,n_2}\sqrt{N_1 N_2}\,,\,\, n_p = h-\bar h=n_1 N_1+n_2 N_2\,.
		\end{equation}
		There is a  point that needs some clarification:  the value of $\tilde J^\pm$: it can be understood by remembering that the heavy states are actually defined by coherent state superpositions of multi-particle states with a different number of elementary constituents and by noting that the state $\tilde J^+_0  \Bigl (L_{-1}^{n_2} \Bigl |\frac{1}{2},-\frac{1}{2}\Bigr \rangle\Bigr ) =  L_{-1}^{n_2} \Bigl |\frac{1}{2},\frac{1}{2}\Bigr \rangle$ is orthogonal to $L_{-1}^{n_1} \Bigl |\frac{1}{2},\frac{1}{2}\Bigr \rangle$ unless $n_1=n_2$; hence, when $n_1=n_2$ acting with $\tilde J^\pm$ on the heavy state gives a state that has non-vanishing overlap with itself, but this does not happen for the non-BPS state with $n_1\not = n_2$. This shows that we have to treat the BPS and non-BPS cases separately in the $\alpha$-class of solutions and we cannot simply set $n_1=n_2$ in the expressions (in contrast, for the $\beta$-class, $n_1=n_2$ is still non-BPS, as explained after \eqref{eq:N1N2b1b2}, and there is no such issue).  In the following we will mostly focus on the non-BPS solutions with $n_1\not=n_2$. 
		
		Comparison of the CFT predictions \eqref{eq:CFTchargesalpha} with the supergravity values computed from the holographic formulas \eqref{eq:angmomhol}, \eqref{eq:hhbar}, requires the dictionary between the CFT variables, $N_1,N_2$, and the gravity ones, $\alpha_1,\alpha_2$, which is only known in the BPS limits, $\alpha_1=0$ or $\alpha_2=0$ (see \eqref{eq:N1N2a1a2}), while the corrections have to be inferred order by order in the double perturbative expansion in $\alpha_1$, $\alpha_2$. Already at zeroth order in both $\alpha_1$ and $\alpha_2$ one sees that the gauge used in the supergravity construction is not the one that describes a proper CFT state. At this order, the non-vanishing gauge fields are constants: 
		\begin{equation}\label{eq:constantguagefields}
		\Phi_1=\Phi_2=\frac{1}{2}\quad , \quad \Psi_1=\frac{n_1}{2}\quad, \quad \Psi_2=\frac{n_2}{2}\,. 
		\end{equation}
		The constant gauge potentials are singular at the origin  and so we must make a gauge transformation to arrange  $\Psi_i(0)=0$, and hence, at this order, $\Psi_i$ vanishes everywhere.  Moreover, the CFT requires, at this order, $J_3=\tilde J_3=0$ and thus, via \eqref{eq:angmomhol}, we must also make a gauge transformation that yields $\Phi_1^{(\infty)}=\Phi_2^{(\infty)}=0$. Hence, one has to apply a gauge rotation of the form \eqref{eq:phaserot} with
		\begin{equation}\label{eq:classicaldeltai}
		\delta_1=\tau+n_1\,\psi\quad,\quad \delta_2=\tau+n_2\,\psi\,,
		\end{equation}
		to cancel the constant parts of the gauge fields. Note than when this transformation acts on the $SO(4)$ vector $\chi_I$, as $\chi\to \mathcal{U} \chi$, it induces the phases $\chi_1+i\chi_2 \sim e^{i \delta_1}$ and $\chi_3+i\chi_4 \sim e^{i \delta_2}$ which are the phases expected for the states $L_{-1}^{n_1}\Bigl |\frac{1}{2},\frac{1}{2}\Bigr \rangle$ and $L_{-1}^{n_2}\Bigl |\frac{1}{2},-\frac{1}{2}\Bigr \rangle$, respectively, based on the values of the energy and momentum. Turning things around, this is the reason why in the supergravity construction one has chosen the constants \eqref{eq:constantguagefields} for the gauge fields, as otherwise the equations of motion depend on $(\tau,\,\psi)$ as well.
		
		The gauge transformation (\ref{eq:classicaldeltai}) has the effect of making the mode numbers of the scalar fields appear explicitly in terms of modes and frequencies.  In particular, we view $\omega_i$ as the classical frequency. At leading order in the large $N$ limit, holographic CFTs reduce to generalized free theories, so the properties of the composite states can be derived by summing up the contributions of the individual constituents. As we will see, the interactions between two individual but mutually non-BPS constituents induce corrections of order $1/N$ in the binding energy $\delta u_{12}$ (see~\eqref{eq:deltau}), so the classical frequency will typically undergo a shift, $\delta \omega_{1,2}$ that is proportional to $N_{2,1} \delta u_{12}$. The field will then have a physical frequency $\hat \omega_i =  \omega_i +\delta \omega_i$.
		
		The ``non-abelian" gauge fields, $\Phi_i$, $\Psi_i$ with $i=3,4$, first appear at order $\alpha_1 \alpha_2$ and are given in \eqref{eq:alphaGen2ndO}. Using the holographic formulas \eqref{eq:angmomhol} to extract the values of the angular momenta from the asymptotic values of these gauge fields we obtain $J^\pm\not=0$ and $\tilde J^\pm=0$, while for the non-BPS states with $n_1\not = n_2$ one expects both $J^\pm$ and $\tilde J^\pm$ to vanish (see \eqref{eq:CFTchargesalpha}). This signals the necessity to perform a further gauge transformation to rotate $J^\pm$ away. The regularity condition that $\Psi_i(\xi=0)=0$ and the requirement for vanishing $\xi$-components of the gauge fields at $\xi=1$ partially constrain the transformation, and a possible choice is
		\begin{equation}\label{eq:nonabegauge}
		\mathcal{U}=\mathbb{1}+\alpha_1\alpha_2\frac{\xi^{n_1+n_2}\big((n_1+n_2)\xi^2-(n_1+n_2+2)\big)}{2(n_1-n_2)(n_1+n_2)^2(n_1+n_2+2)}\begin{pmatrix} 0&0&u_1&0\\0&0&0&u_2\\-u_1&0&0&0\\0&-u_2&0&0\end{pmatrix}\,,
		\end{equation}
		where $u_1=n_1+n_2+2\,n_1\,n_2$ and $u_2=n_1\,(n_1+1)+n_2\,(n_2+1)$. This transforms the gauge fields to
		\begin{equation}
		\begin{aligned}
			&\Phi_3=\Phi_4=\alpha_1\,\alpha_2\frac{n_1^2-n_2^2+(n_1-n_2)\,\xi^{n_1+n_2} \big((n_1+n_2)\xi^2-(n_1+n_2+2)\big)}{4\,(n_1+n_2)^2(n_1+n_2+2)}\,,\\
			&\Psi_3=-\alpha_1\,\alpha_2\frac{(n_2+1)\Big(n_1+n_2+\xi^{n_1+n_2}\big(2+(1-\xi^2) (n_1+n_2)\big)\Big)}{4(n_1+n_2)(n_1+n_2+2)}\,,\\ &\Psi_4=\alpha_1\,\alpha_2\frac{(n_1+1)\Big(n_1+n_2+\xi^{n_1+n_2}\big(2+(1-\xi^2) (n_1+n_2)\big)\Big)}{4(n_1+n_2)(n_1+n_2+2)}\,,\\
			&\Psi_5=\alpha_1\,\alpha_2\frac{(n_1+n_2+2\,n_1\,n_2)}{4(n_1^2-n_2^2)}(1-\xi^2)\,\xi^{n_1+n_2-1}\,,\\
			&\Psi_6=\alpha_1\,\alpha_2\frac{n_1\,(n_1+1)+n_2\,(n_2+1)}{4(n_1^2-n_2^2)}(1-\xi^2)\,\xi^{n_1+n_2-1}\,.
		\end{aligned}
		\end{equation}
		It appears that $\Psi_{5,6}$ are singular at the origin for $n_1=n_2$, but, in fact, the state is then BPS, whereas the transformation above applies only in the non-BPS case. The former is discussed below.
		
		Having obtained the expected values of $J^\pm$ and $\tilde J^\pm$, one should also consider $J^3$ and $\tilde J^3$: discarding terms of order $\alpha_2^2\sim N_2$ and higher, one should have $J^3=\tilde J^3=\frac{N_1}{2}\sim N\frac{\alpha_1^2}{4}$, where we stop at order $O(\alpha_1^2)$ for now.  At this order, the existence of a normalisable solution for $\nu_2$ requires adjusting the constant term in $\Phi_2$ and the asymptotic values of the gauge fields are corrected with respect to the zero-th order values \eqref{eq:constantguagefields} to (see \eqref{eq:alphaGen2ndO})
		\begin{align}\label{eq:constantguagefieldsa12}
		\Phi_1^{(\infty)}&=\frac{1}{2}-\frac{\alpha_1^2}{8}\,,\quad
		\Phi_2^{(\infty)}=\frac{1}{2}+\frac{1}{8}\bigg[-1+\frac{2\,(n_1-n_2)^2}{(n_1+n_2)(n_1+n_2+1)(n_1+n_2+2)}\bigg]\alpha_1^2\,,\notag\\
		\Psi_1^{(\infty)}&=\frac{n_1}{2}\,,\quad \Psi_2^{(\infty)}=\frac{n_2}{2}-\frac{\alpha_1^2}{8}\,. 
		\end{align}
		Then, to obtain the expected values of $J^3$ and $\tilde J^3$, one has to supplement the $SO(4)$ rotation \eqref{eq:nonabegauge} by a gauge transformation of the form of \eqref{eq:phaserot} where the phase $\delta_2$ has to be corrected, with respect to the classical value \eqref{eq:classicaldeltai}, to 
		\begin{equation}
		\delta_2\to (1+\delta\omega_2)\,\tau + n_2\, \psi
		\label{eq:deltaShift}
		\end{equation}
		with
		\begin{equation}\label{eq:deltaomega2first}
		\delta\omega_2= -2\frac{N_1}{N}\left[1-\frac{(n_1-n_2)^2}{(n_1+n_2)(n_1+n_2+1)(n_1+n_2+2)}\right] + \, O\!\left(\frac{N_1}{N} \frac{N_2}{N}, \qty(\frac{N_1}{N})^2 \right)\quad\mathrm{for}\quad n_1\not=n_2\,.
		\end{equation}
		This quantity has an important microscopic meaning: it represents the interaction energy of a particle of type $L_{-1}^{n_2}\Bigl |\frac{1}{2},-\frac{1}{2}\Bigr \rangle$ with the $N_1$ particles of type $L_{-1}^{n_1}\Bigl |\frac{1}{2},\frac{1}{2}\Bigr \rangle$. Thus we can read the binding energy $\delta u_{12}$ between a single particle in the first sector and a single particle in the second sector
		\begin{equation}
			\label{eq:deltau}
			\delta u_{12}= -\frac{2}{N} \left[1-\frac{(n_1-n_2)^2}{(n_1+n_2)(n_1+n_2+1)(n_1+n_2+2)}\right] + O\!\left( \frac{1}{N^2} \right)\;.
		\end{equation}
		As expected, this quantity is symmetric in the exchange of the two constituents and this provides a first check of the result since in the derivation this symmetry was not manifest. In CFT language, this binding energy is related to the anomalous dimension of the non-BPS double-trace operator $:\partial^{n_1} s^{(1)(\frac{1}{2},\frac{1}{2})}_1 \partial^{n_2} s^{(1)(\frac{1}{2},-\frac{1}{2})}_1:$.
	
		Since this energy shift originates from the attractive interaction between oppositely charged particles, one expects $\delta\omega_2< 0$, in agreement with the expression in \eqref{eq:deltaomega2first}, for any $n_1,n_2\ge 0$. One could wonder why $\delta\omega_2$ does not vanish for the BPS state with $n_1=n_2$. The point is that, as we emphasised after \eqref{eq:CFTchargesalpha}, the limit $n_1\to n_2$ is not smooth and the analysis above only applies when $n_1\not=n_2$. When $n_1=n_2$ one has $\tilde J^\pm\not=0$ (see \eqref{eq:CFTchargesalpha}), and this requires modifying the gauge-transformation \eqref{eq:nonabegauge} to 
		\begin{equation}
		\mathcal{U}=\mathbb{1}+\frac{\alpha_1\tau}{4}\begin{pmatrix} 0&0&0&\alpha_2\\0&0&-\alpha_2&0\\0&\alpha_2&0&2\alpha_1\\ -\alpha_2&0&-2\alpha_1&0\end{pmatrix}\,,
		\end{equation}
		which is $\xi$-independent but $\tau$-dependent and it changes the energy shift $\delta\omega_2$ at order $\alpha_1^2$ to
		\begin{equation}
		\delta\omega_2=0 \quad \mathrm{for}\quad n_1=n_2\,,
		\end{equation}
		as expected for a supersymmetric state. 
		
		Both the supergravity and the CFT pictures have a symmetry under the exchange of the states $L_{-1}^{n_1}\Bigl |\frac{1}{2},\frac{1}{2}\Bigr \rangle$ and $L_{-1}^{n_2}\Bigl |\frac{1}{2},-\frac{1}{2}\Bigr \rangle$, which interchanges the labels $1$ and $2$. This implies that the energy of the particles of type $L_{-1}^{n_1}\Bigl |\frac{1}{2},\frac{1}{2}\Bigr \rangle$ is shifted by
		\begin{equation}\label{eq:deltaomega1first}
		\delta\omega_1= -2\frac{N_2}{N}\left[1-\frac{(n_1-n_2)^2}{(n_1+n_2)(n_1+n_2+1)(n_1+n_2+2)}\right] + O\left(\frac{N_1}{N} \frac{N_2}{N}, \qty(\frac{N_2}N)^2\right)\quad\mathrm{for}\quad n_1\not=n_2\,,
		\end{equation}
		when interacting with the $N_2$ particles of type $L_{-1}^{n_2}\Bigl |\frac{1}{2},-\frac{1}{2}\Bigr \rangle$. Moreover the total energy of the non-supersymmetric bound state should be given by the ($1$-$2$) symmetric expression
		\begin{equation}\label{eq:totalenergy}
		h+\bar h=(n_1+1)\,N_1+(n_2+1)\,N_2 -2\frac{N_1 N_2}{N}\left[1-\frac{(n_1-n_2)^2}{(n_1+n_2)(n_1+n_2+1)(n_1+n_2+2)}\right] \,,
		\end{equation}
		up to higher order corrections in $N_1/N$, $N_2/N$. The first two terms describe just the total energy of the particles in the free theory, which is obtained by summing up the free conformal dimension of all constituents. The final term is the attractive interaction energy. Since from the holographic point of view we are describing a multiparticle state with weakly interacting constituents, this interaction energy has to be proportional, at leading order in the $N_i/N\ll 1$ regime, to the numbers of particle $N_1$ and $N_2$ and to the three-dimensional Newtons's constant $G_3\sim 1/N$.
		
				One could try to verify the relation \eqref{eq:totalenergy} by extracting the energy $h+\bar h$ from the three-dimensional metric $ds^2_3$, computed up to order $\alpha_1^2 \alpha_2^2$, using the holographic formulas \eqref{eq:hhbar}. However one faces a difficulty: the prediction \eqref{eq:totalenergy} is expressed in terms of the microscopic numbers $N_1$, $N_2$, while the holographic computation yields $h+\bar h$ as a function of the gravity parameters $\alpha_1$, $\alpha_2$. The comparison thus requires the relation between $(N_1,N_2)$ and $(\alpha_1,\alpha_2)$ up to order $\alpha_1^2 \alpha_2^2$.  Taking into account the ($1$-$2$)-exchange symmetry, this relation could have the general form
		\begin{equation}\label{eq:N1N2mapsecond}
		\frac{N_1}{N}=\frac{\alpha_1^2}{4}+c(n_1,n_2)  \, \alpha_1^2 \,\alpha_2^2 + O\qty(\alpha_1^4,\, \alpha_1^2 \,\alpha_2^4)\quad,\quad \frac{N_2}{N}=\frac{\alpha_2^2}{4}+c(n_2,n_1) \,\alpha_1^2 \,\alpha_2^2  + O\qty(\alpha_2^4,\, \alpha_1^4 \, \alpha_2^2)\,,
		\end{equation}
		where the first term is the BPS value ($\alpha_2=N_2=0$) obtained in~\eqref{eq:alp1a}, while the function $c(n_1,n_2)$ (that should be neither symmetric nor anti-symmetric under the exchange of $n_1$ and $n_2$) describes the first subleading correction when $N_i/N$ are small. The only constraint one has on $c(n_1,n_2)$ is the one coming from the momentum $h-\bar h$:
		\begin{equation}\label{eq:hminushbar}
		h-\bar h =n_1\,N_1+ n_2\,N_2 = \frac{n_1\,\alpha_1^2+n_2\,\alpha_2^2}{4}+ (n_1\,c(n_1,n_2)+n_2\,c(n_2,n_1))\,\alpha_1^2\alpha_2^2 \,+ O\qty(\alpha_1^4, \,\alpha_2^4)\,,
		\end{equation}
		which can again be computed from the gravity solution using \eqref{eq:hhbar}. This lone constraint is not sufficient to determine $c(n_1,n_2)$. One can, however, turn things around\footnote{In \cite{Ganchev:2021ewa} an independent check of a relation analogous to \eqref{eq:totalenergy} was performed: the non-supersymmetric states studied in that article depended on a single parameter $\alpha$ and thus the constraint coming from the momentum charge was sufficient to completely determine the gravity-CFT map. The non-triviality of that check, gives us further motivation for assuming \eqref{eq:totalenergy} in our context.} and assume both \eqref{eq:hminushbar} {\it and} \eqref{eq:totalenergy}, and thus determine $c(n_1,n_2)$. Computationally, this requires obtaining the perturbative solution for general $n_1$ and $n_2$ to at least fourth order. We managed to do this only for the $\alpha$-class of solutions, due to the simplifications of the special locus. We have:
		\begin{align}
		c(n_1,n_2)=\frac{1}{16}&\bigg[\frac{4\,n_1^2}{(n_1+n_2)^2}-\frac{5\,n_1(n_1+2)}{n_1+n_2}+\frac{1}{n_1-n_2}-\frac{2\,(2\,n_1+1)^2}{(n_1+n_2+1)^2}+\frac{4(n_1+1)^2}{(n_1+n_2+2)^2}\notag\\
		&+\frac{2+2\,n_1\big(5\,n_1(n_1+3)+8\big)}{n_1\,(n_1+n_2+1)}-\frac{(n_1+1)\big(5\,n_1\,(n_1+3)+4\big)}{n_1\,(n_1+n_2+2)}\bigg]\notag\\
		&+\frac{(n_1-n_2)^2\big(\psi^{(0)}(n_1+n_2)-\psi^{(0)}(n_1)\big)}{8\,(n_1+n_2)(n_1+n_2+1)(n_1+n_2+2)} ,
		\end{align}
		where $\psi^{(m)}(z)$ is the polygamma function of order $m$. For $m=0$, $\psi^{(0)}(z)$ is also known as the digamma function.
		
		The coefficients $c(n_1,n_2)$ do not have an intrinsic physical meaning, they just encode the dictionary between gravity and CFT, but their knowledge is useful because it allows us to push the computation of the energy shifts, $\delta\omega_1$, $\delta\omega_2$, one step further and include the corrections proportional to $N_1 N_2$:
		\begin{equation}\label{eq:deltaomegasecond}
		\begin{aligned}
			&\delta\omega_1=-2\frac{N_2}{N}\left[1-\frac{(n_1-n_2)^2}{(n_1+n_2)(n_1+n_2+1)(n_1+n_2+2)}\right]+\frac{N_1 N_2}{N^2} \, d(n_1,n_2)\,\\
			&\delta\omega_2=-2\frac{N_1}{N}\left[1-\frac{(n_1-n_2)^2}{(n_1+n_2)(n_1+n_2+1)(n_1+n_2+2)}\right]+\frac{N_1 N_2}{N^2}\, d(n_2,n_1)\,,
		\end{aligned}
		\end{equation}
		where we have again implemented the ($1$-$2$) symmetry and introduced the unknown coefficients $d(n_1,n_2)$. The procedure to determine these unknowns is conceptually the same we used to arrive at \eqref{eq:deltaomega2first}.  One first computes the corrections of order $\alpha_1^2 \alpha_2^2$ to the gauge fields, and in particular to the constant part of $\Phi_1$, $\Phi_2$ (see  \eqref{eq:constantguagefieldsa12}), which is needed to have normalisable solutions for $\nu_1$ (resp.~$\nu_2$) at order $\alpha_1^3 \alpha_2^2$ (resp.~$\alpha_1^2 \alpha_2^3$). One then determines the corrections to the phases, $\delta_1$, $\delta_2$, of the gauge transformation \eqref{eq:phaserot} in such a way as to obtain the CFT-expected values of the angular momenta $J^3$ and $\tilde J^3$ in \eqref{eq:CFTchargesalpha} -- it is at this step that one needs the map \eqref{eq:N1N2mapsecond}. The energy shifts, $\delta \omega_1$, $\delta \omega_2$ in \eqref{eq:deltaomegasecond}, follow from the rotation parameters $\delta_1$, $\delta_2$. As with $c(n_1,n_2)$, we only have an expression for $d(n_1,n_2)$ for generic $n_1$, $n_2$ for the $\alpha$-class of solutions, namely:
		\begin{align}
		&d(n_1,n_2)=-\frac{16\,n_1^4}{(n_1+n_2)^3}+\frac{8\,n_1^3\,(5\,n_1+2)}{(n_1+n_2)^2}+\frac{4\,n_1^2}{n_2\,(n_1^2+3\,n_1+2)^2}-\frac{4\,n_1\,(n_1^2(n_1\,(14\,n_1+31)+26)-3)}{(2\,n_1+1)(n_1+n_2)}\notag\\
		&-\frac{4(3\,n_1+1)^2}{(n_1+1)^2(2\,n_1+n_2+1)}+\frac{4\,(3\,n_1+2)^2(n_1 (n_1+6)+4)}{(n_1+1)^2(n_1+2)^2(2\,n_1+n_2+2)}-\frac{4\,(2\,n_1+1)^4}{(n_1+n_2+1)^3}-\frac{16\,(n_1+1)^4}{(n_1+n_2+2)^3}\notag\\
		&+\frac{4\,(2\,n_1+1)^2(n_1(n_1+1)(2\,n_1\,(3\,n_1+5)+7)+1)}{n_1\,(n_1+1)(n_1+n_2+1)^2}+\frac{16(n_1+1)^3(n_1\,(n_1\,(n_1+4)+6)+2)}{n_1\,(n_1+2)(n_1+n_2+2)^2}\notag\\
		&+\frac{4\,(n_1+1)(n_1\,(n_1\,(n_1\,(n_1\,(n_1\,(2\,n_1\,(3\,n_1\,(10\,n_1+67)+527)+1373)+914)+244)-68)-72)-16)}{n_1^2(n_1+2)^2(2\,n_1+1)(n_1+n_2+2)}\notag\\
		&+\frac{2\,n_1\,(n_1\,(n_1\,(n_1\,(2\,n_1\,(16\,n_1\,(n_1^2-9)-329)-727)-460)-165)-30)-4}{n_1^2(n_1+1)^2(n_1+n_2+1)}\notag\\
		&+\frac{4\,\gamma\,(n_1-n_2)^2(n_1^3+(3\,n_2+4)\,n_1^2+(n_2\,(3\,n_2+4)+2)\,n_1+n_2(n_2(n_2+4)+2))}{(n_1+n_2)^2(n_1+n_2+1)^2(n_1+n_2+2)^2}\notag\\
		&+\bigg[\frac{4\,n_1^4}{n_1+n_2}+\frac{(2\,n_1+1)^4}{n_1+n_2+1}+\frac{4\,(n_1+1)^4}{n_1+n_2+2}\bigg]\bigg(4\,\psi^{(1)}(n_1)-\frac{2\pi^2}{3}\bigg)\notag\\
		&+\frac{4\,(n_1-n_2)^2\,\psi^{(0)}(n_1+n_2)}{(n_1+n_2)(n_1+n_2+1)(n_1+n_2+2)}\bigg(1-\frac{(n_1-n_2)^2}{(n_1+n_2)(n_1+n_2+1)(n_1+n_2+2)}\bigg)\notag\\
		&+\frac{4(n_1-n_2)^2(n_1^3+(3\,n_2+4)\,n_1^2+(n_2\,(3\,n_2+4)+2)\,n_1+n_2\,(n_2\,(n_2+4)+2))}{(n_1+n_2)^2(n_1+n_2+1)^2(n_1+n_2+2)^2}\,\psi^{(0)}(n_1)\notag\\
		&-\frac{4(n_1-n_2)^2(n_1^3+(3\,n_2+2)\,n_1^2+(n_2\,(3\,n_2+8)+2)\,n_1+n_2\,(n_2\,(n_2+2)+2))}{(n_1+n_2)^2(n_1+n_2+1)^2(n_1+n_2+2)^2}\,\psi^{(0)}(n_2)\notag\\
		&+\sum_{p=1}^{n_1}\Big(\frac{(p+n_1-1)^2}{8\,(p-n_1-3)(p-n_1-2)(p-n_1-1)(p+n_2-1)}+\frac{(2\,n_1+p+2)^4}{8\,p^2(p+1)^2(p+2)^2(n_1+n_2+p+2)}\Big),
		\end{align}
		where $\gamma$ is the Euler–Mascheroni constant. Somewhat surprisingly, the result is non-vanishing, implying that the strands of type 2 feel a force proportional to the number of strands of the same type, in the presence of strands of type 1 (and the same with 1 and 2 interchanged, of course).  In other words, the strands of type 1 deform the strands of type 2 in such a way that they are no longer mutually BPS with each other. From a CFT point of view, as the term proportional to $N_1$ in $\delta\omega_2$ is identified with the anomalous dimension of the double-trace operator mentioned after \eqref{eq:deltaomega2first}, the term of order $N_1 N_2$ computes the anomalous dimension of a triple-trace operator of the form  $:\partial^{n_1} \sigma_1^{(\frac{1}{2},\frac{1}{2})} \partial^{n_2} \sigma_1^{(\frac{1}{2},-\frac{1}{2})} \partial^{n_2} \sigma_1^{(\frac{1}{2},-\frac{1}{2})}:$.
		
		\subsection{The \texorpdfstring{$\beta$}{beta}-class}
		
		A similar analysis can be performed for the non-BPS state \eqref{eq:beta1beta2}, whose charges readily follow from its CFT representation:
		\begin{equation}
		J^3 = N_1+N_2\quad ,\quad J^\pm=0\quad ,\quad  \tilde J^3= N_1-N_2\quad ,\quad  \tilde J^\pm=0\quad ,\quad n_p = h-\bar h=2\,n_1\,N_1+2\,n_2\,N_2\,.
		\label{eq:betaHolCharges}
		\end{equation}		
		
		The non-BPS perturbation, at first order in the perturbation parameter $\beta_2$, is controlled by a single field $\lambda_2$ and this greatly simplifies the linear-order analysis: in particular one can study the perturbation at first order in $\beta_2$ and all orders in $\beta_1$ using a WKB approach that is discussed in Section~\ref{sub:perturbation_pureNS}. Here we briefly comment on the double-perturbative approach in $\beta_1$ and $\beta_2$, along the lines of the previous sub-section.
		
		At first order in $\beta_2$, the fact that all the ``non-abelian" gauge fields $\Phi_i$, $\Psi_i$ with $i=3,4$ vanish, trivialises the agreement with the constraints $J^\pm=\tilde J^\pm=0$. A non-trivial input comes from the CFT-gravity matching of $J^3$, $\tilde J^3$. At lowest non-trivial order in $\beta_1$, the asymptotic values of the gauge fields are
		\begin{gather}
		\Phi_1^{(\infty)}=\frac{1}{2}-\frac{\beta_1^2}{4\,(2\,n_1+1)}\quad , \quad \Psi_1^{(\infty)}=\frac{n_1}{2}\quad, \notag\\ \Phi_2^{(\infty)}=\frac{1}{2}- \frac{(n_1+n_2-1)\,\beta_1^2}{4\,(2\,n_1+1)(n_1+n_2+1)}\quad ,\quad \Psi_2^{(\infty)}=\frac{n_2}{2}-\frac{\beta_1^2}{4\,(2\,n_1+1)}\,,
		\end{gather}
		where, in particular, the constant part of $\Phi_2$ is determined by the existence of a normalisable solution for $\lambda_2$. The angular momenta implied by these gauge fields are consistent with the ones predicted by the CFT only after a gauge rotation \eqref{eq:phaserot} with
		\begin{equation}\label{eq:deltaomega2betafirst}
		\delta\omega_2 =-\frac{n_1+n_2}{(2n_1+1)(n_1+n_2+1)}\beta_1^2\approx -4 \frac{N_1}{N}\frac{n_1+n_2}{n_1+n_2+1}\,.
		\end{equation}
		Note that, due to the form of the six-dimensional ansatz, the phase of the field $\lambda_2$ is $2\delta\omega_2$, and thus \eqref{eq:deltaomega2betafirst} represents half of the interaction energy between a particle of type $L_{-1}^{2n_2}|1,-1\rangle$ and $N_1$ particles of type $L_{-1}^{2n_1}|1,1\rangle$. It is a non-trivial consistency check that this interaction energy, when expressed in terms of the microscopic number $N_1$, is symmetric under exchange of $n_1$ and $n_2$, and also that it is always negative, as expected on physical grounds. In CFT terms this energy shift encodes the anomalous dimension of the double-trace operator $:\partial^{n_1} (\sqrt{3} \,s_2^{(2)(1,1)}-\sigma_2^{(1,1)}) \partial^{n_2} (\sqrt{3} \,s_2^{(2)(1,-1)}-\sigma_2^{(1,-1)}):$. Following the same steps explained in the previous sub-section for the $\alpha$-class  of solutions, one can include the corrections of order $\beta_1^2\beta_2^2$ to the energy shift:
		\begin{equation}\label{eq:deltaomega2betasecond}
		\delta\omega_2 = -4 \frac{N_1}{N}\frac{n_1+n_2}{n_1+n_2+1}+ \frac{N_1 N_2}{N^2} \,\tilde d(n_1,n_2) \,,
		\end{equation}
		where the coefficients $\tilde d(n_1,n_2)$ capture the triple-trace anomalous dimension. We have found it technically challenging to compute analytically the coefficients $\tilde d(n_1,n_2)$ for the beta class of solutions. However we can evaluate the energy perturbation $\delta \omega_2$ in complementary regimes by different methods: the WKB approximation of Section~\ref{ss:wkb_approximation}, which is valid for small $\beta_2$ and large $\omega_2$, and the numerical analysis that we will carry out in the next section, where we will be able to take arbitrary values of $\beta_1=\beta_2$ and finite values of the frequencies $\omega_1=\omega_2$.
		
		\section{Numerical Results}
		\label{sec:Nums}
		
		Apart from foreshadowing, in Section \ref{sub:trophies},  the results to be presented in this section, all our  results  so far have been obtained through  perturbative methods. However, we can also solve the full system of non-linear equations numerically and in this section we will outline our scheme for doing so, illustrate the method by computing a strongly non-BPS example, with a ``deep throat,'' and make comparisons with what we already know from perturbation theory.
		
		For simplicity, we will focus on  the $\mathbb{Z}_2$ symmetric $\beta$-class of solutions described in Section \ref{sub:trophies}.  That is, we take $n_1=n_2=n$ and $\beta_1=\beta_2=\beta$. As noted in in Section~\ref{ss:betaClass}, these solutions,  and the dual CFT states, are non-BPS.  We make this simplifying choice,  not because we cannot perform the numerics more generally, but because this is a  scenario in which we know how to map the CFT parameters to bulk supergravity quantities using the holographic dictionary. To elaborate (see also the end of the introductory words to Section~\ref{sec:Holography}), if $n_1\neq n_2$, for either the $\alpha$- or $\beta$-class of solutions, we have different numbers of constituents of the given species in the two sectors: $N_1\neq N_2$. Perturbatively, these are related to the gravity amplitudes that drive the bulk solution via \eqref{eq:N1N2a1a2} and \eqref{eq:N1N2b1b2}. Unfortunately, we cannot obtain exact expressions for $N_{1,2}$ in terms of gravity parameters (or even as an expansion in the gravity amplitudes that we can then attempt to re-sum), as that requires the precise form of $h+\bar{h}$ in terms of the full energy shifts $\delta\omega_{1,2}$, thus making it impossible to read off $N_{1,2}$ from the bulk, numerical solution. 
		
	If one imposes  $\mathbb{Z}_2$ symmetry, then $N_1=N_2$, and so we only need a single relation, which we get from $h-\bar{h}=n_p$,
This is a quantized charge, thus preserved under the breaking of supersymmetry, given by	$n_p=4\,n\,N_1$, and so we can determine $N_1 =N_2$ from the value of $h-\bar{h}$ given by the supergravity solution.  One can then extract the energy shifts, or anomalous dimensions, in terms of constituents CFT states by reading off $h+\bar{h}$ from the supergravity solution.
	
	The equations we will solve are given in Appendix~\ref{app:betaNumericsEqs}, and, specifically,  \eqref{eq:betaZ2Eqs}. These are seven second-order, ordinary differential equations. The reduced number comes about because $n_1=n_2=n$ and $\beta_1=\beta_2=\beta$ imply that $\mu_1=\mu_2$, $\lambda_1=\lambda_2$, $\Phi_1=\Phi_2$ and $\Psi_1=\Psi_2$, as easily seen in perturbation theory. We should note that a linear combination of the $\Phi_1$, $\Psi_1$ and $\kappa$ equations of motion leads to a first-order, differential equation, given in \eqref{eq:BetaZ2firstPsiEq}. This is a genuine equation of motion for one of the fields, but we did not use it in the numerical procedure, as it seemed to have slightly worse convergence properties.
		
\begin{figure}[!ht]
			\centering
			\includegraphics[width=0.75\textwidth]{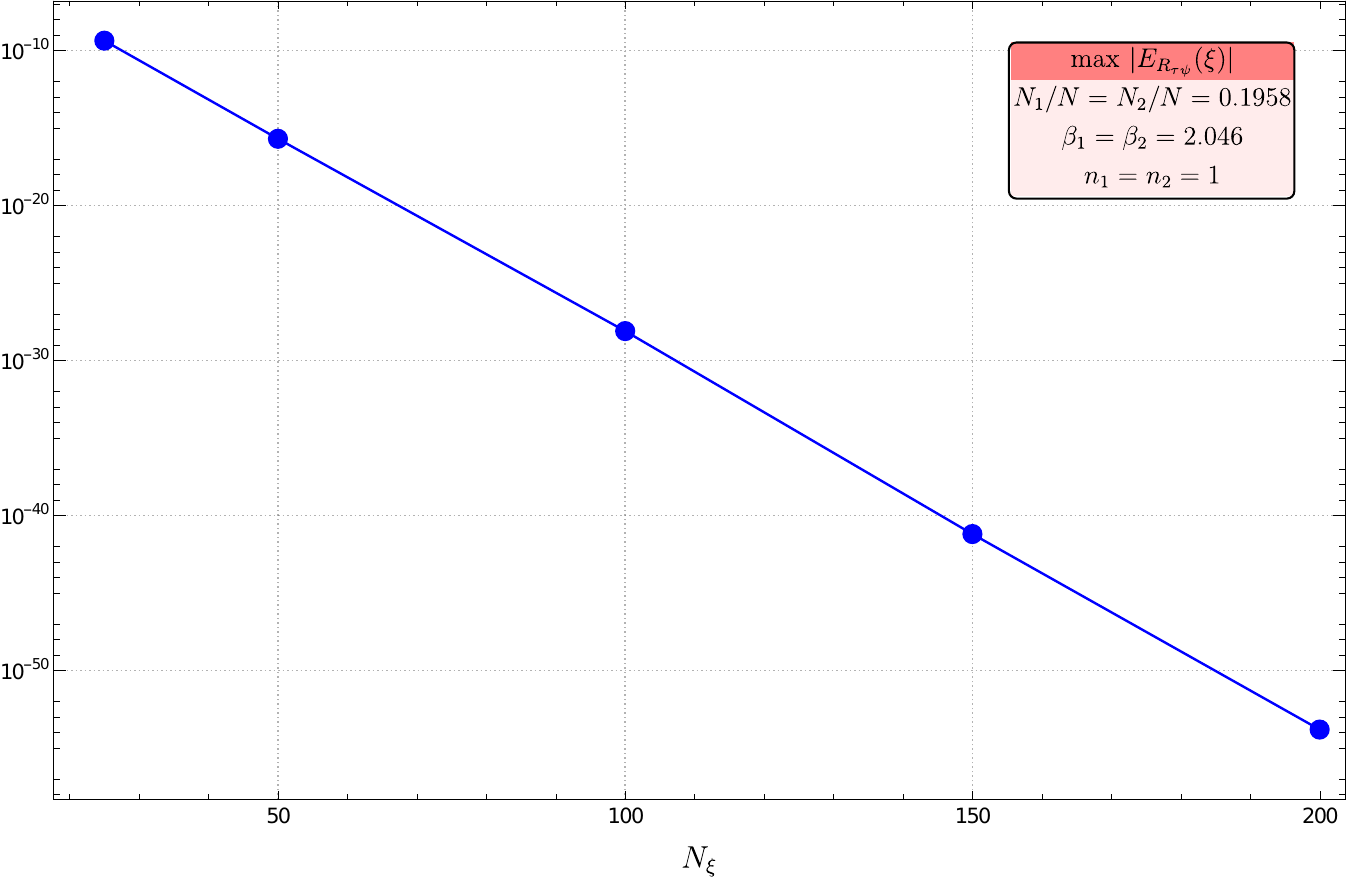}
			\caption{Maximum of the absolute value of the Einstein constraint, $E_{R_{\tau\psi}}$, over the whole integration domain for a highly non-BPS state as a function of the number of points used in the numerics, $N_\xi$. $\omega_1=\omega_2=1$.}
			\label{fig:ConvPlot}
		\end{figure}

		As discussed in Section~\ref{sec:nonBPS}, our system of equations also contains three first-order constraints. The ones coming from the Maxwell sector are automatically satisfied. Only the constraint from Einstein equations, $E_{R_{\tau\psi}}$, remains and it is a constant of motion. We use it to monitor the convergence of our solutions, as seen in Figure~\ref{fig:ConvPlot}, where the maximum value of the constraint over the whole space is plotted as a function of the number of grid points, $N_\xi$, for a highly non-BPS solution. All results presented in this work have $\max\,\lvert E_{R_{\tau\psi}}(\xi)\rvert\leq10^{-50}$. 
		
		We solve for $\{\lambda_1,\,\mu_1,\,\Phi_1,\,\Psi_1,\,\Omega_0,\,\Omega_1,\,\kappa\}$ as functions of the radial variable $\xi\in\[0,1\]$, where $\kappa$ is defined via $k=\xi\,\kappa$. The numerical integration is carried out by using spectral collocation methods on a Chebyshev-Gauss-Lobatto grid and implementing a standard Newton-Raphson iteration (see \cite{Dias:2015nua} for a review of such methods applied to gravity) with increased precision (50 digits). We need to impose appropriate boundary conditions to ensure a well-posed problem. These were discussed in detail in Section~\ref{sub:gen-bcs}, but we will briefly outline them here for completeness.
		
		At the origin, $\xi=0$, regularity enforces Neumann boundary conditions on almost all fields, except\footnote{Remember that $k(\xi=0)\sim\xi^2$ and we have defined $k(\xi)=\xi\,\kappa(\xi)$.} for $\Psi_1$ and $\kappa$:
		\begin{gather}
			\left.\partial_\xi\,X(\xi)\right|_{\xi=0}=0,\quad{\rm for}\,X\in\{\lambda_1,\,\mu_1,\,\Phi_1,\,\Omega_0,\,\Omega_1\},\notag\\
			\kappa(\xi=0)=0,\quad\Psi_1(\xi=0)=\frac{n}{2}.\label{eq:BCOrigin}
		\end{gather}
		In fact, it can be shown that the first $2\,n-1$ derivatives of $\lambda_1$ vanish at the origin, in agreement with expectations from perturbation theory.

		Our spacetime is asymptotically, locally AdS \cite{deHaro:2000xn,Skenderis:2002wp} and at the AdS boundary, $\xi=1$, we are free to choose the conformal class of the boundary metric. As before, eq. \eqref{MinkSliced}, we impose:
		\begin{equation}
			\Omega_0(\xi=1)=1,\quad\Omega_1(\xi=1)=\frac{1}{\kappa(\xi=1)},\quad\left.\partial_\xi\,\kappa\right|_{\xi=1}=1.
		\end{equation}
		The scalars are fixed by demanding that, in addition, we approach the supersymmetric vacuum at the boundary, so that:
		\begin{equation}
			\lambda_1(\xi=1)=0,\quad\mu_1(\xi=1)=0.
		\end{equation}
		
		Our solutions is specified by 2 parameters, $n$ and $\beta$. The former is set at the origin, \eqref{eq:BCOrigin}. In the perturbation theory, we identify $\beta$ with the $2\,n$-th derivative of $\lambda_1$. However, enforcing a boundary condition at the origin with so many derivatives is very hard to implement numerically, especially if we want to access values of $n_1=n_2=n>1$. Therefore, for $\beta_1=\beta_2=\beta$, we have found that numerically it is easiest to use the the value of $\Phi_1$ at infinity as the free parameter. Thus, we set
		\begin{equation}
			\Phi_1(\xi=1)=\tilde{\phi_1},
		\end{equation}
		and use $\tilde{\phi_1}$ to explore the moduli space of solutions for fixed $n$. Nevertheless, we need the value of $\beta$ to compare our numerical results with the perturbative predictions. To extract it precisely we resorted to obtaining our solutions with extended precision of 50 digits, so that we can trust the derivatives.
		
		We do not need to provide a boundary condition for $\Psi_1$ at infinity since the existence of \eqref{eq:BetaZ2firstPsiEq} implies that we have in fact 6 second-order differential equations and 1 first-order equation, and, even though, we do not integrate \eqref{eq:BetaZ2firstPsiEq} explicitly, it can be used to argue that a single boundary condition on $\Psi_1$ at the origin is sufficient, as numerically we are solving for smooth functions.
		
		\subsection{Holographic extraction}
		
		To extract the holographic quantities from the numerical solutions, we first need to expand the fields  in series about the conformal boundary, $\xi=1$, using
		\begin{equation}
			X=\sum_{k=0}^{\infty}\,(1-\xi^2)^k\,q_X^{(k)},\quad{\rm for}\,X\in\{\lambda_1,\,\mu_1,\,\Phi_1,\,\Psi_1,\,\Omega_0,\,\Omega_1,\,\kappa\}.\label{eq:confBdExp}
		\end{equation}
		Some of the coefficients in these expansions  get fixed by the equations of motion, \eqref{eq:betaZ2Eqs}, together with the boundary conditions. The ones that remain unspecified are:
		\begin{equation}
			q_{\lambda_1}^{(1)},\,q_{\mu_1}^{(2)},\,q_{\Phi_1}^{(1)},\,q_{\Psi_1}^{(0)},\,q_{\Omega_0}^{(2)},\,q_{\kappa}^{(0)}.
		\end{equation}
		
		We then write the metric in Fefferman-Graham form for an asymptotically, locally AdS space, \eqref{eq:asympt3D}. The conformal boundary is located at $z=0$. In odd bulk dimensions, (as we have here), there is a logarithmic term in the expansion that is linked to the conformal anomaly of the boundary CFT. However, for a conformally flat, boundary metric, as we have chosen here because we want a holographic $(1+1)$-dimensional field theory,  the conformal anomaly vanishes and the logarithm is not present \cite{deHaro:2000xn}. 
		
		To obtain \eqref{eq:asympt3D} we change coordinates asymptotically using
		\begin{equation}
			\xi=\sum_{k=0}^{\infty}\,a_k\,z^k,
		\end{equation}
		where the coefficients $a_k$ are determined order by order in $z$ by taking our metric ansatz\footnote{Remembering that for the numerics we use $k(\xi)=\xi\,\kappa(\xi)$}, \eqref{genmet1}, plugging in the function expansions \eqref{eq:confBdExp}, applying the above coordinate transformation and matching with \eqref{eq:asympt3D}.   For our purposes we only need:
		\begin{equation}
			a_0=1,\quad a_1=0,\quad a_2=-\frac{1}{2\,q_\kappa^{(0)}},\quad a_3=0,\quad a_4=\frac{1}{8\,\big(q_\kappa^{(0)}\big)^2}.
		\end{equation}
		Finally, we use \eqref{eq:betaHolCharges}, which, because of the  $\mathbb{Z}_2$ symmetry,  reduce to\footnote{$J^{\pm}$ and $\tilde{J}^{\pm}$ don't change.} $J^3=2\,N_1$, $\tilde{J}^3=0$.  These are automatically satisfied since $\Phi_1=\Phi_2$.  We also use  $n_p=h-\bar{h}=4\,n\,N_1$ to determine $N_1/N=N_2/N$ and $\delta\omega_1=\delta\omega_2$ as functions of the gravity parameters.  The result is that we have equal numbers of equations and unknowns, and we find:
		\begin{align}
			\delta\omega~=~ &\delta\omega_1~= \delta\omega_2 ~=~-1+2\big(\tilde{\phi_1}+2\,q_{\Psi_1^{(0)}}\big)+K,\notag\\
			N_1/N~=~ &N_2/N~=~\frac{n}{2}+\frac{K}{4},\label{eq:betaZ2symHoloDic}
		\end{align}
		where
		\begin{equation}
			K=-\frac{1}{\sqrt{3}\,q_{\kappa}^{(0)}}\Bigg[-3 - 8\,\big(q_{\kappa}^{(0)}\big)^4\Big(\big(q_{\lambda_1}^{(1)}\big)^2\tilde{\phi_1}^2+\big(q_{\Phi_1}^{(1)}\big)^2\Big)
			+\big(q_{\kappa}^{(0)}\big)^2\Big(3+12\,n^2-4\,\big(q_{\lambda_1}^{(1)}\big)^2-18\,q_{\Omega_0}^{(2)}\Big)\Bigg]^{\frac12}.
		\end{equation}
		Furthermore, the expression for the total energy in terms of bulk quantities can be determined from \eqref{eq:hhbar} as
		\begin{equation}
			\frac{h+\bar{h}}{N}=\frac{1}{2}\Bigg(1-\frac{1}{\big(q_{\kappa}^{(0)}\big)^2}\Bigg)+n\,\bigg(2\,n+K\bigg).\label{eq:z2Symhphbar}
		\end{equation}
		
		\subsection{Results}
		\label{ss:NumResults}
		
		For the $\beta$-class of solutions,  half-integer values of $n_{1,2}$ are also allowed (see \eqref{eq:nusLambdas}). We have obtained numerical results for six different values: $n_1=n_2=n\in\{\frac{1}{2},\,1,\,\frac{3}{2},\,2,\,\frac{5}{2},\,3\}$ with $\omega_1=\omega_2\in\{1,2\}$. All of them have qualitatively the same behaviour, except for $n=\frac{1}{2}$, which will be discussed further below.
		
		The space of geometries is explored by using the perturbative solutions as seeds to the Newton-Raphson method. We start from $\tilde{\phi_1}\sim\omega_1/2$, which is true\footnote{From perturbation theory we know that for the $\mathbb{Z}_2$ symmetric $\beta$-class we have: $\Phi_1(\xi=1)=\omega_1/2-(n\,\beta^2)/(1+2\,n)^2+\mathcal{O}(\beta^4)$.} for $\beta\ll 1$, and decrease its value, as  $\beta$ increases. In practice, we will present our results as a function of the fraction of single-particle constituents, $N_1/N=N_2/N$, hence we start with a  plot of $N_1/N=N_2/N$ as a function of $\beta_1=\beta_2=\beta$. The latter is read off from the gravity solutions as 
		\begin{equation}
			\beta=\frac{1}{n!}\left.\partial^{2\,n}_{\xi}\lambda_1\right|_{\xi=0},
		\end{equation}
		whereas the former is given by \eqref{eq:betaZ2symHoloDic}. When extracting $\beta$, we make sure that the value we get is stable to $10^{-50}$ when varying the grid size. 
		
\begin{figure}[!ht]
			\centering
			\includegraphics[width=0.75\textwidth]{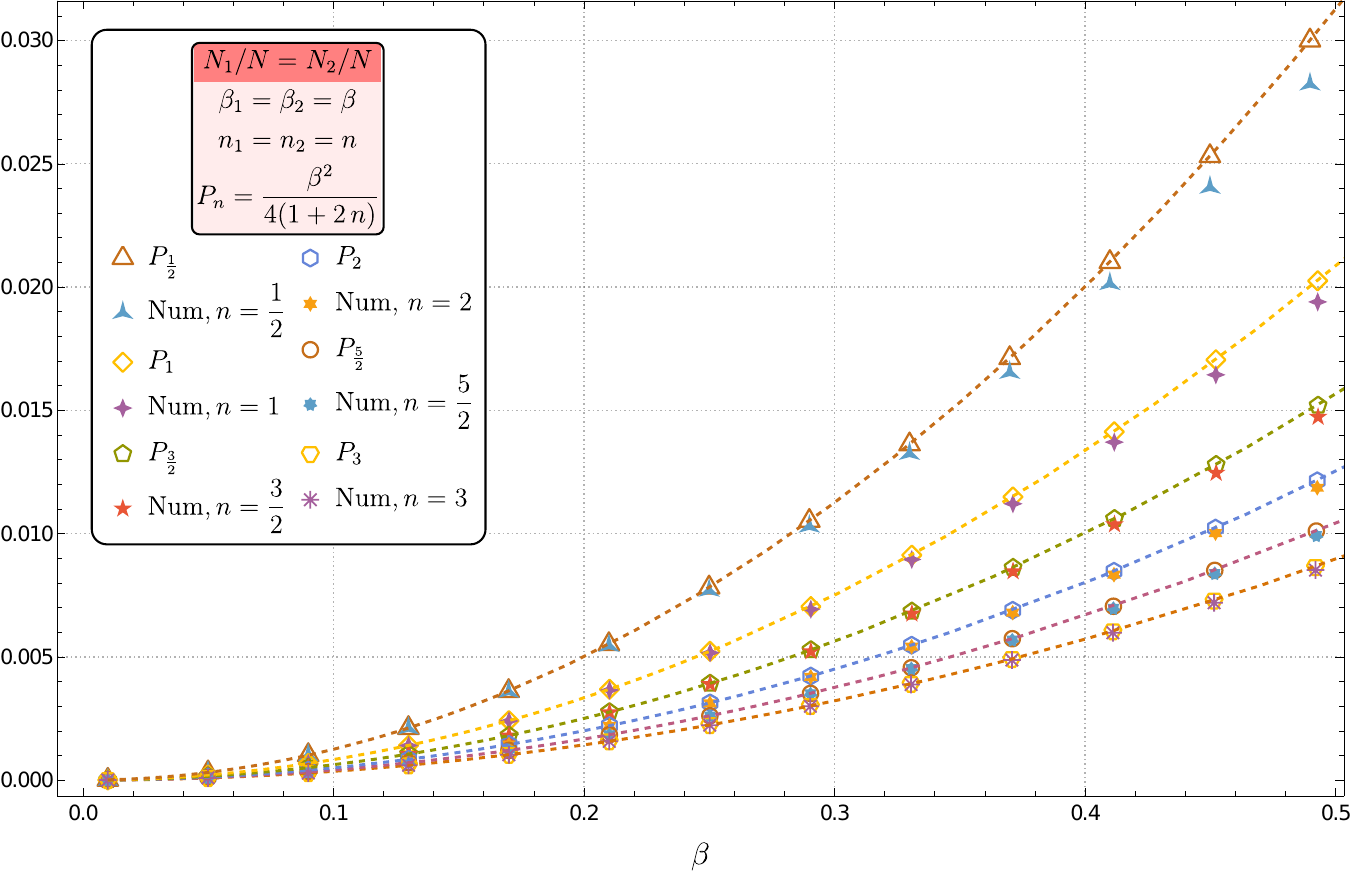}
			\caption{$N_1/N$ as a function of $\beta$ for six different values of $n$ and $\omega_1=\omega_2=1$. We focus on small $\beta$ to compare with the perturbative result, $P_n$, given by dashed curves and additional hollow markers to facilitate the comparison. Solid markers indicate the numerical results.}
			\label{fig:N1Nbeta}
		\end{figure}

		The results, for small $\beta$, are shown in Figure~\ref{fig:N1Nbeta}.  This shows the good agreement of our numerics (solid markers) with the perturbative expression (dashed curves and hollow markers, which are added to make the comparison easier), given as $P_n$ in the plot.  The perturbative results are   obtained from \eqref{eq:N1N2b1b2} by expanding for $\beta_1=\beta_2=\beta\ll1$ and setting $n_1=n_2=n$. As $\beta$ increases the exact and perturbative results deviate significantly.

		The theoretical limit to decreasing $\tilde{\phi_1}$ is the point at which CTCs become present in the spacetime. In the single-sector, BPS solution, Section~\ref{ss:pureNS2}, one can obtain an analytic expression for the bound, \eqref{CTC2}. Here, we monitor the value of $\Omega_1$ at $\xi=1$, which approaches zero, where the bound is attained and changes sign once CTCs are present. That can be easily seen by expanding the left-hand side of \eqref{CTC2} at infinity, using the boundary conditions from Section~\ref{sub:gen-bcs}, resulting in $2\,\Omega_1(1)\,(1-\xi)+\mathcal{O}\big((1-\xi)^2\big)$.  We have pushed the numerics the most for the $n=1$ solutions, reaching $\Omega_1(1)\sim9\times10^{-6}$. 
		
\begin{figure}[!ht]
			\centering
			\includegraphics[width=0.75\textwidth]{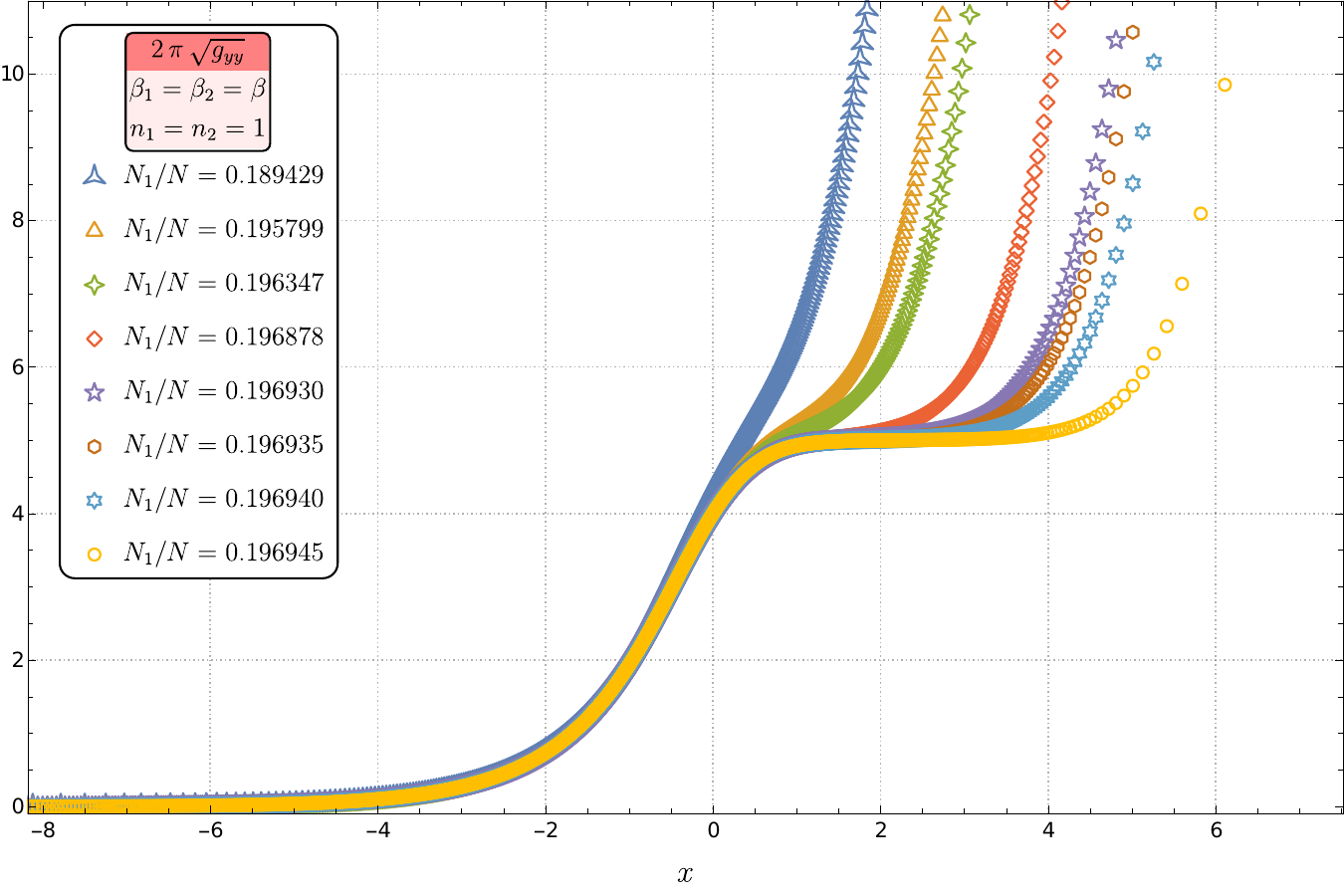}
			\caption{The radius of the $y-$circle $2\,\pi\,\sqrt{g_{yy}}$ for $n=1$ and different values of $N_1/N$, approaching the CTC bound. The x-axis uses a rescaled radial coordinate, $\xi=e^x/\,\sqrt{e^{2\,x}+1}$. $\omega_1=\omega_2=1$.}
			\label{fig:yCircleN1N}
		\end{figure}

Moreover, close to the CTC bound, one sees the development of a long, BTZ-like throat, as depicted in Figure~\ref{fig:yCircleN1N}. As $n$ increases, larger values of $\beta$, or $N_1/N$, become attainable before hitting the bound. Our observations suggest that  one reaches the CTC bound at lower value of $\beta$, or $N_1/N$, in the non-BPS solutions compared to  the BPS solutions.

\begin{figure}[!ht]
			\centering
			\includegraphics[width=0.48\textwidth]{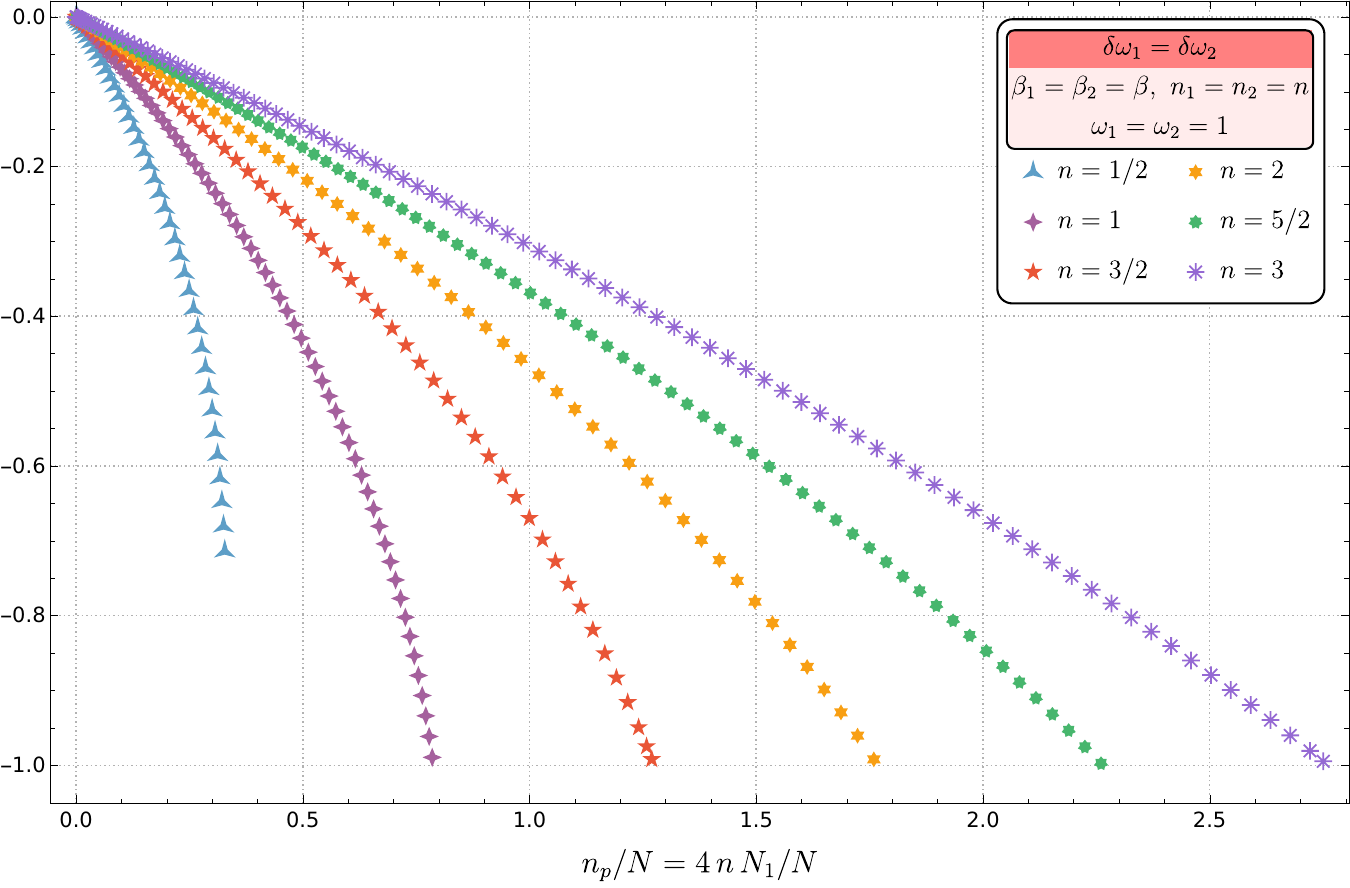}\hspace*{1mm}
			\includegraphics[width=0.48\textwidth]{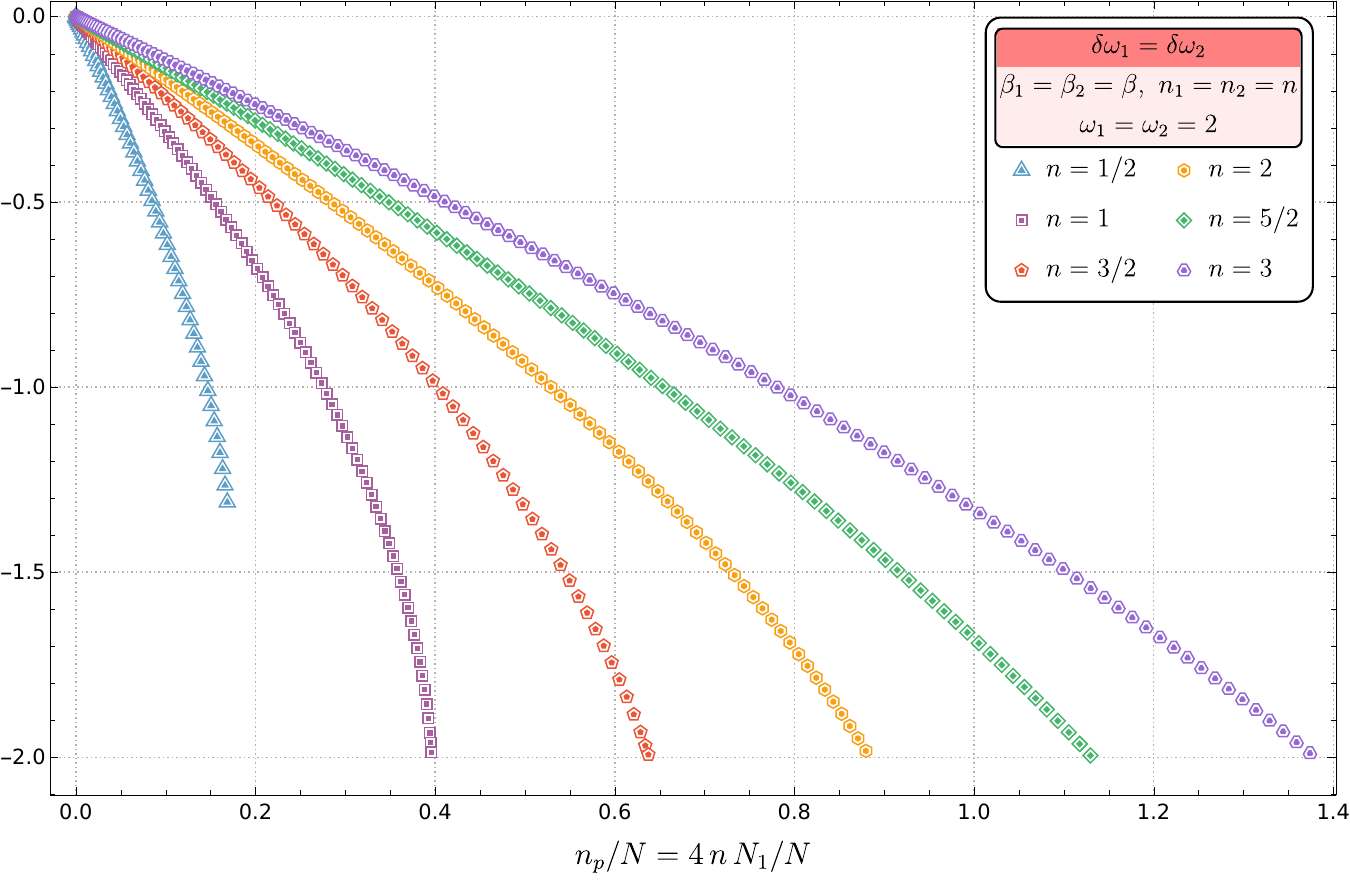}
			\caption{The energy shift $\delta\omega_1=\delta\omega_2$ as a function of the momentum charge $n_p$ for six different values of $n$ and, on the left, $\omega_1=\omega_2=1$ while, on the right, $\omega_1=\omega_2=2$. For $n\geq1$ the curves approach the limiting value $-\omega_1 = -\omega_2$, as we get close to the CTC bound. The first curve, with $n=1/2$, is an exception and our numerics suggest that the limit, $-\omega_i$, will not be reached.}
			\label{fig:omShifts}
		\end{figure}

		The energy shift of our solutions is also  very interesting. A non-zero value indicates that the state is indeed non-BPS. Figure~\ref{fig:omShifts} demonstrates that all our solutions break supersymmetry. One only has $\delta\omega_1=\delta\omega_2=0$ for $\beta=0$ or $n=0$.  The former  is just the AdS$_3$ vacuum, and the latter is a non-trivial supersymmetric deformation of the round supertube carrying no momentum.  In Figure~\ref{fig:omShifts} we plot the energy shifts, or anomalous dimensions, as a function of the momentum charge.  For $n\geq1$ and $\omega_1=\omega_2=1$,  the curves approach $\delta\omega_1=\delta\omega_2=-1$ as $n_p$ grows and our numerics approach the CTC bound.  To test this phenomenon further, we also examined $\omega_1=\omega_2=2$ and found that for  $n\geq1$  the curves approach $\delta\omega_1=\delta\omega_2=-2$.  
		
The fact that $\delta\omega_1=\delta\omega_2$ can limit to $-\omega_1=-\omega_2$  is intriguing because it means that the physical frequencies, $\hat{\omega}_i$ defined in (\ref{hatomdefn}), are going to zero.  In terms of the dual CFT, one should recall that the momentum charge, $n_p = h - \bar h$, is quantized and does not undergo a shift, whereas the energy, $h + \bar h$, is not protected and the ``twist'', $\bar h$ for $h > \bar h$, is given by $N_1 (2\omega+ \delta\omega)$ for $N_1=N_2$.   Thus, while the physical frequency is vanishing, the twist, $\bar h$, still limits to a positive integer value.  

In microstrata, the CTC bound represents the geometry with maximal red-shift to the cap, and for superstrata this is a black-hole limit in which the red-shift becomes infinite.  Thus the decrease of physical frequencies as the throat gets deeper,  and the vanishing of such frequencies in the maximal limit, makes intuitive sense. On the other hand our definition of physical frequencies, which is based on the holographic match of the angular momenta, is not directly related to the physical energy of the heavy state, which should be encoded in the 3D part of the geometry according to the recipe recalled in \eqref{eq:hhbar}. Thus this simple picture of red-shifts and vanishing physical frequencies is a little naive.
 
To elaborate on this point it is useful to look at the WKB results of Section \ref{sub:perturbation_pureNS}, where we considered a superstratum with a very large amplitude and a small non-BPS perturbation of it. The frequency shifts are given by  (\ref{eq:WKBshifts}) and shown in Figure~\ref{fig:resultsWKB}.  As one approaches the CTC bound one has   $\hat \omega_2 \to -1$. The frequency of the perturbation is thus limiting to negative integer value. However, this does not imply that the physical energy becomes negative as a small perturbation cannot wipe out the effects of a large background superstratum (whose physical frequency is  $\hat \omega_1 =  \omega_1 =1$) and so extracting the physics at the CTC bound is going to be more of a challenge. It is tempting to conjecture that in this limit the non-BPS perturbations should combine with some of the BPS sector to create a sector of the theory that is indeed red-shifted to become time independent.  However, such an analysis is well beyond the scope of our current investigation.

We also note that the energy shifts for  $n=1/2$ display a different behaviour in that they do not approach the limit of $-\omega_i$. The numerics for $n=1/2$ are the most challenging, as the gradients in the functions are the steepest, and we are currently not sure if it is possible to reach $-\omega_i$. The closest we have got to the CTC bound for $n=1/2$ is $\Omega_1(1)\sim10^{-4}$ with $\omega_i=1$, which, even if not as good as for $n=1$, has been much harder to achieve. Exploring the phase space close to the CTC bound requires very small steps in $\tilde{\phi_1}$, thus also in $\beta$ (and $N_1/N$), as the gradients in the functions near the asymptotic boundary become steeper and steeper. At the same time, it appears that for smaller $n$ the CTC limit is approached slower. Specifically, the appearance of what we believe is qualitative behaviour in its vicinity - namely the formation of a visibly long throat as in Fig.~\ref{fig:yCircleN1N}, or $\delta\omega_i\to-\omega_i$ might require attaining values of $\kappa(\xi\to1)$ significantly larger than for higher values of $n$. Hence, in combination with the steeper gradients for $n=1/2$ and our current method for generating numerical solutions, obtaining geometries that probe the CTC bound closely enough will take time and resources beyond our current capabilities. Therefore, we cannot determine why the energy shifts for $n=1/2$ should exhibit this different kind of behaviour. Note, however, that we do not have any physical reason to expect such behaviour.
		
\begin{figure}[!ht]
			\centering
			\includegraphics[width=0.75\textwidth]{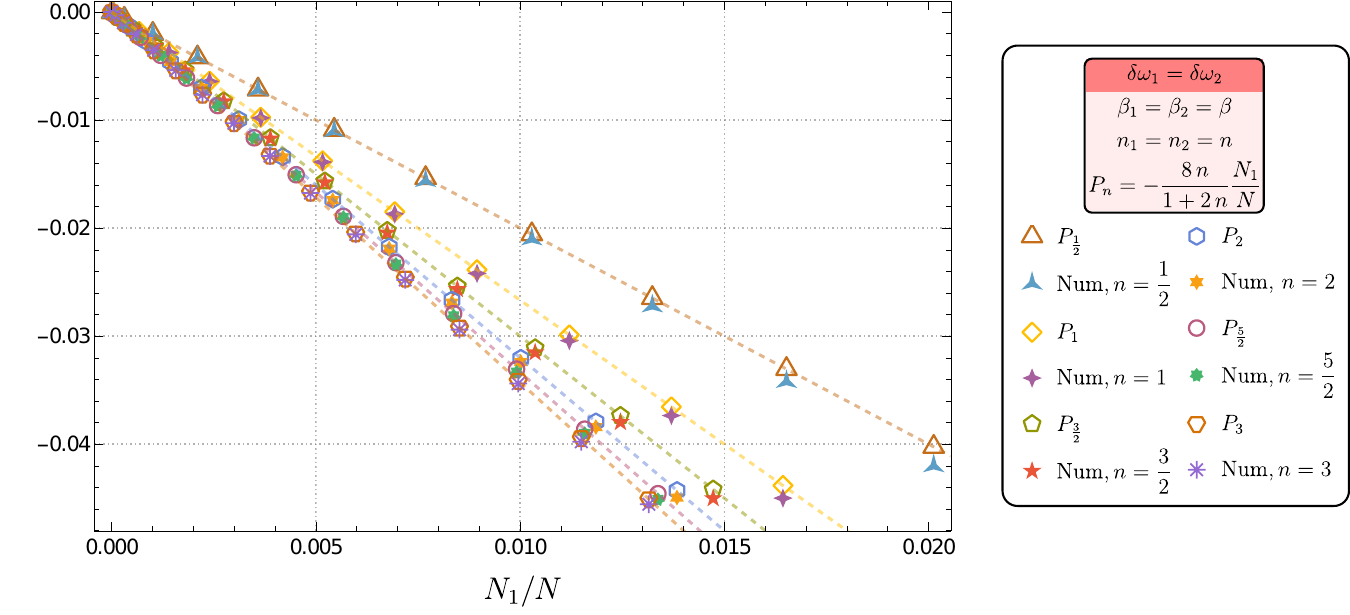}
			\caption{The energy shift $\delta\omega_1=\delta\omega_2$ in comparison with our perturbative results, $P_n$,  given by dashed curves and additional hollow markers to facilitate the comparison, for six different values of $n$ and $\omega_1=\omega_2=1$.  Both graphs use solid markers for the numerical results. $P_n$ is given in \eqref{eq:deltaomega2betafirst} with $n_1=n_2=n$. }
			\label{fig:EnergyShiftPert}
		\end{figure}

Before moving to other observables, the agreement of the numerical data with the predictions from perturbation theory for the energy shift, \eqref{eq:deltaomega2betafirst} with $n_1=n_2=n$, for small values of $N_1/N$ (small $\beta$) is shown in Figure~\ref{fig:EnergyShiftPert}. We have omitted the perturbative results from the main plot, Figure~\ref{fig:omShifts}, for clarity.
		

\begin{figure}[!ht]
			\centering
			\includegraphics[width=0.39\textwidth]{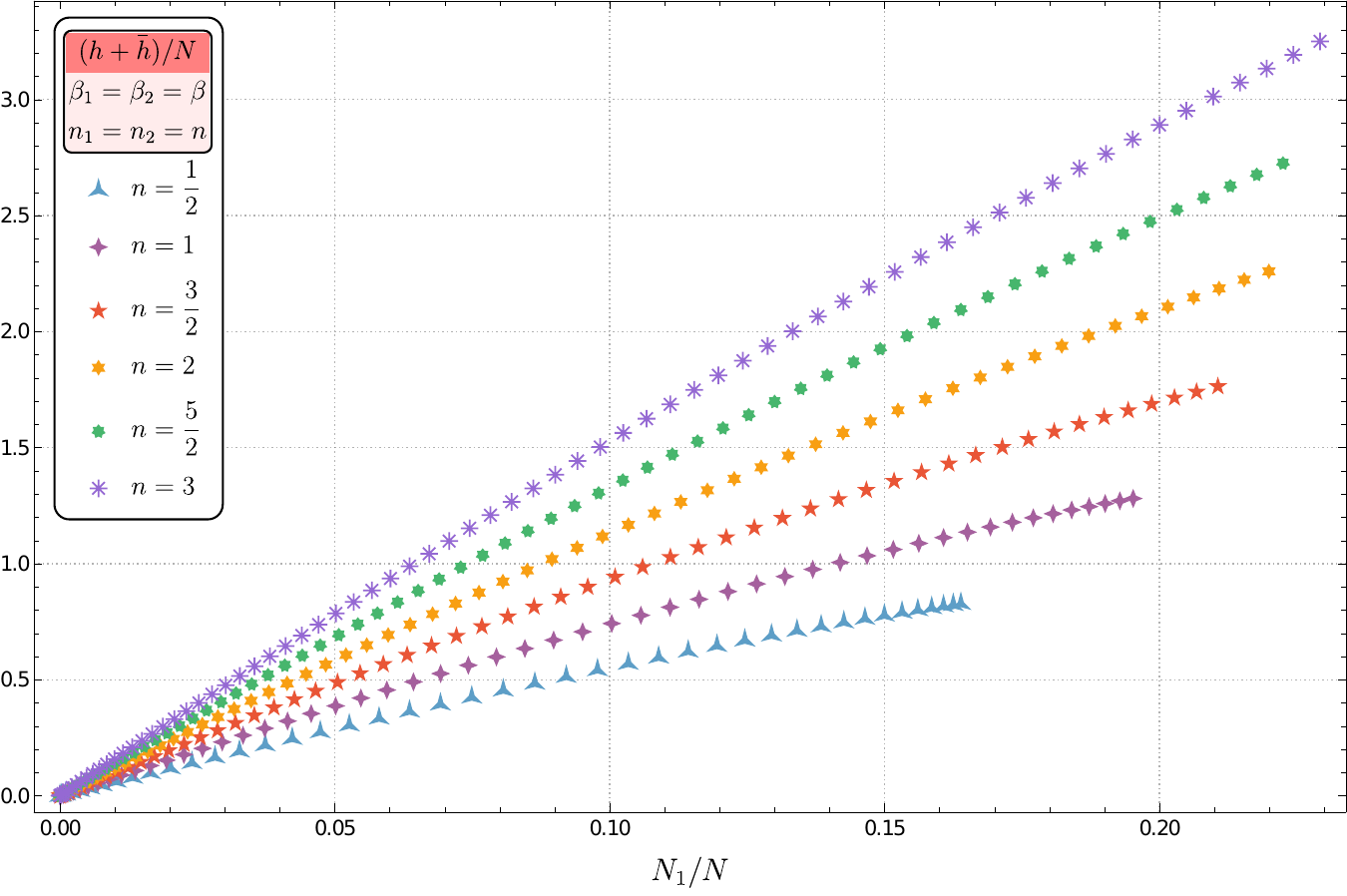}\hspace*{1mm}
			\includegraphics[width=0.57\textwidth]{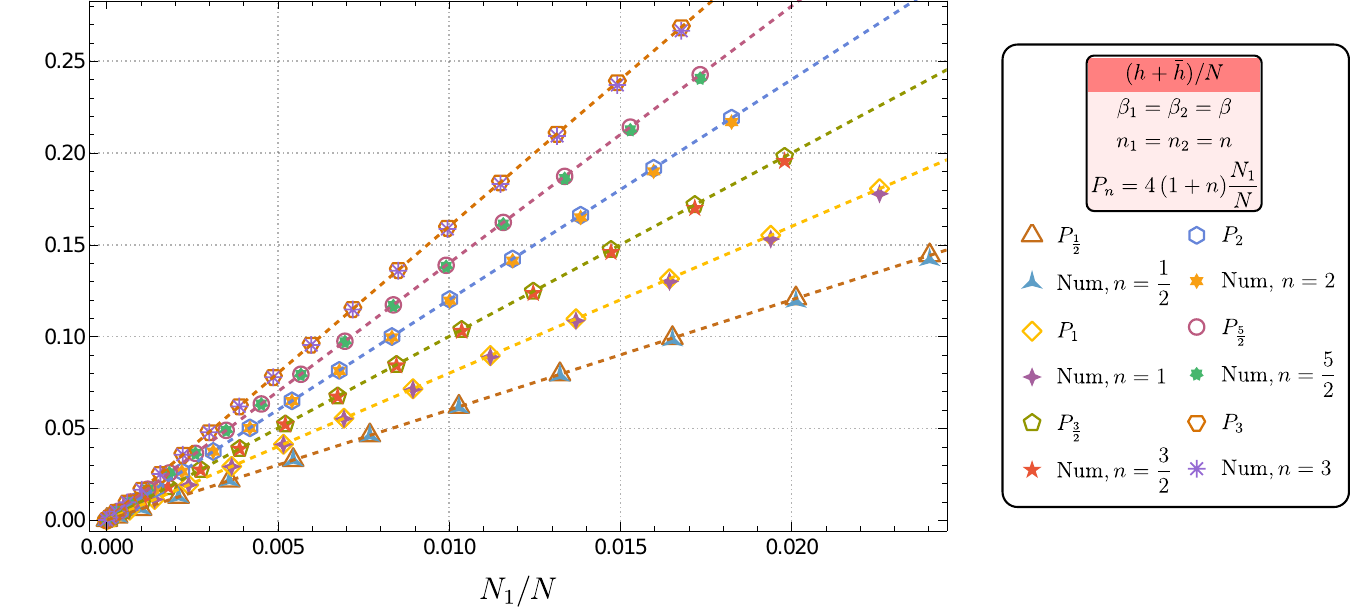}
			\caption{The total energy $(h+\bar{h})/N$ as a function of $N_1/N$ for six different values of $n$ and $\omega_1=\omega_2=1$. On the right we zoom in for small $N_1/N$ and show a comparison with the perturbative result, $P_n$, given by dashed curves and additional hollow markers to facilitate the comparison. Both graphs use solid markers for the numerical results.}
			\label{fig:hphbarN1N}
		\end{figure}

		Finally, we also show a plot of the total energy, \eqref{eq:z2Symhphbar}, of our states, Figure~\ref{fig:hphbarN1N}. The right plot again depicts the excellent agreement with our perturbative results for small values of $N_1/N$, whereas on the left we have all the numerical data.

		\section{Final comments}
		\label{sec:Conclusions}

		We have used both perturbative and numerical methods to construct and analyze several families of microstrata.  The most general such families are characterized by six parameters $(n_1,n_2, \hat \omega_1, \hat \omega_2,\contparama,\contparamb)$, the first four of which are discrete and the last two are continuous ``amplitudes.''  We have matched the CFT data to the supergravity data and extracted the physical frequencies,  $\hat \omega_i$, from the supergravity solutions. The difference between the classical frequencies, $\omega_i$, and the  physical frequencies, $\hat \omega_i$ gives the anomalous dimensions of the states in the CFT.   We have chosen to focus primarily  on solutions that start from $\omega_i=1$, but we have found that it is straightforward to generalize our numerical analysis to other values of $\omega_i$.
		
Apart from showing that microstrata exist as well-behaved, smooth solutions to supergravity, even at large amplitudes at which they approach the CTC bound, there are several other important conclusions coming out of our work.

		\begin{figure}[!ht]
			\centering
			\includegraphics[width=0.9\textwidth]{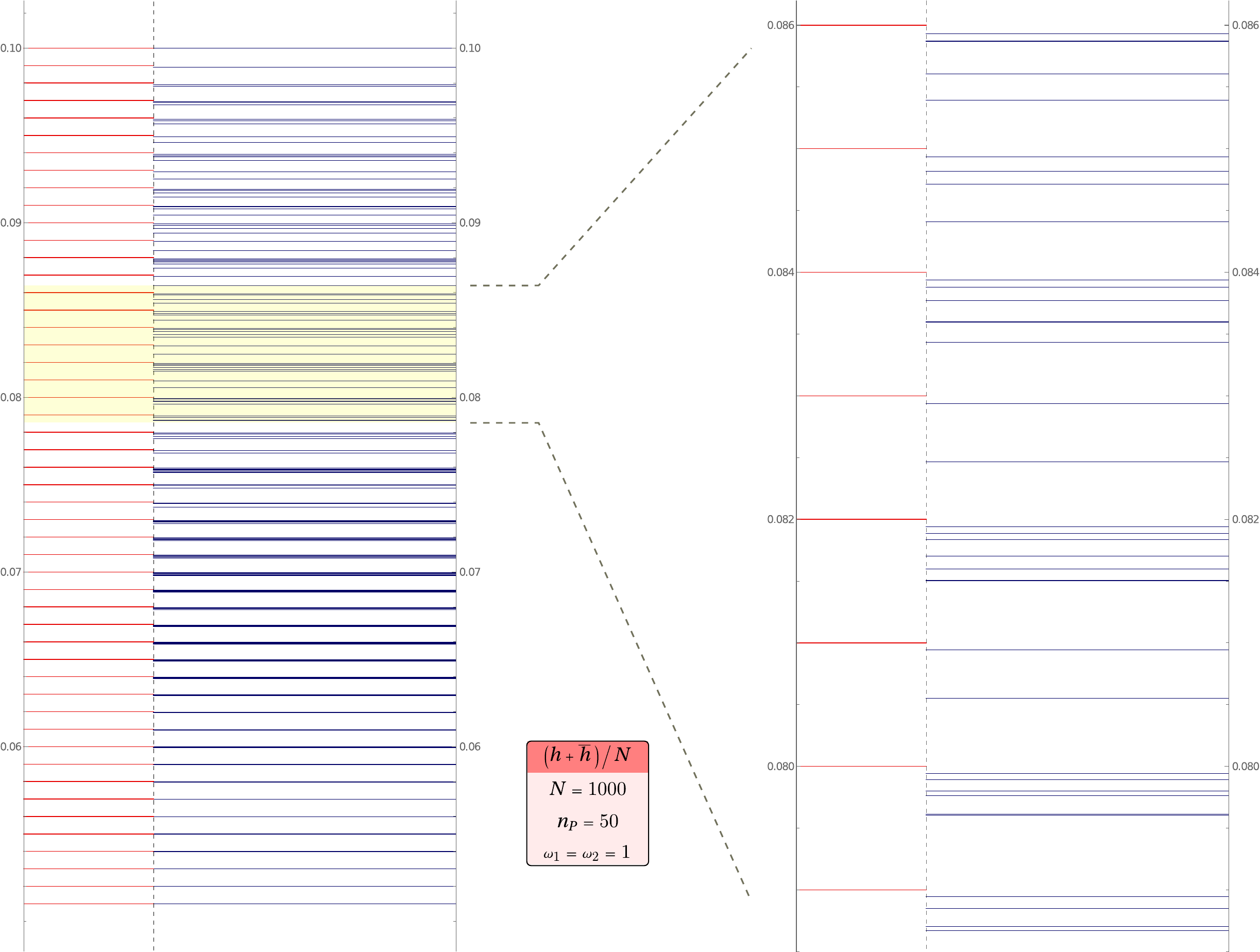}
			\caption{Splitting of the rational CFT degeneracies for the $\alpha$-class solutions. This plot shows the total energies, (\ref{eq:totalenergy}),  of CFT states for $\omega_1=\omega_2=1$, and various values of $N_1$, $N_2$, $n_1$ and $n_2$ but with fixed central charge ($N=1000$) and fixed momentum charge ($n_P = 50$).   The left side of each plot shows (in red) the rational, classical  energy levels, given by the first term in  (\ref{eq:totalenergy}), and the right side of the plot shows the perturbative splitting given by the complete expression in  (\ref{eq:totalenergy}).   The second, right-hand plot shows detail from a part of the spectrum. Note that the perturbative energies are universally shifted below their classical values. }
			\label{fig:spectrumsplitting}
		\end{figure}

We have shown that the normal mode frequencies depend non-linearly on the amplitudes of the excitations and that the anomalous dimensions are generically negative.  Such a lowering of the energy below its classical value reflects the fact that one expects a non-BPS state to become more strongly bound once back-reaction is incorporated.  We have illustrated this in Figure \ref{fig:spectrumsplitting} using our perturbative result   (\ref{eq:totalenergy}) for $\alpha$-class microstrata.  To arrive at this figure we  vary the individual quantum numbers $N_1$, $N_2$, $n_1$ and $n_2$ while keeping the  momentum charge, $n_P$, fixed.  We also fix the the central charge,  $6N$,   and require $n_P \ll N$  so that perturbation theory is valid.   One sees very explicitly how the classical degeneracies of states are then split\footnote{The non-BPS levels are at least doubly degenerate because of the $\ZZ_2$ symmetry under $ 1 \leftrightarrow 2$ sector interchange. There could also be further degeneracy coming from CFT states that lie outside our supergravity analysis.}  and moved universally downwards, revealing how supersymmetry breaking causes a transition away from an evenly spaced rational spectrum to a broader, generic spread of frequencies.  

A more canonical treatment would require one to fix the other thermodynamic variables, like the total angular momentum of the state, however in our very simple family of microstrata this would completely fix the classical energy to a single level.   Were we able to construct richer classes of microstrata, we could  fix all the thermodynamic variables and we expect that the energy level plots would have a structure much like that shown in  Figure \ref{fig:spectrumsplitting}.

The non-linear dependence of the energy shifts on the amplitudes of the fields  suggests, but does not  prove, that there is a transition to chaos.   To see  more evidence of a chaotic spectrum would require a solution with  multiple independent modes.  This would go well beyond our simple Ansatz because turning on  more than one independent frequency, and allowing for independent frequency shifts, when combined with the non-linearities of the equations means that one should expect an infinite range of frequencies to be  activated.  One must therefore find solutions that are fully functions of at least two variables.

There are two natural approaches to this problem. First one can simply embrace the need to solve the system as a function of two variables.  Given the success of the numerical analysis presented here, we expect that this should be quite feasible.    An intriguing alternative is to use perturbation theory about one of our microstrata, and controlling the introduction of new frequencies order by order in perturbations. Indeed, there is one example that could potentially produce some very interesting  physics: perturbing away from the special locus.

Our $\alpha$-class ``Q-ball'' solutions make critical use of the 2:1 phase locking between two of the scalar fields in the Ansatz.  We then find that regularity requires that the amplitudes of these two scalars must be locked together on a {\it special locus}.   In supergravity, we believe (but have not proven) that moving away from the special locus will result in a ``de-tuning'' of the frequencies of these two different fields, with the emergence, through non-linear interactions, of all the non-trivial combinations of frequencies of these underlying modes.  The  expression (\ref{eq:specialLoc})  for  the special locus  suggests  very natural perturbative parameters: $\epsilon_i = \beta_i+ \frac{\alpha_i^2}{4}$.  New frequencies would then be introduced order by order in $\epsilon_i$.  

 The CFT characterization of the special locus in terms of multi-particle states defined through the standard OPE is one of the primary results of this paper. As discussed in Section \ref{sec:heavystates}, this is a novelty with respect to the CFT description of the superstrata of Section~\ref{sec:doublestratum} (see \cite{Shigemori:2020yuo} for a recent review of such solutions): in these superstrata multi-particle states are defined in terms of the orbifold locus by making symmetrized products of constituents that act on different copies of the CFT.  To move off the this class of ``standard'' superstrata one must take into account situations in which  the elementary constituents act more than once {\it on the same CFT factor}. When exactly the combination corresponding to the OPE of single-particle operators is used, then one reaches the special locus. In general, different definitions of the multi-particle states involves operators with a distinctly different set of traces over the Hilbert space, and it is likely that such CFT operators with  different trace structures will develop different anomalous dimensions.  If so, then one should expect the non-BPS supergravity description of states that are not on the special locus to develop a range of frequencies corresponding to all the  different trace structures of the operators that make up a multi-particle state that only involves symmetrized products of single constituents. This is why we expect that deviating from the special locus in supergravity must, of necessity, result in a spread of frequencies that go well outside the Q-ball Ansatz.  It would be extremely interesting to try to track this process, using perturbation theory in $\epsilon_i$,   in both the CFT and in supergravity, 

In this paper we have also chosen to focus on solutions with $\omega_1 = \omega_2 =1$. We made this choice because we expected the solutions to be relatively simple because they only involve left-moving momentum excitations.   We also suspect that some of these  solutions may be  accessible to using almost-BPS methods in supergravity.  Based on experience  \cite{Ganchev:2021pgs,Ganchev:2021ewa} (and some limited calculations done while developing the work presented here), the perturbative and numerical construction of solutions for other values of $\omega_1, \omega_2$ does not pose a significantly higher level of difficulty.  It would, of course, be extremely interesting to construct a much more general time-dependent solution, and see how it relaxes towards a ground state.
 
Our numerical work and WKB analysis has enabled us to construct solutions very close to the CTC bound.  The expectation is that such solutions should closely approximate a black hole and thus have a large red-shift along the throat. In our numerical  examples  there were an equal number of BPS and non-BPS elements ($N_1 = N_2$), and we found that $\hat \omega_i  =  \omega_i+\delta\omega_i \to 0$.  However, in our WKB analysis of non-BPS excitations around a superstratum, we found $\hat \omega_2  \to -1$.  It is intriguing that these limiting values are integers, however, one would like a deeper interpretation  in terms of black-hole-like behavior.   With our limited families of examples, it is difficult to come up with a compelling conjecture and so it would be very interesting to gather more data   from a broad range of solution that lie near the CTC bound. One would also like to understand the CTC bound from   the CFT perspective.  While this is probably very challenging for generic non-BPS states, one might be able to gain insight from a systematic study of non-BPS perturbations around superstrata that near the CTC bound.

 There are also interesting questions about how the results we present here  connect to other areas of research.  
 
 Within the microstate geometry programme, one of the imperatives is to uplift these results to six-dimensional (and thence to IIB) supergravity.  There are several motivations for this.   First, the consistent truncation to three dimensions rests on restricting to some of the simplest  modes on the $S^3$.  It would be very interesting to see if the methodology, like the Q-ball Ansatz, can be extended to other modes on the $S^3$, and then look at the holography and perhaps find new special loci.  Of even more interest is that it is only after performing such an uplift that  one can look for asymptotically-flat microstrata.  That is, the consistent truncation is locked  to AdS$_3$ asymptotics, but some really interesting physics should emerge from coupling  microstrata to flat space-times.  We expect such geometries to be  time dependent because they should decay by some analogue of  ``Hawking Radiation.''  This is especially true of states with $\omega_i >1$ because such CFT states  contain both left-moving and right-moving quanta, whose collisions will create states in the gravity multiplet that are no longer trapped in an AdS ``box.'' As a first step one might use perturbation theory to perform some tunneling calculations like those in \cite{Chowdhury:2007jx,Chakrabarty:2015foa,Eperon:2016cdd,Chakrabarty:2019ujg,Bena:2020yii}.    It is also intriguing to conjecture that the states presented here, with   $\omega_i = 1$, and hence only left-moving momenta, will also be unstable when coupled to flat space, but the instability will be akin to that of the JMaRT solution \cite{Chowdhury:2007jx,Chowdhury:2008bd,Chowdhury:2008uj,Avery:2009tu,Chakrabarty:2015foa} and its ergo-region instability.
 
 The uplift and coupling to flat space will also facilitate   all manner of string probes, like those of  \cite{Martinec:2020cml, Ceplak:2021kgl, Martinec:2023plo}.  Moreover, the fact that our $\beta$-class solutions are S-dual to excitations that live purely in the NS sector of IIB supergravity means that one might also be able to use world-sheet methods  \cite{Martinec:2017ztd,Martinec:2019wzw,Brennan:2020bju,Martinec:2020gkv,Martinec:2022okx} to explore these solutions at the stringy level.
 
It is also important to note that a generic ``Q-ball'' Ansatz is the mechanism that underlies the construction of Bose stars.  The important distinction, however, is that the scalar is the only mechanism to support a Bose star, and the  mass of the star is entirely determined by the mass of the scalar.     While Bose stars are interesting in their own right, they have very few microstates, especially compared to a black hole of the same mass, and so cannot begin to address the black-hole information problem.   The whole point behind microstate geometries is that they are primarily supported by fluxes \cite{Gibbons:2013tqa}, and provide a framework that can be used to construct geometric duals of vast numbers of black-hole microstates. \cite{Bena:2013dka, Bena:2022ldq,Bena:2022rna}.  Moreover, in microstate geometries, the ``Q-ball'' Ansatz is simply a mathematical device to find a simple, solvable system of equations and it is not an essential part of the overall infrastructure. We cannot help but feel that the depth and range of Bose stars might be greatly extended by an infusion of some of the essential elements that underpin microstate geometries, such as higher-dimensions and magnetic fluxes.
 
 It is also evident from our work here that the supergravity solutions are  outstripping our ability to perform the dual CFT calculations.  The problem is simply that once supersymmetry is broken, correlators are no longer protected by supersymmetry and computations using the orbifold CFT are of limited value in the black-hole regime.  One really needs a new infusion of ideas and methods.  In this context, it is important to remember that the supergravity theory requires large $N$, and perhaps there are some as yet undiscovered large-$N$ methods that might be found.  Perhaps there could be some conformal bootstrap techniques that could be used to compute some of the anomalous dimensions we computed here?  It is also possible that one might make progress with CFT correlators at large frequencies, $\omega$, where WKB works well in supergravity.

What is evident from the work presented here,    is that our microstrata represent an important starting point for a more systematic exploration of non-BPS black-hole microstructure within the framework of microstate  geometries, fuzzballs and holography.
 
		\section*{Acknowledgments}
		\vspace{-2mm}
		The work of NPW is supported in part by the DOE grant DE-SC0011687. The work of BG, AH and NPW is supported in part by the ERC Grant 787320 - QBH Structure. RR is partially supported by the UK EPSRC grant ``CFT and Gravity: Heavy States and Black Holes'' EP/W019663/1 and the STFC Consolidated Grants ST/T000686/1.


		\appendix

		\section{Reduced equations on special locus}
		\label{app:spLocusEqs}
		
		The special locus in supergravity is introduced in Section \ref{ss:speciallocus}. For completeness, we will restate the relations characterising it, namely:
		\begin{align} \label{eq:splocus_algebraic_def}
			\mu_{1,2}&=\lambda_{1,2}=\frac{1}{2}\log\Big(1-\frac{1-\xi^2}{2}\nu_{1,2}^2\Big),\notag\\
			m_5&=-\frac{1}{2}(1-\xi^2)\,\nu_1\,\nu_2,\quad m_6=0,
		\end{align}
		which reduce the matrix $m$ in equation \eqref{mmatrix-axgauge} to
		\begin{equation}
			m ~=~\begin{pmatrix}
				1-\frac{1}{2}(1-\xi^2)\,\nu_1^2  &0  & -\frac{1}{2}(1-\xi^2)\,\nu_1\,\nu_2 & 0\\
				0 & 1 & 0 & 0\\
				-\frac{1}{2}(1-\xi^2)\,\nu_1\,\nu_2 & 0 & 1-\frac{1}{2}(1-\xi^2)\,\nu_2^2 & 0\\
				0 & 0 & 0 & 1 
			\end{pmatrix} \,,
			\label{mmatrix-spLoc}
		\end{equation}
		with only one non-trivial eigenvalue, given by
		\begin{equation}
			m_0=1-\frac{1}{2}\,\chi_I \chi_I=1-\frac{1}{2}(1-\xi^2)\big(\nu_1^2+\nu_2^2\big) \,.
		\end{equation}
		Moreover, $\chi$ is a null vector of $\widetilde{F}$
		\begin{equation}
			\widetilde F_{\mu \nu}{}^{IJ} \,  \chi_J ~=~ 0 \,,
		\end{equation}
		which results in the following algebraic relation between the gauge fields
		\begin{equation}
			\Phi_4\,\Psi_2+\Phi_3\,\Psi_1-\Phi_1\,\Psi_3-\Phi_2\,\Psi_4=0.
		\end{equation}
		
		Substituting the relations above in the equations of motion leads to a reduced set of ``almost BPS'' equations. The procedure is tedious, so we will only present the final results. It will prove convenient to define the following complex vectors:
		\begin{equation}
			\mathcal{V}^{F} ~=~ \mqty(F_1 + i F_4 \\ 0 \\ -F_3 + i F_2 \\ 0)\,, \qquad 
			\mathcal{V}^{\Phi} ~=~ \mqty(\Phi_1 + i \Phi_4 \\ 0 \\ -\Phi_3 + i \Phi_2 \\ 0)\,.
		\end{equation}
		where the functions $F_j$ have been defined in \eqref{FHdefns}. We will also define
		\begin{equation}
			\Lambda=2\,m_0=2-(1-\xi^2)\big(\nu_1^2+\nu_2^2\big), \qquad H_0 = \frac{\xi^2 \, \Omega_0^2}{(1-\xi^2)^2 \Omega_1^2} \,.
		\end{equation}
		
		The gauge fields all satisfy first-order linear ODEs:
		\begin{gather}
			\xi \partial_\xi (\Phi_1 + i \Phi_4) =\frac{\sqrt{1-\xi^2} \,\nu_2}{\Lambda}\,\Omega_1 (i \chi^{\,}_I) \mathcal{V}^{F}_I,
			\qquad
			\xi \partial_\xi (\Phi_3 - i \Phi_2) =\frac{\sqrt{1-\xi^2}\,\nu_1}{\Lambda}\,\Omega_1 (i \chi^{\,}_I) \mathcal{V}^{F}_I,\notag
			\\[1.5ex]
			\xi \partial_\xi (\Psi_1 + i \Psi_4) =\frac{\sqrt{1-\xi^2}\,\nu_2}{\Lambda}\,\Omega_1 (i \chi^{\,}_I ) \qty(\frac{k}{1-\xi^2} \mathcal{V}^{F}_I + H_0\mathcal{V}^{\Phi}_I) ,\notag
			\\[1.2ex]
			\xi \partial_\xi (\Psi_3 - i \Psi_2) =\frac{\sqrt{1-\xi^2}\,\nu_1}{\Lambda}\,\Omega_1 (i \chi^{\,}_I ) \qty(\frac{k}{1-\xi^2} \mathcal{V}^{F}_I + H_0\mathcal{V}^{\Phi}_I) .\notag
		\end{gather}
		
		The above equations also imply:
		\begin{align}
			\nu_1\,\Phi_1'-\nu_2\,\Phi_3'&=0,\qquad \nu_2\,\Phi_2'+\nu_1\,\Phi_4'=0,\notag\\
			\nu_1\,\Psi_1'-\nu_2\,\Psi_3'&=0,\qquad \nu_2\,\Psi_2'+\nu_1\,\Psi_4'=0.
		\end{align}
		The equation for $k$ is also first-order and linear:
		\begin{equation}\label{eq:spLock}
			\Omega_1 \, \xi \partial_\xi\bigg(\frac{k}{1-\xi^2}\bigg)+8\big(\Phi_1\,\Phi_2+\Phi_3\,\Phi_4-\kappa\big) H_0 =0,
		\end{equation}
		where $\kappa$ is a constant of integration, that can be determined order by order in the perturbative expansion using the regularity constraints. One can also derive a first-order equation for $H_0$, and through it the metric fields $\Omega_0$ and $\Omega_1$:
		\begin{equation}\label{eq:spLocw0w1}
			\Omega_1 \, \xi \partial_\xi \log (H_0)
			- 8 \big(\Phi_1\,F_2+\Phi_2\,F_1+\Phi_3\,F_4+\Phi_4\,F_3\big) - 16 \kappa \frac{k}{1-\xi^2} + 
			\varpi=0,
		\end{equation}
		where $\varpi$ is another constant of integration.  Notice the similarity of these two equations \eqref{eq:spLock} and \eqref{eq:spLocw0w1} with \eqref{BPS3}, where $\varpi$ plays the role of $\eta_\psi$.  The metric sector of the equations does not fully reduce to first-order, hence one has to complement \eqref{eq:spLocw0w1} with  the second-order equation of motion for $\Omega_1$:
		\begin{align}
			2 \, &\frac{\xi \partial_\xi\qty(\xi \partial_\xi \Omega_1)}{\Omega_1 H_0} \,-\, 2\, \frac{\Lambda + 2}{\Lambda}\, \Omega_1^2 \,+\, H_0^{-2} \qty(\xi\partial_\xi \frac{k}{1-\xi^2})^2 -\frac{8}{\Lambda}\abs{\chi_I \mathcal{V}^{\Phi}_I}^2 = 0.
		\end{align}
		
		Finally, we need equations for $\nu_{1,2}$. One can derive a Wronskian type of constraint for them:
		\begin{equation}
			\frac{\xi(1-\xi^2)\Omega_1}{\Lambda}\big(\nu_2\,\nu_1'-\nu_1\,\nu_2'\big)=4\,\big(\Phi_4\,\Psi_1-\Phi_1\,\Psi_4+\Phi_3\,\Psi_2-\Phi_2\,\Psi_3\big),
		\end{equation}
		which is to be supplemented by one second-order differential equation from the equations of motion of $\nu_1$ and $\nu_2$, that we can write in terms of the eigenvalue $\Lambda$:
		\begin{align}
			\xi \partial_\xi (\xi \partial_\xi \Lambda) &
			+ \frac{3\Lambda - 4}{2 \Lambda (2-\Lambda)} \qty(\xi \partial_\xi \Lambda)^2
			+ \qty(\xi \partial_\xi \log\Omega_1) \qty(\xi \partial_\xi \Lambda)
			- 8 \,\abs{\chi^{\,}_I \mathcal{V}^{F}_I}^2 + 8 H_0 \,\abs{\chi^{\,}_I \mathcal{V}^{\Phi}_I}^2
			\\
			& + 32 \frac{\Lambda^2}{(2-\Lambda) \Omega_1^2} \qty(\Phi_1 \Psi_4 + \Phi_2 \Psi_3 - \Phi_3 \Psi_2 - \Phi_4 \Psi_1)^2
			+ 2 (2-\Lambda) \Omega_1^2 H_0 ~=~ 0\,,
		\end{align}

		\subsection{Linear equations}
		\label{app:linear_special_locus}
		
		We can simplify these equations by looking at a particular linearisation. We start with the special locus \eqref{eq:singleSectorBPSspec} in the $(1,0,n)$ sector parametrized by $\alpha_1$, that we treat as a background, and linearise the equations at first order in $\alpha_2$. The fields that receive corrections are: $\nu_2, m_5, m_6, \Phi_3, \Phi_4, \Psi_3, \Psi_4$. The fields $m_5$ and $m_6$ are fixed by the algebraic equations \eqref{eq:splocus_algebraic_def}, and for the rest we find a system of coupled equations:
		\begin{subequations}
			\begin{equation}
				\frac{\xi(1-\xi^2)\,\Omega_1}{4(2-(1-\xi^2)\nu_1^2)}\,(\nu_2 \nu'_1-\nu_1 \nu'_2)=\Phi_4\Psi_1+\Phi_3\Psi_2-\Psi_3\Phi_2-\Psi_4\Phi_1\,,
			\end{equation}
			\begin{equation}
				\Phi'_3=-\frac{(1-\xi^2)\nu_1 \,\Omega_1}{\xi(2-(1-\xi^2)\nu_1^2)}\,\left(\nu_2F_2+\nu_1 F_4\right)\,,
			\end{equation}
			\begin{equation}
				\Phi'_4=\frac{(1-\xi^2)\,\nu_1\, \nu_2\,\Omega_1}{\xi(2-(1-\xi^2)\nu_1^2)}\,F_1=-\frac{\nu_2}{\nu_1} \,\Phi_2'\,,
			\end{equation}
			\begin{equation}
				\Psi'_3=-\frac{(1-\xi^2)\nu_1 \,\Omega_1}{\xi(2-(1-\xi^2)\nu_1^2)}\,\left[\frac{k}{1-\xi^2} \qty(\nu_2 F_2 + \nu_1 F_4) + H_0\qty(\nu_2 \Phi_2 + \nu_1 \Phi_4)\right]\,,
			\end{equation}
			\begin{equation}
				\Psi'_4=\frac{(1-\xi^2)\nu_1\, \nu_2\,\Omega_1}{\xi(2-(1-\xi^2)\nu_1^2)}\,
				\qty(\frac{k}{1-\xi^2} F_1 + H_0\Phi_1)
				=-\frac{\nu_2}{\nu_1}\,\Psi'_2\,.
			\end{equation}
		\end{subequations}

		\section{Uplift to six dimensions}
		\label{app:uplift}
		
		The solutions of the three-dimensional gauged supergravity can be uplifted to type IIB supergravity compactified on a four-dimensional compact space $\mathcal{M}$, which can be either $\mathbb{T}^4$ or $K3$. The latter is the appropriate framework to interpret the solutions as bound states of D1 and D5 branes and to perform the holographic analysis of sections \ref{sec:CFT} and \ref{sec:Holography}. The compact space $\mathcal{M}$ plays a trivial role in our solutions and one can conveniently focus on the remaining six-dimensional directions, which we describe, following \cite{Romans:1986er,Kanitscheider:2007wq,Rawash:2021pik}, by the six-dimensional Einstein metric $ds^2_6$, the three-forms $G^{(I)}$ and the scalars $\phi^{(mr)}$. In our solutions the only non-trivial fields sit in the gravity multiplet, which contains the three-form $G^{(5)}$, and two tensor multiplets, which add two more three-forms, $G^{(6)}$, $G^{(7)}$, and two scalars, $\phi^{(56)}$, $\phi^{(57)}$. From the ten-dimensional perspective the non-trivial fields are: the RR forms, $C_0$, $C_2$, $C_4$, the NSNS two-form, $B_2$, the dilaton, $\Phi$, and the volume of $\mathcal{M}$, $V_4$; since the RR four-form has only a component along $\mathcal{M}$, plus the six-dimensional component needed for self-duality, we will simply denote by $C_4$ the scalar giving its $\mathcal{M}$-component; we will also denote by $C_6$ the two-form in six-dimensions obtained by reducing on $\mathcal{M}$ the six-form dual to $C_2$. The four scalars, $\phi$, $V_4$, $C_0$, $C_4$, are related by
		\begin{equation}\label{eq:c4v4}
			C_4 = \frac{e^{2\Phi} C_0}{e^{2\Phi} C_0^2+1}\,,\quad V_4^2 =  \frac{e^{2\Phi}}{e^{2\Phi} C_0^2+1}\,.
		\end{equation}
		The relation between the ten-dimensional and six-dimensional quantities is
		\begin{equation}\label{eq:PhQ1}
			\phi^{(56)}=\frac{1}{2}\left[ \sqrt{\frac{Q_5}{Q_1}\frac{C_4}{C_0}}- \sqrt{\frac{Q_1}{Q_5}\frac{C_0}{C_4}}\,\right]\,,\quad \phi^{(57)} = \sqrt{C_0 C_4}\,,
		\end{equation}
		\begin{equation}\label{eq:PhQ2}
			G^{(5)}=\frac{Q_1 G^{(1)}+Q_5 G^{(2)}}{2 Q_1 Q_5}\,,\quad G^{(6)}=\frac{Q_1 G^{(1)}-Q_5 G^{(2)}}{2 Q_1 Q_5}\,,\quad G^{(7)} = -\frac{G^{(4)}}{\sqrt{Q_1 Q_5}}\,,
		\end{equation}
		with 	
		\begin{equation}
			G^{(1)}=\frac{1}{2} dC_2\,,\quad G^{(2)}=\frac{1}{2} d C_6\,,\quad G^{(4)} = -\frac{1}{2}dB_2\,;
		\end{equation} 
		$Q_1$ and $Q_5$ are the D1 and D5 supergravity charges that determine the magnetic ($S^3$) components of the three-forms at infinity: $G^{(1)}|_{S^3} \to Q_5$, $G^{(2)}|_{S^3} \to Q_1$.
		
		The general uplift formulas were given in \cite{Mayerson:2020tcl}: we would like to make these formulas explicit for the generalised microstratum truncation introduced in Section \ref{sub:qball}; we will work in the axial gauge specified by the functions in \eqref{functionlist2}. It is convenient to define
		\begin{equation}
			\Gamma_i=e^{\lambda_i} \cos^2\varphi_i+e^{-\lambda_i} \sin^2\varphi_i \quad(i=1,2)\,,
		\end{equation}
		\begin{equation}
			\begin{aligned}
				\Delta =e^{\mu_1}\Gamma_1 \sin^2\theta+e^{\mu_2}\Gamma_2 \cos^2\theta +2 \left(m_5 \cos\varphi_1 \cos\varphi_2 + m_6 \sin\varphi_1\sin\varphi_2 \right) \sin\theta\cos\theta\,,
			\end{aligned}
		\end{equation}
		\begin{equation}
			\mathrm{det} (m) = (e^{\mu_1+\mu_2}-e^{-\lambda_1-\lambda_2}m_5^2)(e^{\mu_1+\mu_2}-e^{\lambda_1+\lambda_2}m_6^2)\,,
		\end{equation}
		\begin{equation}\label{eq:Xd}
			X=\sqrt{1-\xi^2} \,(\nu_1 \cos\varphi_1 \sin\theta + \nu_2 \cos\varphi_2 \cos\theta)\,.
		\end{equation}
		Then the six-dimensional Einstein metric is
		\begin{equation}\label{eq6dem}
			ds^2_6 = - \left(\mathrm{det} (m) \right)^{-1/2}\Delta^{1/2} \,ds^2_3+ R^2_{AdS} \left(\mathrm{det} (m) \right)^{1/2}\Delta^{-1/2} \left(G_{\alpha\beta}\, \mathcal{D}y^\alpha\, \mathcal{D} y^\beta\right)\,,
		\end{equation}
		where $ds^2_3$ is defined in \eqref{genmet1} and the components of the $S^3$ metric are
		\begin{subequations}
			\begin{equation}
				\begin{aligned}
					G_{\theta\theta} &= e^{\mu_2} \left[\frac{e^{-\lambda_1} \cos^2\varphi_1}{e^{\mu_1+\mu_2}-e^{-\lambda_1-\lambda_2}m_5^2}+\frac{e^{\lambda_1}\sin^2\varphi_1}{e^{\mu_1+\mu_2}-e^{\lambda_1+\lambda_2}m_6^2}\right] \cos^2\theta \\
					&+ e^{\mu_1} \left[\frac{e^{-\lambda_2} \cos^2\varphi_2}{e^{\mu_1+\mu_2}-e^{-\lambda_1-\lambda_2}m_5^2}+\frac{e^{\lambda_2}\sin^2\varphi_2}{e^{\mu_1+\mu_2}-e^{\lambda_1+\lambda_2}m_6^2}\right] \sin^2\theta \\
					&+  2\left[\frac{e^{-\lambda_1-\lambda_2} \,m_5 \cos\varphi_1 \cos\varphi_2}{e^{\mu_1+\mu_2}-e^{-\lambda_1-\lambda_2}m_5^2}+ \frac{e^{\lambda_1+\lambda_2} \,m_6 \sin\varphi_1 \sin\varphi_2}{e^{\mu_1+\mu_2}-e^{\lambda_1+\lambda_2}m_6^2}\right] \sin\theta\cos\theta\,,
				\end{aligned}
			\end{equation}
			\begin{equation}
				G_{\varphi_1\varphi_1}=e^{\mu_2} \left[\frac{e^{-\lambda_1}\sin^2\varphi_1}{e^{\mu_1+\mu_2}-e^{-\lambda_1-\lambda_2}m_5^2}+\frac{e^{\lambda_1}\cos^2\varphi_1}{e^{\mu_1+\mu_2}-e^{\lambda_1+\lambda_2}m_6^2}\right]\sin^2\theta\,,
			\end{equation}
			\begin{equation}
				G_{\varphi_2\varphi_2}=e^{\mu_1} \left[\frac{e^{-\lambda_2}\sin^2\varphi_2}{e^{\mu_1+\mu_2}-e^{-\lambda_1-\lambda_2}m_5^2}+\frac{e^{\lambda_2}\cos^2\varphi_2}{e^{\mu_1+\mu_2}-e^{\lambda_1+\lambda_2}m_6^2}\right]\cos^2\theta\,,
			\end{equation}
			\begin{equation}
				\begin{aligned}
					G_{\theta\varphi_1} &= e^{\mu_2} \left[-\frac{e^{-\lambda_1}}{e^{\mu_1+\mu_2}-e^{-\lambda_1-\lambda_2}m_5^2}+\frac{e^{\lambda_1}}{e^{\mu_1+\mu_2}-e^{\lambda_1+\lambda_2}m_6^2}\right] \sin\varphi_1\cos\varphi_1\sin\theta\cos\theta \\
					&+\left[-\frac{e^{-\lambda_1-\lambda_2} \,m_5 \sin\varphi_1 \cos\varphi_2}{e^{\mu_1+\mu_2}-e^{-\lambda_1-\lambda_2}m_5^2}+ \frac{e^{\lambda_1+\lambda_2} \,m_6 \cos\varphi_1 \sin\varphi_2}{e^{\mu_1+\mu_2}-e^{\lambda_1+\lambda_2}m_6^2}\right] \sin^2\theta\,,
				\end{aligned}
			\end{equation}
			\begin{equation}
				\begin{aligned}
					G_{\theta\varphi_2} &= e^{\mu_1} \left[\frac{e^{-\lambda_2}}{e^{\mu_1+\mu_2}-e^{-\lambda_1-\lambda_2}m_5^2}-\frac{e^{\lambda_2}}{e^{\mu_1+\mu_2}-e^{\lambda_1+\lambda_2}m_6^2}\right] \sin\varphi_2\cos\varphi_2\sin\theta\cos\theta \\
					&+\left[\frac{e^{-\lambda_1-\lambda_2} \,m_5 \cos\varphi_1 \sin\varphi_2}{e^{\mu_1+\mu_2}-e^{-\lambda_1-\lambda_2}m_5^2}- \frac{e^{\lambda_1+\lambda_2} \,m_6 \sin\varphi_1 \cos\varphi_2}{e^{\mu_1+\mu_2}-e^{\lambda_1+\lambda_2}m_6^2}\right] \sin^2\theta\,,
				\end{aligned}
			\end{equation}
			\begin{equation}
				G_{\varphi_1\varphi_2}=-\left[\frac{e^{-\lambda_1-\lambda_2} m_5\,\sin\varphi_1\sin\varphi_2}{e^{\mu_1+\mu_2}-e^{-\lambda_1-\lambda_2}m_5^2}+\frac{e^{\lambda_1+\lambda_2} m_6\,\cos\varphi_1\cos\varphi_2}{e^{\mu_1+\mu_2}-e^{\lambda_1+\lambda_2}m_6^2}\right]\sin\theta\cos\theta\,,
			\end{equation}
		\end{subequations}
		and the $S^3$ one-forms are
		\begin{equation}
			\mathcal{D} y^\alpha = d y^\alpha -g_0 \, K^\alpha_{IJ} \,\tilde A^{IJ}\,,
		\end{equation} 
		with the $S^3$ Killing vectors ($K^\alpha_{IJ}=-K^\alpha_{JI}$)
		\begin{equation}\label{eq:killingvecbis}
			\begin{aligned}
				&K_{12}=\frac{\partial}{\partial \varphi_1}\,,\,\,K_{34}=\frac{\partial}{\partial \varphi_2}\,,\\
				&K_{13}=\cos\varphi_1\cos\varphi_2 \,\frac{\partial}{\partial \theta}-\sin\varphi_1\cos\varphi_2 \cot\theta\,\frac{\partial}{\partial \varphi_1}+\cos\varphi_1\sin\varphi_2\tan\theta\, \frac{\partial}{\partial \varphi_2}\,,\\
				&K_{14}=\cos\varphi_1\sin\varphi_2 \,\frac{\partial}{\partial \theta}-\sin\varphi_1\sin\varphi_2 \cot\theta\,\frac{\partial}{\partial \varphi_1}-\cos\varphi_1\cos\varphi_2\tan\theta\, \frac{\partial}{\partial \varphi_2}\,,\\
				&K_{23}=\sin\varphi_1\cos\varphi_2 \,\frac{\partial}{\partial \theta}+\cos\varphi_1\cos\varphi_2 \cot\theta\,\frac{\partial}{\partial \varphi_1}+\sin\varphi_1\sin\varphi_2\tan\theta\, \frac{\partial}{\partial \varphi_2}\,,\\
				&K_{24}=\sin\varphi_1\sin\varphi_2 \,\frac{\partial}{\partial \theta}+\cos\varphi_1\sin\varphi_2 \cot\theta\,\frac{\partial}{\partial \varphi_1}-\sin\varphi_1\cos\varphi_2\tan\theta\, \frac{\partial}{\partial \varphi_2}\,.
			\end{aligned}
		\end{equation}
		All the scalars can be reconstructed from the dilaton and axion, which are given by
		\begin{equation}\label{eq:asxD}
			e^{2\Phi}=\frac{Q_1}{4 Q_5}\,\frac{(X^2+2\Delta)^2}{\Delta}\quad,\quad C_0 = \sqrt{\frac{2Q_5}{Q_1}}\,\frac{X}{X^2+2\Delta}\,.
		\end{equation}
		Formulas for the three-forms are considerably more cumbersome and are not needed in this article. We will only give the results that are used in the holographic computation of Sections \ref{sub:onequat} and \ref{sub:oneeig}, i.e. the components, $G^{(I)}|_{S^3}$, of the three-forms along the $S^3$ (divided by the volume of the round $S^3$) for the restricted ansatz with $\mu_2=\lambda_2=\nu_2=m_5=m_6=\Phi_3=\Psi_3=\Phi_4=\Psi_4=0$:
		\begin{subequations}\label{eq:GS3}
			\begin{equation}
				G^{(1)}|_{S^3}=\frac{Q_5\,e^{\mu_1}}{\Delta}\left[\frac{e^{\lambda_1}+e^{-\lambda_1}}{2}-\frac{e^{\mu_1}(e^{2\lambda_1}\cos^2\varphi_1+e^{-2\lambda_1}\sin^2\varphi_1)-e^{\mu_2}\Gamma_1}{\Delta}\sin^2\theta\right]\,,
			\end{equation}
			\begin{equation}
				\begin{aligned}
					G^{(2)}|_{S^3}=&\,\,Q_1-\frac{Q_1}{2} (1-\xi^2)\nu_1^2 \Bigl[e^{\mu_2}\left(\frac{\cos2\theta}{\Delta}-\frac{e^{\mu_1} \Gamma_1 -e^{\mu_2}}{\Delta^2}\sin^2\theta\cos^2\theta \right)\cos^2\varphi_1\\
					&-\frac{1}{4} \frac{\partial}{\partial \varphi_1} \left( \frac{\sin(2\varphi_1)}{\Delta}\right)(e^{\mu_1-\lambda_1}\sin^2\theta+e^{\mu_2}\cos^2\theta) \Bigr]\,, 
				\end{aligned}
			\end{equation}
			\begin{equation}
				G^{(7)}|_{S^3}=X\frac{e^{\mu_1}}{2\sqrt{2} \Delta}\left[e^{-\lambda_1}-2\frac{(e^{\mu_1}+e^{\mu_2}\Gamma_1)\sin^2\theta+e^{\mu_2}(e^{\lambda_1}+e^{-\lambda_1})\cos^2\theta)}{\Delta} \right]\,.
			\end{equation}
		\end{subequations}
		
		\section{\texorpdfstring{$\mathbb{Z}_2$}{Z2} symmetric \texorpdfstring{$\beta$}{beta}-class equations for numerics}
		\label{app:betaNumericsEqs}

		Here are the equations for the $\mathbb{Z}_2$ symmetric case of the $\beta$-class of solutions - that is, $n_1=n_2=n$, $\beta_1=\beta_2=\beta$ and $\mu_1=\mu_2$, $\lambda_1=\lambda_2$, $\Phi_1=\Phi_2$ and $\Psi_1=\Psi_2$. For numerical convenience, we have defined $k=\xi\,\kappa$. The equations follow:
		{\small
		\begin{gather}
			E_{\mu_1}\equiv\xi\,(1-\xi^2)^2\,\Omega_0^2\,\Omega_1\,\partial_\xi\big(\xi\,\Omega_1\,\mu_1'\big)-4\,e^{2\,\mu_1}(1-\xi^2)^2\,\Omega_1^2\big(\xi\,\kappa\,\Phi_1'-(1-\xi^2)\,\Psi_1'\big)^2\notag\\+4\,e^{2\,\mu_1}\,\xi^2(1-\xi^2)^2\Omega_0^2\,\Phi_1'^2 +4\,e^{-4\,\mu_1}\xi^2\,\Omega_0^4\,\Omega_1^2\,\big(1-e^{2\,\mu_1}\big)=0,\notag\\[2ex]
			E_{\lambda_1}\equiv\xi\,(1-\xi^2)^2\,\Omega_1\,\partial_\xi\big(\xi\,\Omega_1\,\lambda_1'\big)-2\,\sinh\big(2\,\lambda_1\big)\Big(4\,\Omega_1^2\,\big(\xi\,\kappa\,\Phi_1-(1-\xi^2)\,\Psi_1\big)^2-4\,\xi^2\Omega_0^2\,\Phi_1^2\Big)=0,\notag\\[2ex]
			E_{\Phi_1}\equiv-4\,\xi^2\,\sinh\big(\lambda_1\big)^2\,\Phi_1\,\Omega_0^4\,\Omega_1+e^{2\,\mu_1}\,\xi\,(1-\xi^2)^2\,\Omega_0^2\,\Big(-\xi\,\Phi_1'\,\Omega_1'+\Omega_1\,\big((1+2\,\xi\,\mu_1')\,\Phi_1'+\xi\,\Phi_1''\big)\Big)\notag\\
			-(1-\xi^2)\,\Omega_1^2\,\Big[-2\,\xi\,\Omega_0^2\,\big(\xi\,\kappa\,\Phi_1'-(1-\xi^2)\,\Psi_1'\big)+e^{2\,\mu_1}\,\Omega_1\,\big((1+\xi^2)\,\kappa+\xi\,(1-\xi^2)\,\kappa'\big)\big(\xi\,\kappa\,\Phi_1'-(1-\xi^2)\,\Psi_1'\big)\Big]=0,\notag\\[2ex]
			E_{\Psi_1}\equiv-\kappa\,\Omega_1^2\,\Big[-2\,\xi\,\Omega_0^2\,\big(\xi\,\kappa\,\Phi_1'-(1-\xi^2)\,\Psi_1'\big)+e^{2\,\mu_1}\,\Omega_1\,\big((1+\xi^2)\,\kappa+\xi\,(1-\xi^2)\,\kappa'\big)\big(\xi\,\kappa\,\Phi_1'-(1-\xi^2)\,\Psi_1'\big)\Big]\notag\\
			+e^{2\,\mu_1}\,\Omega_0\,\Phi_1'\,\Big[-\xi\,(1-\xi^2)\,\Omega_0\,\Omega_1\,(\kappa+\xi\,\kappa')+2\,\xi\,\kappa\,\Big(\xi\,(1-\xi^2)\,\Omega_1\,\Omega_0'+\Omega_0\,\big(\Omega_1-\xi\,(1-\xi^2)\,\Omega_1'\big)\Big)\Big]\notag\\
			-e^{2\,\mu_1}\,(1-\xi^2)\,\Omega_0\,\Big[2\,\xi\,(1-\xi^2)\,\Omega_1\,\Psi_1'\,\Omega_0'+\Omega_0\Big(\Psi_1'\,\big(\Omega_1\,(1+3\,\xi^2-2\,\xi(1-\xi^2)\,\mu_1')-\xi\,(1-\xi^2)\,\Omega_1'\big)\notag\\ -\xi\,(1-\xi^2)\,\Omega_1\,\Psi_1''\Big)\Big]
			-2\,\xi^2\,\Omega_0^4\,\Phi_1'-4\,\xi\sinh\big(\lambda_1\big)^2\,\Psi_1\,\Omega_0^4\,\Omega_1=0,\notag\\[2ex]
			E_{\Omega_0}\equiv2\,\xi\,\Omega_1\Big[-\xi(1-\xi^2)^2\,\Omega_1\,\Omega_0'^2+\Omega_0^2\big(4\,\xi\,\Omega_1+(1-\xi^4)\,\Omega_1'\big)+(1-\xi^2)^2\,\Omega_0\big(\Omega_0'(\Omega_1+\xi\,\Omega_1')+\xi\,\Omega_1\Omega_0''\big)\Big]\notag\\
			-\Omega_1^4\,\big((1+\xi^2)\kappa+\xi(1-\xi^2)\kappa'\big)^2+32\,\Omega_0^2\,\Omega_1^2\,\sinh\big(\lambda_1\big)^2\big(\xi\,\kappa\,\Phi_1-(1-\xi^2)\,\Psi_1\big)^2+8\,(1-\xi^2)^2\xi^2\,\Omega_0^2\,e^{2\,\mu_1}\,\Phi_1'^2\notag\\
			+4\,e^{-4\,\mu_1}\,\xi^2\,\Omega_0^4\,\Omega_1^2\,\big(1-2\,e^{2\,\mu_1}\big)=0,\notag\\[2ex]
			E_{\Omega_1}\equiv-32\,\xi^2\,\sinh\big(\lambda_1\big)^2\,\Phi_1^2\,\Omega_0^4+4\,e^{-4\,\mu_1}\,\xi^2\,\Omega_0^4\,\Omega_1^2\,\big(1-2\,e^{2\,\mu_1}\big)+\Omega_1^4\big((1+\xi^2)\,\kappa+\xi\,(1-\xi^2)\,\kappa'\big)^2\notag\\
			-8\,(1-\xi^2)^2\,\Omega_1^2\,e^{2\,\mu_1}\,\big(\xi\,\kappa\,\Phi_1'-(1-\xi^2)\,\Psi_1'\big)^2+2\,\xi\,(1-\xi^2)^2\Omega_0^2\,\Omega_1\,\big(\Omega_1'+\xi\,\Omega_1''\big)=0,\notag\\[2ex]
			E_{\kappa}\equiv32\,\xi\,\sinh\big(\lambda_1\big)^2\,\Phi_1\,\big(\xi\,\kappa\,\Phi_1-(1-\xi^2)\,\Psi_1\big)\,\Omega_0^3+8\,\xi\,(1-\xi^2)^2\,\Omega_0\,e^{2\,\mu_1}\Phi_1'\,\big(\xi\,\kappa\,\Phi_1'-(1-\xi^2)\,\Psi_1'\big)\notag\\
			-(1-\xi^2)\,\Omega_1\,\Big[2\,\xi\,\Omega_1\,\big((1+\xi^2)\,\kappa+\xi\,(1-\xi^2)\,\kappa'\big)\Omega_0'+\Omega_0\big(-3\,\xi\,\big((1+\xi^2)\kappa+\xi\,(1-\xi^2)\,\kappa'\big)\,\Omega_1'\notag\\
			-(1-\xi^2)\,\Omega_1\,\big(-\kappa+\xi\,(\kappa'+\xi\,\kappa'')\big)\big)\Big]=0.\label{eq:betaZ2Eqs}
		\end{gather}
		}
		These are the equations we integrate numerically. However, one can show that a linear combination of $E_{\Phi_1}$, $E_{\Psi_1}$ and $E_{\kappa}$ reduces to a first order equation, namely:
		\begin{equation}
			\frac{(1-\xi^2)^2}{\xi}\frac{\Omega_1^3}{\Omega_0^2}\,\partial_\xi\bigg(\frac{\xi\,\kappa}{1-\xi^2}\bigg)+8\frac{e^{2\,\mu_1}\,(1-\xi^2)^2\,\Omega_1\,\Big(\frac{\xi\,\kappa}{1-\xi^2}\,\Phi_1'-\Psi_1'\Big)}{\xi\,\Omega_0^2}\Phi_1+8\,\Phi_1^2-c_\kappa=0,\label{eq:BetaZ2firstPsiEq}
		\end{equation}
		where $c_\kappa$ is a constant of integration. Equation \eqref{eq:BetaZ2firstPsiEq} generalizes to the $n_1\neq n_2$ $\beta$-class and we treat it as the equation of motion for the $\Psi_1$ gauge field, implying that the latter is governed by a first-order differential equation. It is not related to the 3 constraint equations, discussed at the beginning of Section~\ref{sec:nonBPS}. The latter reduce to a single constraint for the $\beta$-class(for general $n_{1,2}$), as the ones coming from the Maxwell equations are automatically satisfied. Only the Einstein constraint remains non-trivial and we omit it here, as it is rather cumbersome. We use it to monitor the convergence of our numerics.

		
		\begin{adjustwidth}{-1mm}{-1mm} 

			\bibliographystyle{utphys}
			
			\bibliography{microstates}       

		\end{adjustwidth}

	\end{document}